\documentclass[aps,prd,twocolumn,showpacs,superscriptaddress,groupedaddress,nofootinbib]{revtex4-1}  

\RequirePackage{ae}
\RequirePackage{bm} 
\RequirePackage{color}
\RequirePackage{amssymb}
\RequirePackage{amsmath}
\RequirePackage{amsfonts}
\RequirePackage{mathtools}
\RequirePackage{graphicx}
\RequirePackage{slashed}
\RequirePackage{wasysym}
\RequirePackage{setspace}
\RequirePackage{subfigure}
\RequirePackage{ulem}
\RequirePackage{placeins}

\RequirePackage{tabularx}
\RequirePackage{algpseudocode}
\RequirePackage{nicefrac}
\RequirePackage{bbm}
\RequirePackage{calligra}
\RequirePackage{microtype}
\RequirePackage{flafter}

\RequirePackage{ctable}
\usepackage{capt-of}

\newcommand{\dns}[1]{} 

\newcommand*{\I}{ {\rm i} }
\newcommand*{\ee}{ {\rm e} }

\DeclareMathAlphabet{\mathcalligra}{T1}{calligra}{m}{n}

\newlength{\figlen}
\newlength{\figlenB}
\newlength{\figlenC}
\newlength{\figlenD}
\newlength{\tablen}
\newcount\Bool	
\def\eq$#1${\begin{multline}#1\end{multline}}

\newcommand{\txt}[2]{
  \ifnum\Bool=1 {#1} 
  \else	{#2}
  \fi
}

\begin{document} \sloppy

\title{On the effect of time-dependent inhomogeneous magnetic fields on the particle momentum spectrum in electron-positron pair production}

\author{Christian Kohlf\"urst}
\email{c.kohlfuerst@hzdr.de}
\affiliation{Helmholtz-Zentrum Dresden-Rossendorf, Bautzner Landstra{\ss}e 400, 01328 Dresden, Germany}

\begin{abstract}

Electron-positron pair production in spatially and temporally inhomogeneous electric and magnetic fields is studied within
the Dirac-Heisenberg-Wigner formalism (quantum kinetic theory) through computing the corresponding Wigner functions. 
The focus is on discussing the particle momentum spectrum regarding 
signatures of Schwinger and multiphoton pair production.  Special emphasis is put on studying the impact of a strong dynamical magnetic field
on the particle distribution functions.
As the equal-time Wigner approach
is formulated in terms of partial integro-differential equations an entire section of the manuscript 
is dedicated to present numerical solution techniques applicable to Wigner function approaches in general.

 
\end{abstract}

\maketitle

\section{Introduction}

The creation of matter out of the vacuum is one of the most exciting concepts of high-energy physics.
In particular strong-field quantum electrodynamics (QED) is perfectly suited to study the generation of matter through energy,
because it provides a comparatively clean setting \cite{Heisenberg:1935qt, Schwinger:1951nm, Sauter:1931zz}.


The main issue is to create the right laboratory conditions as it takes
extremely strong field strengths in order to see any form of signal of quantum vacuum 
non-linearities \cite{Lundstrom:2005za, Gies:2017ezf, Dinu:2013gaa, Dinu:2014tsa}.
However, as laser technology has significantly advanced over the last years probing the quantum vacuum
in an earth-based laboratory seems to be in reach \cite{Heinzl:2008an, Ringwald:2001ib, Marklund:2008gj}. In fact, the research field has recently gained interest  
to the extent that upcoming laser facilities already prepare for high-intensity experiments
\cite{LasersB, EPPlasma, Turcu:2015cca}. 

In the context of particle creation, it has been shown already that matter can indeed be created in a lepton-photon collider. It was demonstrated that highly energetic
electrons, if subjected to huge electric fields, emit photons in the multi-keV regime which, in turn, interact with the light beam itself eventually forming an electron-positron pair
\cite{Burke:1997ewA, Burke:1997ewB}.
On the other hand, the Schwinger effect the depletion of the quantum vacuum in constant background fields, has been predicted 90 years ago \cite{Sauter:1931zz}, but still remains untested.
Only recently it was suggested that both mechanism could be utilized together in a multi-beam scenario leading, in theory, to a tremendous increase in the creation rate \cite{PhysRevLett.101.130404}. 
For further reviews on pair production in general see Refs. \cite{reviewA, reviewC, reviewE}.

Due to the highly complex nature of the subject, only a few analytical results can be found \cite{Narozhnyi, Dunne:2004nc, Brezin:1970xf}. 
Hence, we have to rely on numerical approaches in order to describe pair production accurately.
Over the timespan of approximately $20$ years we have seen the rise of an abundance of computational methods, all performing very well in certain parameter 
regions \cite{Blinne:2015zpa,Jansen:2016crq,Kluger:1992md, Acosta:2019bvh}.
One common trait, however, was that they worked best for homogeneous electric fields as it was argued that the chances to produce particles should be highest in the vicinity of
the focus of two crossed beams where the magnetic field vanishes. In particular the so-called quantum kinetic theory (QKT) proved to be very successful
under these special circumstances \cite{Smolyansky:1997fcA, Smolyansky:1997fcB, Smolyansky:1997fcC, Smolyansky:1997fcD}.
Only recently, numerical solvers have been improved to the point that it is now possible to take into account the spatial inhomogeneity
of laser beams \cite{Ruf:2009zz,Hebenstreit:2011wk, Gies:2005bz, Aleksandrov:2016lxd, Aleksandrov:2018uqb, Lv:2017qpx, Jiang:2014bwa, Ababekri:2019dkl} beyond any 
semi-classical approximations \cite{SemiClassA, SemiClassC, SemiClassF}. 
Performing calculations with these tools revealed additional features of strong field pair production;
particle self-bunching \cite{Hebenstreit:2011wk}, ponderomotive effects \cite{Kohlfurst:2017hbd} and spin-field interactions \cite{Kohlfurst:2017git}. Furthermore
it was shown that disregarding the spatial finiteness of a laser pulse can lead to spurious effects in the particle spectrum \cite{Aleksandrov:2017mtq, Lv:2018wpn}.

In order to fully account for the spatial inhomogeneity of background fields we rely on the so-called Dirac-Heisenberg-Wigner (DHW) formalism \cite{Vasak:1987umA, Vasak:1987umB, BB}.
Its major advantage is its flexibility, because the formalism allows to study the pair production process for any given 
background field. Once the structure of electric and magnetic fields has been established the formalism acts as a black box automatically
incorporating all possible production and interactions mechanisms. Furthermore, as the DHW formalism is deduced a priori from a Lagrangian level no models or assumptions are needed.
The only drawback is that a computational solver based on the Wigner formalism is generally very resource-hungry, as one solves for the time evolution of a multi-dimensional partial 
differential equation. To overcome these issues in order to make simulations feasible novel computational techniques have to be applied.  

Although various articles on the Wigner formalism and its applications have been published in recent years \cite{Chernodub:2017bbd, Wang:2019moi, Sheng:2017lfu, Gorbar:2017awz}, 
a compendium containing methods and techniques to successfully treat the corresponding equations of motion with computational methods is not available. 
The intention of this article therefore is to introduce various computational solution techniques that are applicable to solving them. 
In this context, we discuss caveats of the formalism and bring up ways to overcome them. This is done in addition
to discussions on the solvers' effectiveness and overall performance including considerations on noise suppression.


We apply all these novel techniques in order to compute pair production in a regime where tunneling as well as absorption effects are important.
In particular, we study the impact the magnetic field has on the production yield and thus on the particles' momentum spectrum.
Furthermore, as the whole creation process is very sensitive to variations in the background field, e.g., a small change in one parameter
can mean a different creation mechanism becomes favored thus heavily altering the spectrum, we rigorously analyze slightly different
field configurations eventually improving our understanding of magnetic field effects in particle creation scenarios in general.

In short, this manuscript is organized as follows. We first introduce a quite general model of a background field, which serves as the connection point for 
the discussion on the characteristics of 
the various production mechanism, see section \ref{Sec:PP}. In section \ref{Sec:DHW} we introduce the DHW formalism and in section \ref{Sec:Numerical} we
describe how to solve the governing set of equations of motion efficiently. In the main part of this manuscript, section \ref{Sec:Results},
the numerical results are discussed.
Eventually, a conclusion is given in section \ref{Sec:Conclusion}.

Throughout this paper we use natural units $\hbar=c=1$ and express all quantities in terms
of the electron mass $m$ ($= m_{e^+} = m_{e^-}$).

\section{Motivation}
\label{Sec:PP}

      \begin{figure}[t]
      \begin{center}
	\includegraphics[width=\figlenB]{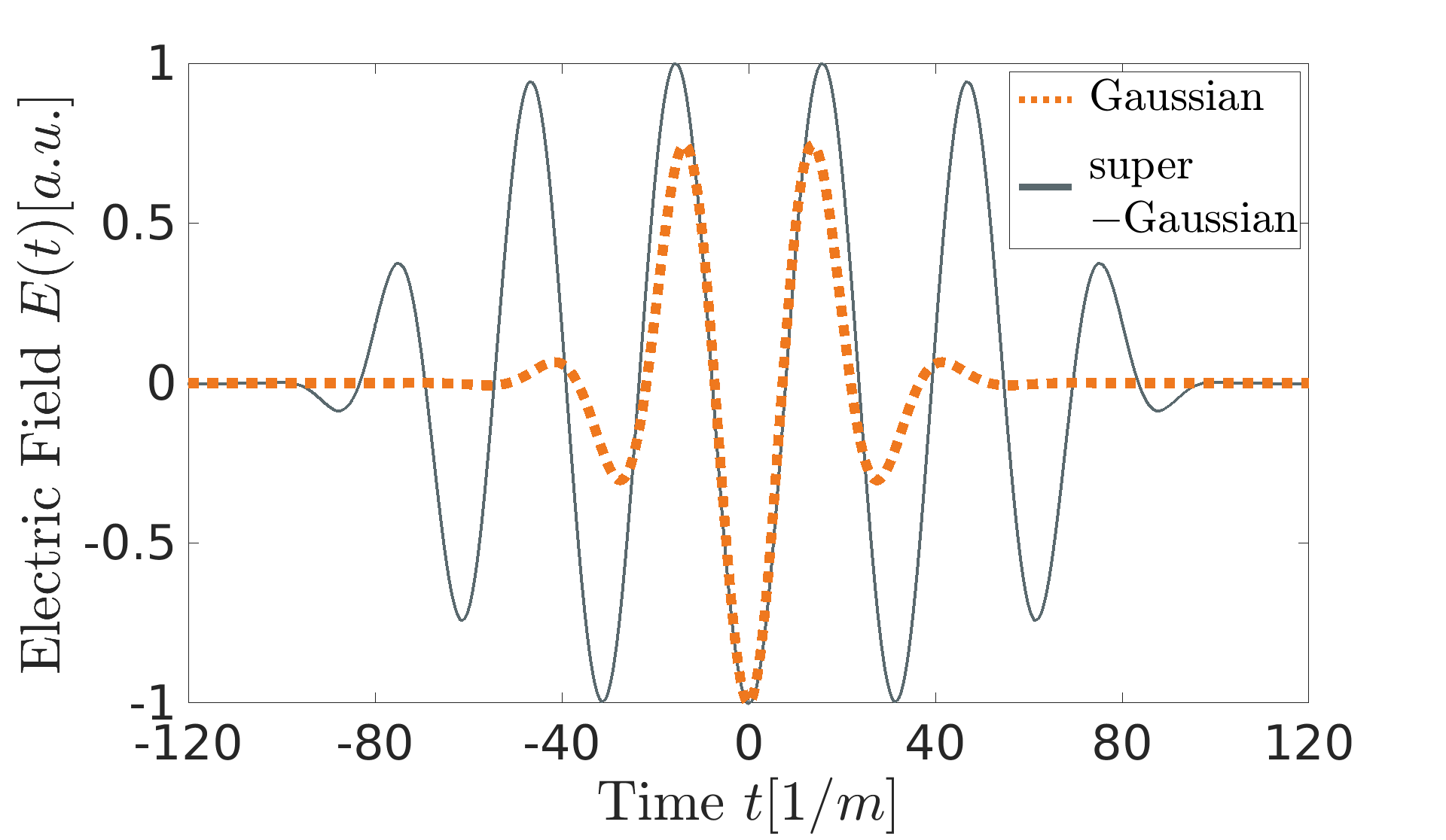} 		
      \end{center}
      \caption{Illustration of the two different temporal envelopes used in the manuscript. For pulse lengths of $\tau=25m^{-1}$ (Gaussian)
      and $\tau=75m^{-1}$ (super-Gaussian) we find that around $\tau \approx 125m^{-1}$ both fields take on similar values.}
      \label{fig:Fields}
      \end{figure} 

A virtual electron-positron pair is formed due to the omnipresence of vacuum fluctuations, where  
the scale is determined by the Compton time 
$\tau_C = 1/m_{e^-} \approx 10^{-21}$s and Compton length of an, in this case, electron $\lambda_C = 1/m_{e^-} \approx 10^{-12}$m.

The main task is to prevent these two particles from recombining. In this way, a virtual pair is turned into two separate particles. 
In order to do so, a strong electromagnetic field is employed acting on the charge carriers. There are now two
different ways to pump energy into the system. Either via providing enough work to allow for tunneling (Schwinger effect)
or due to the absorption of energy given by the individual photons forming the background field (multiphoton pair production). These effects
are not mutually exclusive, thus depending on the background field any combination of mechanisms is possible.

The probabilities for such an event to happen scale with the critical field strength
\begin{equation}
 E_{\rm cr} = \frac{m_e^2}{e} \approx 1.3 \times 10^{16} \, \mathrm{V/cm}.
\end{equation}
It is defined as the minimal work that has to be done by the field over one Compton length in order to create one pair $2 e E_{\rm cr} \lambda_C = m_{e^-} + m_{e^+}$.
The factor of $2$ is due to the fact, that electrons and positrons have opposite charge thus they are accelerated in opposite directions, too.

\subsection{Model for the fields}
\label{Sec:Model}

The ultimate goal of this article is to study the consequences of applying an additional magnetic field to the particle creation process and investigate how the different creation mechanism
are affected by it. As we want to smoothly interpolate between Schwinger and multiphoton pair production we employ a toy model of the form of
\begin{multline}
 \mathbf A (t,z) = \\
   \frac{\varepsilon}{\omega} \ \exp \left( -\frac{z^2}{\lambda^2} \right) \ \exp \left( -\frac{t^j}{\tau^j} \right) \sin \left( \omega t \right) \mathbf{e}_x \label{equ:A}
\end{multline}
and derive electric and magnetic fields accordingly
\begin{multline}
 \mathbf E (t,z) = -\partial_t \mathbf A (t,z) = \\
   \frac{\varepsilon}{\omega} \ \exp \left( -\frac{z^2}{\lambda^2} \right) \ \exp \left( -\frac{t^j}{\tau^j} \right) \hspace{2.5cm} \\
   \times \frac{\Big( jt^{j-1} \sin \left(\omega t \right) -\omega \tau^j \cos \left( \omega t \right) \Big)}{\tau^j} \ \mathbf{e}_x, \label{equ:E} 
\end{multline}
\begin{multline}
 \mathbf B (t,z) = \boldsymbol \nabla \times \mathbf A (t,z) = \\
  -\frac{\varepsilon}{\omega} \ \exp \left( -\frac{z^2}{\lambda^2} \right) \ \exp \left( -\frac{t^j}{\tau^j} \right) 
   \ \frac{ 2z \sin \left( \omega t \right) }{\lambda^2} \ \mathbf{e}_y. \label{equ:B}
\end{multline}
Here, $\varepsilon$ gives the peak field strength in terms of the critical field strength $m^2/e$, $\omega$ states the field frequency and $\tau$
describes the pulse length. Moreover, we have decided to use a variable envelope function, $\exp \left( -\dfrac{t^j}{\tau^j} \right)$, in order to have
better control over the number of cycles in the field. This choice makes it indeed very easy to switch from few-cycle to prolonged many-cycle pulses, c.f. Fig. \ref{fig:Fields}.
The parameter $\lambda$, on the other hand, controls the extent of the peak in spatial direction and indirectly also defines the peak strength of the magnetic field,
where a smaller $\lambda$ accounts for higher magnetic field strengths.

Technically, Eqs. \eqref{equ:A}-\eqref{equ:B} describe a head-on collision of two identical, linearly polarized laser beams 
within a quasi-dipole approximation. To be more specific, we have two non-propagating pulses $\ee^{\pm \I k_z z} \approx 1 + \mathcal O \left(k_z \right)$ 
each consisting of left- and right-handed waves that form a temporally and spatially localized interference pattern mimicking the region of highest intensity
in a realistic setup. In this context, we can therefore interpret the background field as incoming spherical waves with angular momentum $j=1$ (electric dipole). 
Moreover, the electric field \eqref{equ:E} still accounts for the individual photon energies in the form of the field frequency $\omega$.

Naturally, one were also to assume that photons would transfer linear momentum to the created particles. 
Our model \eqref{equ:A}, however, does not take into account the photon's linear momentum
as we are only interested in basic particle behavior in presence of a strong magnetic field. In order to allow for a non-zero photon momentum one would have to go beyond
the dipole approximation and examine pair production in, e.g., a standing wave
\begin{equation}
 A (t,z) \sim \sin \left(\omega t \right) \cos \left(\omega z \right).
\end{equation}
One of the reasons we have decided to not use such a model is that a standing wave pattern creates the opportunity for simultaneous pair production at a multitude of 
locations. This would render any analysis on specific details of the production process unclear, because in a phase-space approach every bunch of particles would
be stacked on top of each other leading to a confusing particle spectrum.

Given the definition of the background field \eqref{equ:E}-\eqref{equ:B} we can already 
discriminate between the Schwinger effect and multiphoton pair production using the Keldysh parameter \cite{Keldysh}
\begin{align}
 \gamma = \frac{m \omega}{e\varepsilon} \begin{cases} \ll 1 \qquad {\rm Schwinger} \\ \gg 1 \qquad {\rm Multiphoton} \end{cases}
\end{align}
As $\gamma$ was first introduced for infinitely long fields it does not take the temporal envelope into account. 
In our case, this means that in addition to the choice of $\varepsilon$ and $\omega$ the shape of the envelope function 
is a major contributing factor.

\subsection{Regimes of Pair Production}    
\label{Sec:Regimes}

In the following, we will briefly introduce the different regimes of pair production (Schwinger, multiphoton and intermediate) and
discuss their characteristic signatures in the momentum spectrum.

\subsubsection{The Schwinger effect}
\label{Par:Schwinger}

The Schwinger regime applies when a strong background field is employed over a sufficiently long period of time in order to allow the particle to tunnel
through the Coulomb barrier. As we work exclusively with linearly polarized fields we can adopt the notion in 
Refs. \cite{Casher:1979gwB, Casher:1979gwA} (describing the conceptually similar $q \bar q$-string breaking) yielding
\begin{equation}
 P \left(t , z, p_z \right) = \frac{e a \left(t , z \right)}{4 \pi^2} \exp \left(- \frac{\pi \left(m^2 + p_z^2 \right)}{e a\left(t , z \right)} \right), \label{equ:P}
\end{equation}
where $a=\sqrt{ \frac{1}{2} \big( \lvert B^2-E^2 \rvert - \left( B^2-E^2 \right) \big) }$.
The key aspects of the probability rate $P \left(t , z, p_z \right)$ are, that (i) particles are produced at times when the electric field is strong and
(ii) that these particles have zero initial parallel momentum $p_x$. A non-vanishing initial transversal momentum $p_z$ is possible. However, it comes 
with a penalty in terms of likeliness, see the exponential term $m^2 + p_z^2$. 

After creation, these particles are exposed to strong forces due to the extreme strength of the background field. As we are primarily interested in
the momentum spectrum at final times (when the electromagnetic field is switched off) we have to account for these field-particle interactions. The simplest
way to achieve such a description is by applying a single-trajectory formalism, where electrons/positrons are considered to be semi-classical point-like particles, which
follow classical paths \cite{Kohlfurst}. Excluding any kind of interaction between generated particles, thus naturally ignoring quantum interferences and particle collisions, these
trajectories can then be tracked and evaluated. 

Based on the works in Refs.~\cite{Silenko:2007wi, MengWenA, MengWenB}, in Ref.~\cite{Kohlfurst:2017git} such a 
semi-classical formalism has been established\footnote{In this specific article we neglect spin-orbit interactions
due to its small effect.}
\begin{alignat}{5}
 & \frac{\partial z}{\partial t} &&= p_z \left(t\right) , \label{equ:sin1} \\
 & \frac{\partial p_x}{\partial t} &&= eE \left(t,z(t) \right) &&- \frac{p_z \left(t\right) \ eB \left(t,z(t) \right)}{\gamma \left(t\right)}, \label{equ:sin2} \\
 & \frac{\partial p_z}{\partial t} &&= &&+\frac{p_x \left(t\right) \ eB \left(t,z(t) \right)}{\gamma \left(t\right)} + s \frac{\partial_z eB \left(t,z(t) \right)}{\gamma \left(t\right)}, \label{equ:sin3}
\end{alignat}
with the Lorentz factor $\gamma (t) = \sqrt{1 + p_x^2 + p_z^2}$ and the particle spin $s$.

We can already deduce that in case of a quasi-homogeneous configuration and therefore a vanishing magnetic field the particle trajectory can 
be easily calculated
\begin{align}
 p_x(t) &= A(t_0) - A(t), \\
 p_z(t) &= p_{z,0} = {\rm const},
\end{align}
where $t_0$ gives the particle's time of creation.

If the temporal structure of the electric field exhibits multiple peaks  
it essentially acts as a double-slit in time reflecting the particle's quantum statistics, see Refs. \cite{Kaminski:2018ywj,Hebenstreit:2009km} for a more profound discussion. 
Alternatively, one can interpret the final distribution as the sum over trajectories with the particles initial time of existence as starting points \cite{Kohlfurst:2017git}. 
Quantum interferences are then obtained through summing up the quantum phases these particles have picked up when exposed to the background field \cite{Becker}.

If a strong magnetic field is present, however, these trajectories are altered significantly. Speaking in terms of Eqs.~\eqref{equ:sin1}-\eqref{equ:sin3},
as soon as a particle has acquired a parallel momentum $p_x$, the magnetic force automatically converts it to transversal momentum $p_z$.
As such the particle is pushed into direction $\pm z$. This happens the quicker the stronger the magnetic field is. Interestingly,
if the particle leaves the strong-field region at times of high acceleration, it cannot be pushed back once it reaches
regions of low field strengths. As a result this particle is basically unaffected by variations of the background field at later times.

The spin force term $F_S = s \ \partial_z eB \left(t,z(t) \right)$ in Eq.~\eqref{equ:sin3} is an interesting addition to the classical Lorentz force equations,
because it allows for symmetry-breaking phenomena. In the vein of Ref. \cite{Kohlfurst:2017git}, we have decided to work in $2+1$ dimensions in order to keep the volume of the phase-space on a numerically manageable level. 
As a side effect, there are multiple ways to define QED (reducible or irreducible representation). To keep the calculations as simple as possible
we have opted for one of the $2-$spinor representation. As a result, the spin in our calculations is fixed, thus a spatially rapidly varying magnetic field can indeed break the symmetry in 
transversal direction.

\subsubsection{Multiphoton pair production}
\label{Par:Multi}

Contrary to the Schwinger effect, multiphoton pair production relies on a sufficient energy transfer from photons to virtual pairs. Moreover, electrons and positrons quiver within
an oscillating background field resulting in an increased potential energy. Only if the amount of transfered energy exceeds the pairs rest energy, a particle and an antiparticle is formed. 
As energy has to be conserved, the surplus of photon energy is directly converted to the particle's initial kinetic momentum. All in all, we can define the effective energy of a particle as \cite{Otto:2014ssa}
\begin{multline}
 {\cal E} \left(p_x, p_z \right) = \\
 \frac{\omega}{2 \pi} \int_{-\pi/\omega}^{\pi/\omega} {\rm d}t' \ \sqrt{ m^2 + \left( p_x + \frac{e\varepsilon}{\omega} \sin \left(\omega t' \right) \right)^2 + p_z^2}.
\end{multline}
Correspondingly, energy conservation reads $2 {\cal E} \left(p_x, p_z \right) = n\omega$ with $n$ being the number of absorbed photons.

\subsubsection{Intermediate regime}

For few-cycle pulses with rapidly varying time-dependency or for background fields exhibiting a Keldysh parameter of $\gamma \approx 1$ neither
Schwinger nor multiphoton effects dominate. In such a case either no distinct signatures of Schwinger nor multiphoton effects are visible or 
the particle spectrum shows characteristics of both.
Conceptually, it is, for example, easily possible, that a virtual particle pair absorbs $n$ photons and is simultaneously pulled apart by the strong background field
resulting in interferences showing up in the particle spectrum. 
Interactions of this kind are difficult to calculate, thus they make the perfect case study to demonstrate the power of kinetic approaches. 

On a side note, the so-called dynamically assisted mechanism, where photon absorption is used to effectively lower the threshold for pair production, is quite similar. 
The main difference is that the enhancement effect is much more pronounced, 
because multiple fields are fine-tuned in order to give
a strong signal in the spectrum \cite{PhysRevLett.101.130404, Torgrimsson:2017cyb, Torgrimsson:2016ant, Panferov:2015yda}. 

\section{DHW formalism}
\label{Sec:DHW}

The DHW formalism is a very general method that allows to study particle creation within arbitrary electromagnetic fields. In the context of this article,
this means we only have to develop one single numerical algorithm to study the spectra produced through few-cycle as well as many-cycle pulses. 

As providing a detailed introduction into phase-space methods in general is not the goal of this manuscript, we only state the most important steps in deriving the DHW formalism here.
We recommend Ref. \cite{Vasak:1987umA} for a more profound introduction into quantum transport theories. Additional information on quantum kinetic theories
in the context of particle creation can be found in Refs. \cite{Smolyansky:1997fcA, Smolyansky:1997fcC}. A complete derivation of the equations of
motion has been performed in Ref. \cite{Kohlfurst:2015zxi}.
%

To save computational costs later, but without disregarding magnetic fields, we introduce the QED Lagrangian in $2+1$ dimensions 
\begin{multline}
{\mathcal L} \left( \Psi, \bar{\Psi}, A \right) = \\
\frac{1}{2} \left( \I \bar{\Psi} \gamma^{\mu} \mathcal{D}_{\mu} \Psi - \I \bar{\Psi} \mathcal{D}_{\mu}^{\dag} \gamma^{\mu} \Psi \right) 
 -m \bar{\Psi} \Psi - \frac{1}{4} F_{\mu \nu} F^{\mu \nu}, \label{equ:Lag}
\end{multline}
where $\mathcal{D}_{\mu} = \left( \partial_{\mu} +\I e A_{\mu} \right)$ and $\mathcal{D}_{\mu}^{\dag} = \left( \overset{\leftharpoonup} {\partial_{\mu}} -\I e A_{\mu} \right)$. 
Additionally, we have the vector potential $A_{\mu}$, the electromagnetic field strength tensor $F_{\mu \nu} = \partial_{\mu} A_{\nu} - \partial_{\nu} A_{\mu}$ 
and the spinor fields $\Psi$ and $\bar{\Psi}$. 
In an attempt to further reduce the computational overhead we use only one $2$-spinor basis\footnote{We perform all calculations in the $xz$-plane, thus we label the third matrix $\gamma^3$.} 
\begin{equation}
\gamma^0 = \begin{pmatrix} 1 & 0\\ 0 & -1 \end{pmatrix},\ 
\gamma^1 = \begin{pmatrix} 0 & \I \\ \I & 0 \end{pmatrix},\
\gamma^3 = \begin{pmatrix} 0 & 1\\ -1 & 0 \end{pmatrix}
\end{equation}
effectively cutting numerical costs in half.
Varying with respect to $\psi$ ($\bar \psi$) we obtain the (adjoint) Dirac equation
\begin{align}
  \left(\I \gamma^{\mu} \partial_{\mu} - e \gamma^{\mu} A_{\mu} - m \right) \Psi &= 0, \label{equ:Dirac1} \\
  \bar{\Psi} \left(\I \overset{\leftharpoonup} {\partial_{\mu}} \gamma^{\mu} + e \gamma^{\mu} A_{\mu} + m \right) &= 0. \label{equ:Dirac2}
\end{align}

The backbone of the formalism is the density operator
\begin{multline}
 \hat {\mathcal C}_{\alpha \beta} \left( r , s \right) = \\
 \mathcal U \left(A,r,s 
\right) \ \left[ \bar {\Psi}_\beta \left( r - s/2 \right), {\Psi}_\alpha \left( r + 
s/2 \right) \right],
\end{multline}
where we introduced the center-of-mass coordinate $r$ and the relative coordinate $s$.
The Wilson line factor 
\begin{equation}
 \mathcal U \left(A,r,s \right) = \exp \left( \mathrm{ie} \int_{-1/2}^{1/2} d 
\xi \ A \left(r+ \xi s \right) \ s \right)
\end{equation}
is essential to guarantee gauge invariance. The covariant Wigner operator is then obtained via a 
Fourier transform with respect to the relative coordinate $s$
\begin{equation}
 \hat{\mathcal W}_{\alpha \beta} \left( r , p \right) = \frac{1}{2} \int d^4 s \ 
\mathrm{e}^{\mathrm{i} ps} \  \hat{\mathcal C}_{\alpha \beta} \left( r , s 
\right). \label{equ:W}
\end{equation}
In the following we will drop the indices in $\hat{\mathcal W}$ to improve readability.

In a rather lengthy calculation, see Ref. \cite{Vasak:1987umA, Kohlfurst:2015zxi} for details, we can combine the Wigner operator \eqref{equ:W} with
Eqs. \eqref{equ:Dirac1} and \eqref{equ:Dirac2} to generate the operator equations
\begin{alignat}{3}
 & \left( \frac{1}{2} D_{\mu}  - \I P_{\mu}  \right) \gamma^{\mu} \hat{\mathcal W} \left( r , p \right) && = - &&\I \hat{\mathcal W} \left( r , p \right), \label{equ:W1} \\
 & \left( \frac{1}{2} D_{\mu}  + \I P_{\mu}  \right) \hat{\mathcal W} \left( r , p \right) \gamma^{\mu}  && = &&\I \hat{\mathcal W} \left( r , p \right), \label{equ:W2} 
\end{alignat}
with the pseudo-differential operators
\begin{alignat}{4}
 & D_{\mu}  && = \partial_{\mu}^r - e &&\int_{-1/2}^{1/2} d \xi \ && \hat F_{\mu \nu} \left( r - \I \xi \partial^p \right) \partial_p^{\nu}, \\
 & P_{\mu}  && = p_{\mu} - \I e && \int_{-1/2}^{1/2} d \xi \ \xi \ && \hat F_{\mu \nu} \left( r - \I \xi \partial^p \right) \partial_p^{\nu}.
\end{alignat}
Due to the operator nature of the equations above, it is difficult to use them in a numerical simulation.

In order to obtain computational feasible equations of motion we therefore introduce a mean-field (Hartree) approximation
\begin{equation}
 \langle \Phi | \hat F^{\mu \nu} \left( r \right) | \Phi \rangle \approx F^{\mu \nu} \left( r \right).
\end{equation}
Consequently, the operator-valued electromagnetic field strength tensor $\hat F^{\mu \nu}$ is treated as a C-number field $F^{\mu \nu}$. 
In fact, a direct evaluation of $\langle \Phi | \hat F^{\mu \nu} \left( r \right) \ \hat{\mathcal C} \left( r , s \right) | \Phi \rangle $ would introduce couplings
to an arbitrary number of $n$-body terms (BBGKY hierarchy). 
The Hartree approximation is in principle a truncation at one-body level effectively turning an infinite hierarchy of coupled differential equations into a closed system, c.f. Ref. \cite{Vasak:1987umB} for an in-depth analysis.

We proceed by taking the vacuum expectation value of Eqs. \eqref{equ:W1} and \eqref{equ:W2}. 
As a result of the Hartree approximation
\begin{equation}
 \langle \Phi | \hat F^{\mu \nu} \left( r \right) \ \hat{\mathcal C} \left( r , s \right) | \Phi \rangle 
  \approx F^{\mu \nu} \left( r \right) \langle \Phi | \hat{\mathcal C} \left( r , s \right) | \Phi \rangle,
\end{equation}
Eqs. \eqref{equ:W1} and \eqref{equ:W2} are transformed into equations of motion for the covariant Wigner function
\begin{equation}
 \mathcal W \left( r , p \right) = \langle \Phi | \hat{\mathcal W} \left( r , p \right) | \Phi \rangle.
\end{equation}

Working directly with the covariant Wigner function is problematic, because it requires solving the equations of motion at all points in time at once. This is certainly not
feasible in a real-time simulation, thus we choose a definite frame to project on equal-time effectively reformulating everything to an initial-value problem. 
In this way, we obtain the equal-time Wigner function
\begin{equation}
 {\mathbbm w} \left( t, \mathbf{x} , \mathbf{p} \right) = \int \frac{dp_0}{2 \pi} \mathcal W \left( r , p \right),
\end{equation}
which we decompose into Dirac bilinears
\begin{equation}
 \mathbbm{w} \left( t, \mathbf x, \mathbf p \right) = \frac{1}{2} \left( \mathbbm 1 \ \mathbbm s + \gamma_{0} \mathbbm v^{0} + \gamma_{1} \mathbbm v^{1} + \gamma_{3} \mathbbm v^{3} \right). \label{equ:wigner}
\end{equation}
This makes it easy to interpret the quantities $\mathbbm s$ as mass density, $\mathbbm{v}_0$ as charge density and $\mathbbm v^{1},~ \mathbbm v^{3}$ as the current density. 

Finally, we obtain a coupled set of differential equations for the equal-time Wigner coefficients, cf. Refs. \cite{Kohlfurst:2015zxi},
\begin{alignat}{6}
  & D_t \mathbbm{v}_0 && + D_x \mathbbm{v}^1 && + D_z \mathbbm{v}^3 && && &&= 0, 
\label{eqn1_1} \\  
  & D_t \mathbbm{s}  && && && -2 \Pi_x \mathbbm{v}^3 && +
 2 \Pi_z \mathbbm{v}^1 &&= 0,  \label{eqn1_2} \\  
  & D_t \mathbbm{v}^1 && +D_x \mathbbm{v}_0 &&  && && -2 \Pi_z
 \mathbbm{s} && = -2 \mathbbm{v}^3,  \label{eqn1_3} \\  
  & D_t \mathbbm{v}^3 && && +D_z \mathbbm{v}_0 && +2\Pi_x
 \mathbbm{s} && &&= 2 \mathbbm{v}^1, \hspace{0.7cm}  \label{eqn1_4} 
\end{alignat} 
with the pseudo-differential operators
\begin{alignat}{8}
  & D_t && = \quad && \partial_{t} && + e && \int_{-1/2}^{1/2} d\xi \, && \mathbf{E} \left(t, 
\mathbf{x}+ \textrm {i} \xi \boldsymbol{\nabla}_p \right) && \cdot && \boldsymbol{\nabla}_p,
\label{eqn2_1} \\
 & D_x && = \quad && \partial_x && + e && \int_{-1/2}^{1/2} d\xi \, && B \left( t, \mathbf{x}+\textrm {i} \xi \boldsymbol{\nabla}_p \right) && && \partial_{p_z},
\label{eqn2_2} \\
 & D_z && = \quad && \partial_z && - e && \int_{-1/2}^{1/2} d\xi \, && B \left( t, \mathbf{x}+\textrm {i} \xi \boldsymbol{\nabla}_p \right) && && \partial_{p_x},
\label{eqn2_3} \\
  & \Pi_x && = \quad && p_x && - \textrm {i} e && \int_{-1/2}^{1/2} d\xi \,
\xi \, && B \left( t, \mathbf{x}+\textrm {i} \xi \boldsymbol{\nabla}_p \right) && && \partial_{p_z}, \label{eqn2_4} \\
  & \Pi_z && = \quad && p_z && + \textrm {i} e && \int_{-1/2}^{1/2} d\xi \,
\xi \, && B \left( t, \mathbf{x}+\textrm {i} \xi \boldsymbol{\nabla}_p \right) && && \partial_{p_x}. \label{eqn2_5} 
\end{alignat}
The initial conditions are given by the vacuum solution
\begin{equation}
  \mathbbm{s}_{vac} \left(\boldsymbol{p} \right) = -\frac{2}{\sqrt{1 +
\boldsymbol{p}^2}},\ 
  \mathbbm{v}_{vac}^{1,3} \left(\boldsymbol{p} \right) = -\frac{2
\boldsymbol{p}}{\sqrt{1 + \boldsymbol{p}^2}}. \label{equ:vac}
\end{equation}

\section{Numerical solution techniques}
\label{Sec:Numerical}


One of the aims of this manuscript is to demonstrate how to solve the equations of motion \eqref{eqn1_1}-\eqref{eqn1_4} effectively. Although a brute-force calculation might be
successful, running such simulations is computationally expensive and numerically quite unstable. To combat these issues we have developed various solution techniques that
certainly help to speed-up the calculations while reducing the computational load and still maintaining a high accuracy. It is worth mentioning that all methods and numerical recipes
introduced in this manuscript can be easily applied to other system (${\rm QED}_{1+1}, \ldots$), too.

\subsubsection{Preconditioning}

In order to solve the differential equations \eqref{eqn1_1}-\eqref{eqn1_4} for a configuration featuring fields of the form of Eqs.~\eqref{equ:E} and \eqref{equ:B} 
multiple preconditioning steps are in order. At first we introduce
modified Wigner components \cite{Hebenstreit:2011wk}
\begin{equation}
 \mathbbm{w}^v = \mathbbm{w} - \mathbbm{w}_{vac}
 \label{equ:red}
\end{equation}
such that we obtain a set of inhomogeneous partial differential equations with vanishing initial conditions.
In this way, we pro-actively ensure that all components fall off at the boundary already minimizing truncation errors.

Numerical feasibility can then be significantly improved by establishing the relation
\begin{equation}
 \mathbf{p} = \mathbf{q} - e\mathbf{A} (t,z),
\end{equation}
where $\mathbf{A} (t,z)$ is the vector potential defined previously \eqref{equ:A}.
In this way, we take into account so-called "minimal coupling". More specifically, we switch from kinetic momenta $\mathbf{p}$ to canonical momenta $\mathbf{q}$. 
Naturally, the derivative terms have to be transformed, too
\begin{align}
 \partial_t &\to \partial_t - e \mathbf{E} (t,z) \cdot \boldsymbol{\nabla}_{q},\\ 
 \partial_z &\to \partial_z + eB (t,z) \ \partial_{q_x}.
\end{align} 
The most prominent advantage of this transformation is the change in the time-development operator
\begin{multline}
D_t \to {\mathcal D}_t = \partial_{t} + \Delta_E = \\ 
 \partial_t + e \int_{-1/2}^{1/2} d\xi \, \big( \mathbf{E} \left(t, z + \textrm {i} \xi \partial_{q_z} \right) - \mathbf{E} \left(t, z \right) \big) \cdot \boldsymbol{\nabla}_{q}. \label{equ:Dt}
\end{multline}
As is immediately obvious from a Taylor expansion of the derivative operator
\begin{equation}
\Delta_E \approx \left( \sum_{n=0}^\infty \ \frac{(-1)^n \ \mathbf{E}^{(2n)} (t,z)}{4^n (2n+1) (2n)!} - \mathbf{E} (t,z) \right) \cdot \boldsymbol{\nabla}_{q},
\end{equation}
where $\mathbf{E}^{(2n)}$ denotes the $2n$-th derivative with respect to $z$,
the leading term in Eqs.~\eqref{equ:Dt} is of the order of ${\mathcal O} \big( E'' (t,z) \big)$. Hence, quasi-homogeneous quantum kinetics is already incorporated
in the transformed equations.
In particular, in the limit of a locally constant vector potential, $\mathbf{E} \left(t, z + \textrm {i} \xi \partial_{q_z} \right) \approx \mathbf{E} (t,z)$ and 
$B \left(t, z + \textrm {i} \xi \partial_{q_z} \right) \approx 0$, 
the momentum derivatives vanish, thus one is left
with a partial differential equation in $t$ and $z$. If one employs a local constant field approximation (LCFA)
the derivatives with respect to $z$ vanish, too \cite{Hebenstreit:2011wk}.

Eventually, we change the geometry of the domain we work in. The equations \eqref{eqn1_1}-\eqref{eqn1_4} are defined on an open domain
$z \in (-\infty, +\infty)$. Due to the change of variables \eqref{equ:red} and the fact, that we use localized fields \eqref{equ:E} the region of interest is clearly finite.
Hence, we truncate the length of the domain $z \in [ -L_z, +L_z ]$ to minimize memory requirements. The same holds for the momentum variables $q_x$ and $q_z$.

\subsubsection{Background fields}

In the code, we only specify the time-dependence of the electric field analytically. 
Time dependence of the magnetic field and thus of the vector potential are calculated numerically by adding one term of the form of $\frac{d A}{dt} = - E(t)$ to the system of differential equations.\\

In previous studies \cite{Hebenstreit:2011wk, Kohlfurst:2017git} an analytical form of the differential operators
has been used. Such an approach is highly inflexible and one has to deal with cancellation effects in the anti-derivative. A better way is to perform
the integration with respect to $\xi$ numerically and store the result in a lookup table.

\subsubsection{Spectral solver}

\begin{figure}[t]
      \begin{center}
	\includegraphics[width=\figlenB]{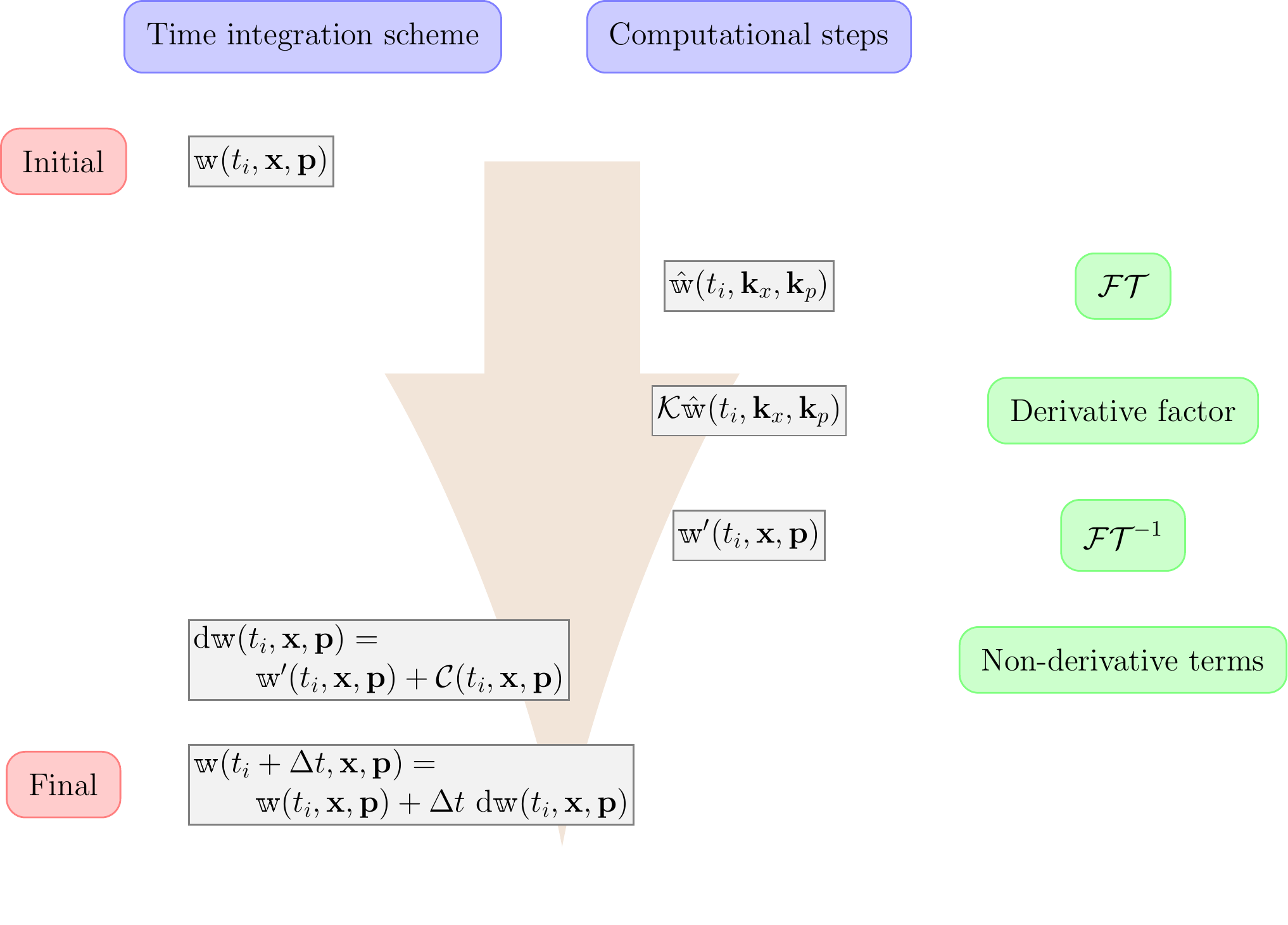} 
      \end{center}
      \caption{Sketch of the integration scheme employed to find solutions for the set of equations of motion. 
        At every time step $t_i$ derivatives with respect to coordinates ${\mathbf x}$ and momenta ${\mathbf p}$ are calculated via pseudospectral methods.
        Using a Fourier transform ${\mathcal FT}_{{\mathbf x,p}}$ derivative operators are converted into multiplicative factors. 
        Advancing in time by one step $\Delta t$
        is done in coordinate-momentum space.} 
      \label{fig:RK}
      \end{figure}

In order to solve the equations of motion a combination of a spectral solver (phase-space) and a finite-difference solver (time dependency) is used. To be more specific,
we employ pseudo-spectral methods at every time step to achieve optimal accuracy at minimal grid size in the phase-space domain. 
The general idea of using a pseudo-spectral solver is to solve the system of equations in phase-space, but evaluate derivatives via
multiplications in Fourier space, c.f. Fig.~\ref{fig:RK}. In this way, exponential accuracy can be achieved \cite{Boyd}.
Time integration is done using a Dormand-Prince Runge-Kutta integrator of order 8(5,3) \cite{NR}.  

A generic derivative is calculated in the following way
\begin{equation}
 \mathcal{FT}^{-1} \left[ \mathcal{FT} \left[ \frac{d^n}{dx^n} f \left( x \right) \right] \right] = \mathcal{FT}^{-1} \left[ \left( \I k \right)^n \hat f \left( k \right) \right], \label{equ:FT}
\end{equation}
where $\hat f \left( k \right)$ denotes a Fourier transformed quantity. Derivatives with respect to $z$ and $q_x$ are evaluated in this way.
Pseudo-differential operators \eqref{eqn2_1}-\eqref{eqn2_5} and therefore derivatives with respect to $q_z$ are evaluated differently. For a generic function 
$G \left( z + \I \xi \partial_{q_z},t \right) = G_t \left( t \right) G_x \left( z + \I \xi \partial_{q_z} \right)$ we write
\begin{multline}
  \mathcal{FT}^{-1} \ \left[ \mathcal{FT} \ \left[ G \left( z + \I \xi \partial_{q_z},t \right) \mathbbm{w}^v \left( z, q_x, q_z \right) \right] \right] = \\
 G_t \left( t \right) \ \mathcal{FT}^{-1} \ \left[ G_x \left( z - \xi k_{q_z} \right) \hat{\mathbbm{w}}^v \left( z, q_x, k_{q_z} \right) \right]. \label{equ:FT2}
\end{multline}
In contrast to previous studies \cite{Kohlfurst:2017hbd,Kohlfurst:2017git} this technique has been applied to evaluate derivatives of the inhomogeneities, too.

The use of fast Fourier transforms (FFTs) accelerates the simulations tremendously \cite{Boyd, Frigo05thedesign}. Only downside is that we have to have Fourier basis functions namely
$\sin \left(k x \right)$ and $\cos \left( k x \right)$.
This choice of basis functions requires (i) an equidistant discretization of the domain and (ii) periodic boundary conditions. 
Alternatively, Chebychef polynomials can be used, which have the advantage that they can be used together with a non-periodic grid. However, they demand
a discretization where the point density is highest at the boundaries. As we wanted to have a high resolution at the center of the domain to capture as many physical
effects as possible we chose equidistant point sampling.

A further improvement on lattice point distribution is given by the fact, that we can in principle introduce any bijective transformation of the Eqs. \eqref{eqn2_1}-\eqref{eqn2_5}
and then solve a similar set of equations of motion. The solution to the original problem is then simply given by re-transforming the results once the simulation has finished. 
In our case, we introduce the transformations
\begin{align}
 q_x &= \frac{2 L_q}{\pi } \textrm{arctan} \left( \frac{1}{\alpha_{q}} \, \textrm{tan} \left( \frac{\pi}{2 L_q} \, \tilde q_x \right) \right), \\
 z &= \frac{2 L_z}{\pi } \textrm{arctan} \left( \frac{1}{\alpha_{z}} \, \textrm{tan} \left( \frac{\pi}{2 L_z} \, \tilde z \right) \right),
\end{align}
where $L_q$ ($L_z$) gives the length of the domain in $q_x$ ($z$)-direction and $\alpha_{q}$ ($\alpha_{z}$) controls the strength of the
deformation. In this way, we can (i) achieve a much better convergence rate and (ii) we are still allowed to use a customized rasterization of the domain. 
The derivative operators transform accordingly
\begin{align}
 \partial_{q_x} &= \\ &\left( \alpha_{q} \ \textrm{cos} \left( \frac{\pi}{2 L_q} \, \tilde q_x \right)^2 + \frac{1}{\alpha_{q}} \ \textrm{sin} \left( \frac{\pi}{2 L_q} \, \tilde q_x \right)^2 \right) \partial_{\tilde q_x}, \notag \\
 \partial_z &= \\ &\left( \alpha_{z} \ \textrm{cos} \left( \frac{\pi}{2 L_z} \, \tilde z \right)^2 + \frac{1}{\alpha_{z}} \ \textrm{sin} \left( \frac{\pi}{2 L_z} \, \tilde z \right)^2 \right) \partial_{\tilde z}. \notag 
\end{align}
Keep in mind, however, that the grid in $\tilde z$ and $\tilde q_x$ is regular, the grid in the physical domain $(z, q_x)$ is not. 
Regarding periodic boundary conditions, we observe that due to the transformation to modified Wigner components \eqref{equ:red} all quantities fall off for larger values of $\tilde z$, $\tilde q_x$ or $q_z$.
We can thus truncate the domain at points where the coefficients supposedly vanish yielding a finite domain of size $L_{\tilde z} \times L_{\tilde q_x} \times L_{q_z}$.
Then we set $\tilde z_1 = \tilde z_{N_z}$, $\tilde q_{x,1} = \tilde q_{x,N_{q_x}}$
and $q_{z,1} = q_{z,N_{q_z}}$, where $N_z$, $N_{q_x}$ and $N_{q_z}$ give the total number of grid points in $\tilde z$, $\tilde q_x$ and $q_z$, respectively. 

While momentum and spatial derivatives have to be evaluated at every single time step (often multiple times), the overall time evolution can be calculated in a straight-forward way, see Fig.~\ref{fig:RK}. 
First, a starting point $t_{\rm vac}$ of the simulation has to be determined. Ideally, the vector potential as well as the fields are still zero at this point such that one is left
with the pure vacuum state. However, background fields may have infinite support making a truncation of the time domain necessary. To minimize the error
due to such a truncation, while not needlessly increasing computing time, we have set the initial times to $t_{\ast} = -6\tau$ (Gaussian envelope) and $t_{\ast} = -2\tau$ (super-Gaussian).

The actual time stepping $t_{i+1} = t_i + \Delta t$ might be done through explicit or implicit methods. We have chosen a higher-order explicit stepper, because it seems to be
the ideal compromise between accuracy and run time. Specifically, we neither have to allocate additional memory due to the overhead in terms of inverting the derivative matrix nor do we 
have to solve the corresponding algebraic equations. On the other hand, we still obtain high accuracy using an 8th-order method with error-correction. 

As we are only interested in the particle's momentum spectrum, simulations terminate once the background fields vanish, here $t^{\ast} = -t_{\ast}$.

\subsubsection{Filters}

Due to the fact, that we want to apply the procedure introduced in Eq.~\eqref{equ:FT2} only trivial transformations in $q_z$ are feasible.
Consequently, we have to use an equidistant grid in $q_z$, thus we are automatically limited to a small domain $L_{q_z}$. 
In turn, we have to account for the fact that the Wigner coefficients may not fall off
sufficiently fast, thus the requirement for periodic boundary conditions could be in conflict with domain truncation. 
More specific, there could be a significant discrepancy between coefficients defined at the now neighboring points $q_{z,N_{q_z}-1},~ q_{z,1}$ and 
$q_{z,2}$ resulting in Gibbs phenomenon induced effects, e.g., an artificial growth of the coefficients $\mathbbm{w}$. 

To avoid the creation of spurious patterns or a
wrong sampling of high-frequency modes so-called anti-aliasing methods are employed, see Refs. \cite{Boyd, GOTTLIEB199281} for a more detailed analysis.
First, we introduce an artificial dampening factor at every time step, which is
fairly easy to implement as we only have to modify the prefactor of the $\partial_{q_z}$-derivative to
\begin{equation}
 w_{q_z} = \exp \left( -\left( \frac{q_{z,i}}{q_{z,1}} \right)^{2 N_w} \right),
\end{equation}
with $N_w \in \mathbbm{N}$.

Secondly, to resolve any additional problems regarding high-order modes we cut the upper part of the Fourier spectrum at the end of a simulation (at
the last time step $t_f$, but before evaluating any observables).
The combination of applying these filters proved to work excellently, as it was possible to avoid the appearance of any non-physical terms in the final distribution functions.
On a side note, in previous calculations \cite{Kohlfurst:2017git} the highest order modes have been terminated at each
time step. However, that technique seemed to fail when applied to spatially strongly inhomogeneous, long pulsed fields as spurious artifacts started to show up
at long run times.

Unfortunately, for configurations that require a long runtime, none of these methods work sufficiently well. In such a case it seems as if the only even remotely successful
way is to sacrifice some resolution in direction of $q_z$ and expand the total domain size. As the source of these artificial oscillations is the jump at the boundary, any
artifacts associated with it will start at $\pm L_{q_z}$ and then propagate towards the center. If the domain, however, is large enough, the time it takes for these 
oscillations to reach the real particle distributions is so long that we can safely stop the simulation at times $t_f$, where we still have an intact particle signal. Of course,
before evaluating any observables the Wigner components at all higher transversal momenta have to be artificially set to zero. 

\subsubsection{Observables}

The DHW formalism only allows to solve for the Wigner coefficients $\mathbbm{w}$. As we are, however, mainly interested in observables we have to
rearrange these coefficients to obtain particle distribution functions. To keep the discussion simple, 
we only state the particle number density at asymptotic times $t_f$ ($\mathbf A (t \to \infty,z) \to 0$)
\begin{align}
f \left( z, q_x, q_z \right) = \frac{\mathbbm{s}^v + q_x
\mathbbm{v}^{v,1} + q_z \mathbbm{v}^{v,3}}{\sqrt{1+q_x^2+q_z^2}}.
 \label{equ:n}
\end{align}
Accordingly, the particles momentum spectrum is given by
\begin{equation}
 f \left( q_x, q_z \right) = \int {\rm d}z ~ f \left( z, q_x ,q_z \right)
 \label{equ:nn}
\end{equation}
as well as
\begin{align}
 f \left( q_x \right) &= \int {\rm d} q_z ~ f \left( q_x ,q_z \right), \\
 f \left( q_z \right) &= \int {\rm d} q_x ~ f \left( q_x ,q_z \right).
 \label{equ:nnn}
\end{align}
The total particle number is then given by
\begin{equation}
 N = \int {\rm d}q_x ~ {\rm d}q_z ~ f \left( q_x ,q_z \right).
 \label{equ:NN}
\end{equation}
Note, that we have already transformed the data obtained from the simulation to ordinary spatial coordinates and canonical momenta $\{z, q_x, q_z \}$, because
it is easier to discuss the results in terms of a physical basis.

The crucial point in evaluating Eqs. \eqref{equ:n}-\eqref{equ:NN} is given by the fact, that the particle signal is generally only acquired after summing up
all relevant Wigner components. The reason is, that each of $\mathbbm{s}^{v}$, $\mathbbm{v}^{v,1}$ and $\mathbbm{v}^{v,3}$ not only holds the information of the particle spectrum 
but also the information for additional observables, c.f. Ref. \cite{Vasak:1987umA}. As a consequence, the fraction of the particle spectrum on a Wigner component might be lower
than 1$\%$. Hence, when calculating particle distributions one technically extracts sub-leading contributions, which turns this final step into a delicate process. 

As the particle yield is naturally higher the further the electric field is extended we have to normalize the rates to allow for a fair comparison between
results with different $\lambda$. The problem is that in principal Schwinger effect, multiphoton pair production and assistance mechanism all demand different normalization.
The compromise was to normalize parallel $f(p_x)$ as well as transversal $f(p_z)$ distribution functions in terms of semi-classical expectation values 
\begin{equation}
 N_{\rm cl} = \int \ {\rm d}t \ {\rm d}z \ a(t,z)^{\frac{3}{2}} \ e^{-\frac{m^2\pi}{e a(t,z)}} 
 \label{equ:Ncl}
\end{equation}
with $a=\sqrt{ \frac{1}{2} \big( \lvert B^2-E^2 \rvert - \left( B^2-E^2 \right) \big) }$ due to its simplicity and its capability of taking into account
the electric field's spatial finiteness as well as the magnetic field's attribute to suppress pairs to form. In the limit of pure Schwinger pair production Eq. \eqref{equ:Ncl} is
even exact. Moreover, signatures of multiphoton particle creation generally extent non-linearly in transversal direction thus we display such results in $2$d contour plots,
where no particular normalization is needed.

\section{Results}
\label{Sec:Results}

One big goal of this manuscript is to provide some insight into particle creation and, consequently, particle dynamics
in high-intensity electromagnetic background fields. 
We discuss the final particle momentum spectra with respect to the different creation mechanisms and show how to interpret certain features by simple means. 
On a side note, the full set of data in terms of contour plots of the particle distribution function $f \left( p_x, p_z \right)$ is attached
at the end of the manuscript, c.f. section \ref{Sec:Appendix}.

\subsection{Tunneling-dominated pair production}
\label{Sec:Schwinger}

To obtain a clear qualitative picture of particle creation via tunneling we employ slowly varying, few-cycle fields with
a peak field strength of $e\varepsilon = 0.5m^2$ as well as $e\varepsilon = 0.2m^2$, respectively.

      \begin{figure*}[t]
      \begin{center}
	\includegraphics[width=\figlenB]{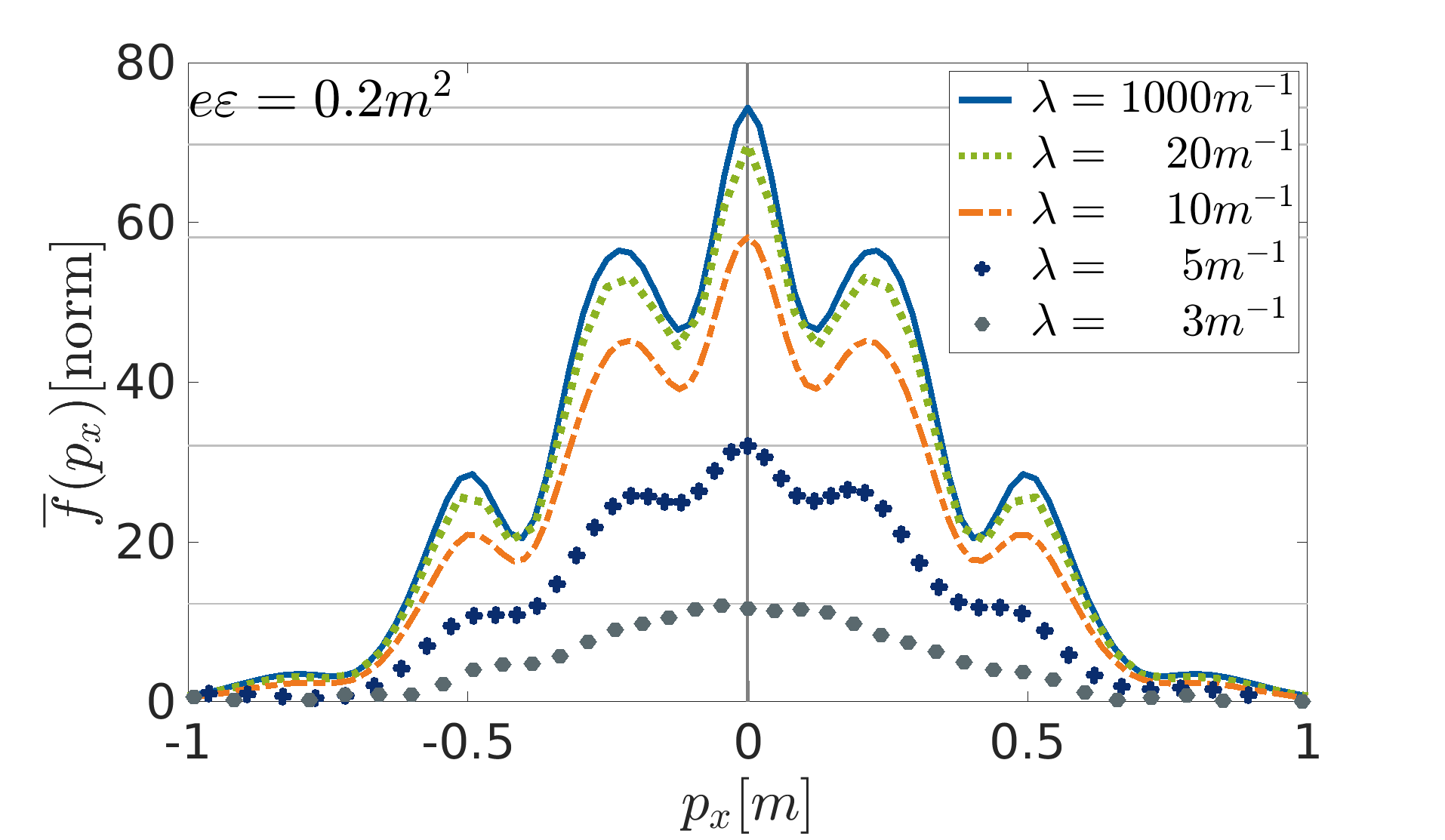} 
	\includegraphics[width=\figlenB]{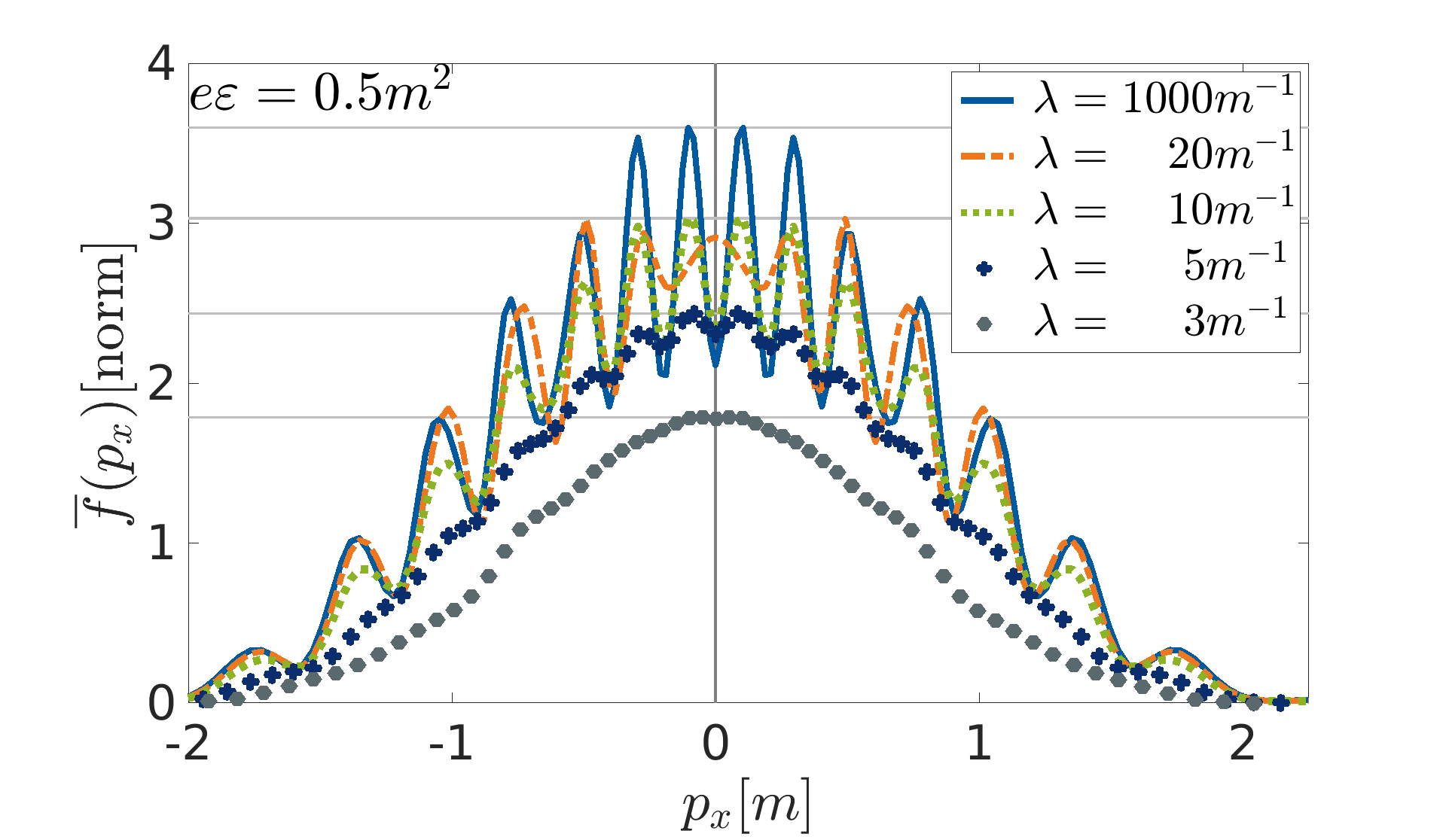} 	
	\includegraphics[width=\figlenB]{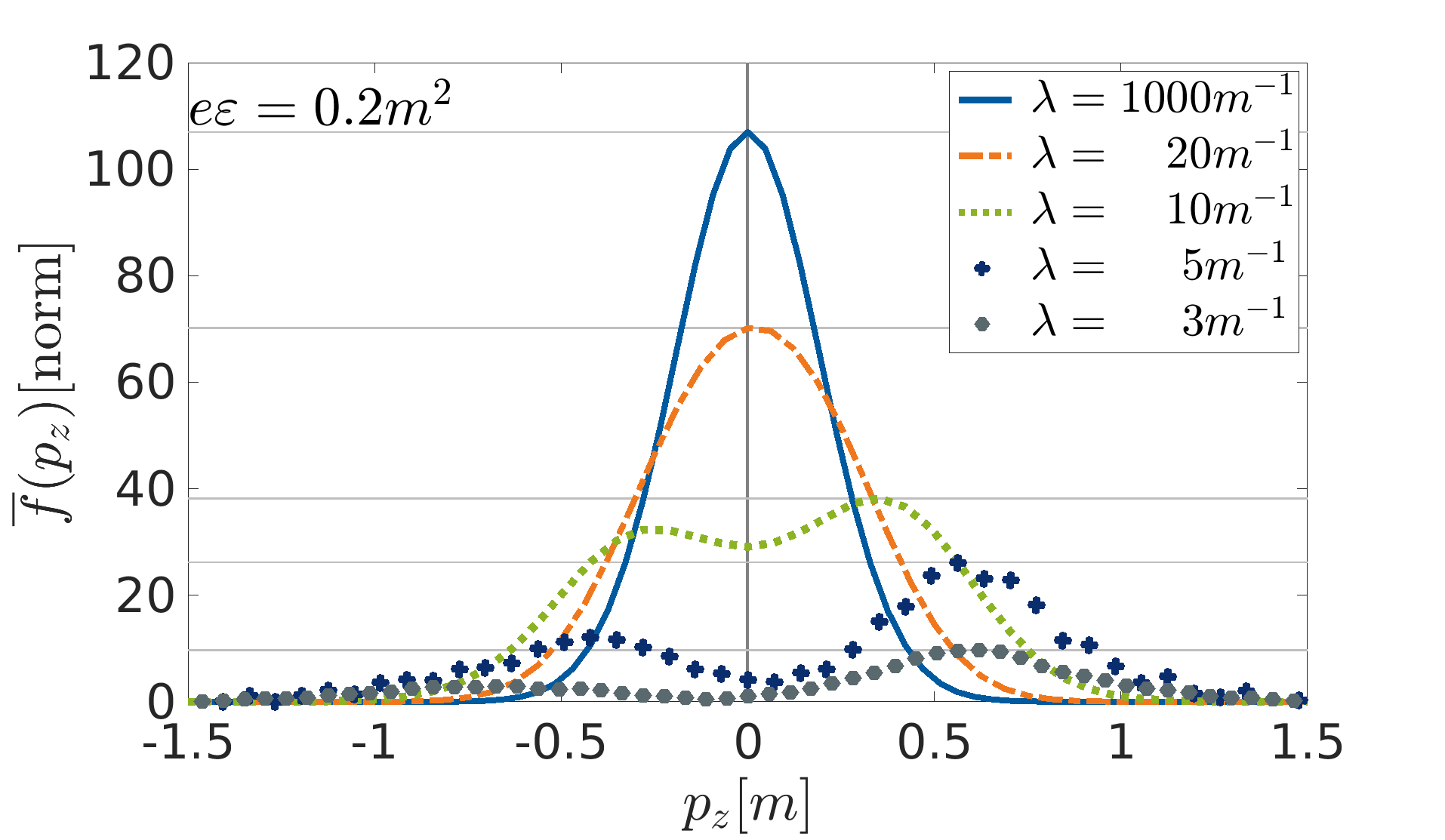} 
	\includegraphics[width=\figlenB]{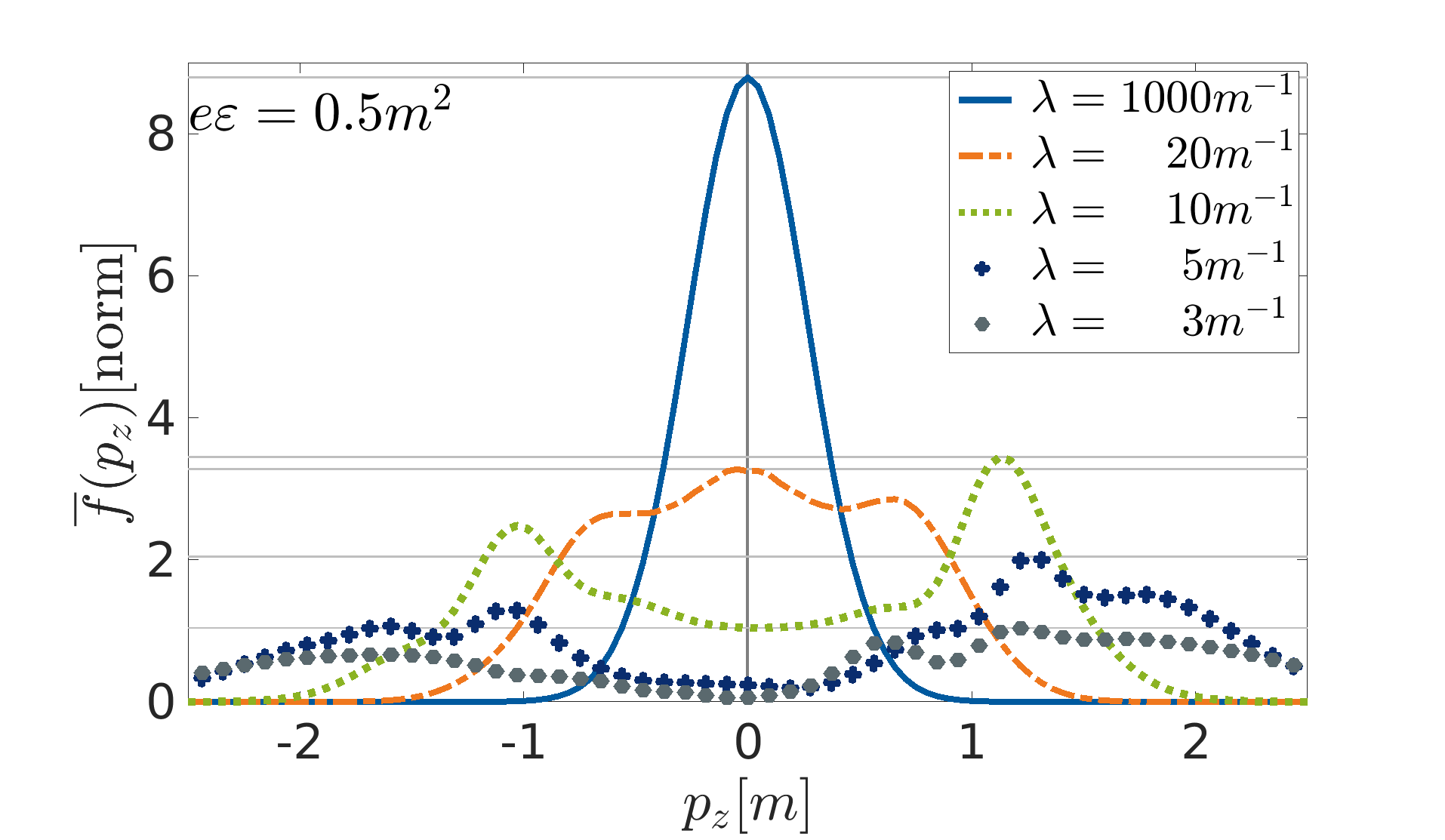} 	
      \end{center}
      \caption{Particle distribution functions $f(p_x)$ (top) and $f(p_z)$ (bottom) normalized by the expected production rate in local constant field approximation.
      Due to the fact, that the field is polarized along ${\mathbf e}_x$ quantum interferences show up in parallel direction $p_x$ only. The smaller $\lambda$ 
      the higher the magnetic peak field strength and, in turn, the stronger particles are accelerated along $p_z$-direction. As for $\lambda=3m^{-1}$ particle bunches do not 
      occupy the same region in phase-space any more, the interference pattern vanishes. 
      Both simulations feature a temporal pulse length of $\tau = 25m^{-1}$ and a frequency of $\omega = 0.2m$.} 
      \label{fig:fpxpz0}
      \end{figure*}       

Both field configurations have in common that particles are predominantly produced via tunneling. Reason is that, although the Keldysh parameter ($\gamma = 1$, $\gamma = 0.4$) 
might indicate that both particle creation mechanism are important, the total pulse length is quite short $\tau=25m^{-1}$ and 
the envelope function does not allow for many sub-cycles. As a result multiphoton effects are suppressed, thus 
the spectra can be nicely described through a semi-classical single-trajectory picture. 
Nevertheless, the temporal variation in the field is
sufficiently strong such that one cannot observe the ``pure'' Schwinger effect, see the deviations in the normalized total yield in Fig.~\ref{fig:fpxpz0}. 

The simple man's model introduced in Ref. \cite{Kohlfurst:2017git} states that particles are produced in regions where $E (t,z)^2 - B(t,z)^2>0$ holds (if $\mathbf{E} \cdot \mathbf{B}=0$) and
the higher the effective field strength $E (t,z)^2 - B(t,z)^2$ the higher the chance for pair production. 
For a field with a Gaussian envelope, pulse length $\tau = 25m^{-1}$ and frequency $\omega = 0.2m$ only three peaks are really capable of producing a sizable amount of particles. 
More specifically, given that $z=0$ the field is strongest at $t=0$, while at $t=\pm 13.7m^{-1}$ it
reaches $\sim 75\%$ of its maximum value. Due to the Schwinger effect's exponential suppression by the effective field strength every other peak can be considered as minor, 
thus they can be safely neglected in the further discussion. 
 
In a quasi-homogeneous calculation ($\lambda=1000m^{-1}$), see the solid blue curves in Fig.~\ref{fig:fpxpz0}, 
we obtain a broad peak superposed by oscillations in $f(p_x)$ and an exponentially declining distribution in $p_z$. 
Assuming, that for this configuration we can neglect the magnetic field and consider the electric field as homogeneous in $z$, we can solve the set of equations \eqref{eqn1_1}-\eqref{eqn1_4} analytically.
As particles in the simple man's model \eqref{equ:P}-\eqref{equ:sin3} are created with zero initial parallel momentum $p_{x,i}$
the final particle momentum $p_{x,f}$ solely depends on the strength of the vector potential at
the time of creation $t_0$. Hence, we can easily calculate the reference points for the final distribution. Unsurprisingly, pair creation at $t_0=0$ translates
into a peak at $p_{x,f}=0$. At $t=\pm 3.25m^{-1}$ the electric field still has $75\%$ of its maximal strength translating into a 90 \% smaller chance for
pair production \eqref{equ:P}. Consequently, if a particle was created at $t_0=3.25m^{-1}$ 
it would obtain a final momentum of $p_{x,f}=2.98 ~ \varepsilon/E_{\rm cr} ~ m$, thus roughly determining the point where the particle density has fallen off by 90 \%.
Particles created at the two side maxima of the electric field 
($t=\pm 13.7m^{-1}$) acquire a final momentum of $p_{x,f}=\pm 1.44 ~ \varepsilon/E_{\rm cr} ~ m$. 
Hence, particle bunches stemming from main and side peaks in $E(t,z)$ are clearly overlapping. Moreover, 
one would expect these particles to carry different phase information due to their different times of creation, c.f. Refs.~\cite{SemiClassA} 
for a quantum field theoretical explanation and Refs.~\cite{Becker,Salieres902} for discussions in the context of atomic ionization. 
Adding up the individual contributions then automatically results in quantum interferences. 

As $\lambda$ decreases the magnetic field strength rises and thus the overall impact
of the magnetic field increases. Pictorially speaking, the magnetic field between the three main peaks in $E(t,z)$ acts as an accelerator in direction of $p_z$. However, as $B(t,z)$ 
is oscillating in space and time, particles created at subsequent peaks 
are accelerated in opposite directions. As a result, those particles do not share the same phase space
at final times any more. Averaging over particle paths with similar phase information, however, does not result in quantum interference ultimately leading to a smooth distribution function,
c.f. the grey dotted curve in Fig.~\ref{fig:fpxpz0}. 

The distribution function $f(p_z)$ can be very well understood in the same way assuming that (i) particle creation is exponentially suppressed with 
$\exp \left( -\dfrac{m^2 + p_z^2}{e\varepsilon} \right)$, see Eq.~\eqref{equ:P}, and (ii) it takes a  strong magnetic field to deflect the particles significantly.
In case of a vanishing magnetic field the transversal particle spectrum $f(p_z)$ shows a Gaussian distribution, see Fig.~\ref{fig:fpxpz0}. The absence of quantum interferences is given due to the fact that
although all particles have picked up a phase, the distribution of the phase information only varies in direction of $p_x$. Hence, when summing over the phases in direction
of $p_z$ no interference pattern appears.

For strong magnetic fields the peak at $p_z=0$ splits into two weakly pronounced
peaks at $p_z \approx \pm 0.6m$ ($e\varepsilon=0.2m^2$) and $p_z \approx \pm 1.25m$ ($e\varepsilon=0.5m^2$). 
These two peaks are not equal in height clearly favoring the ones at $p_z>0$, respectively. The reason for this asymmetry lies in our choice of
representation as this has intrinsically fixed the particle spin, see Sec. \ref{Sec:DHW}. 
Hence, despite the fact that $A \left(t,z \right)$ is symmetric in $z$ performing all calculations for only one $2$-spinor basis and,
therefore, neglecting half of all electrons and positrons automatically results in an uneven particle distribution in phase-space. 
As a consequence the spatially varying magnetic field introduces a net force in one transversal momentum direction \eqref{equ:sin1}-\eqref{equ:sin3}. 

In summary, smooth particle distributions superposed by quantum interferences are clear signatures of the Schwinger regime. 
Furthermore, the momentum spectrum can be understood under the assumptions, that the effective field strength $E (t,z)^2 - B(t,z)^2$
determines the chances for particle creation and particles, once created, follow semi-classical trajectories.

\subsection{Absorption-dominated pair production}
\label{Sec:Multi}

In a multi-cycle field the virtual pair can obtain energy from the background field via absorbing photons. If the total energy gain is higher than the production threshold
an electron-positron pair is created. The decisive quantities are the number and energy of the absorbed photons $n \omega$ as well as the particle's rest energy
plus a modification factor due to the oscillatory motion of the particle, see Ref. \cite{Kohlfurst:2013ura}. Furthermore,
if an $l$-photon process could occur there is also the chance for $l+s$-photon absorption. In such a case, particles are created with higher kinetic momenta and
a different angular momentum profile \cite{Yang:1948nq}. As a result, 
the final particle distribution is given by multiple ellipses of fixed energy ${\cal E}_n$, where $n$ gives the number of photons initially involved in the process. 

      \begin{figure}[t]
      \begin{center}
	\includegraphics[trim=3.5cm 0cm 3.5cm 0cm,clip,width=\figlenB]{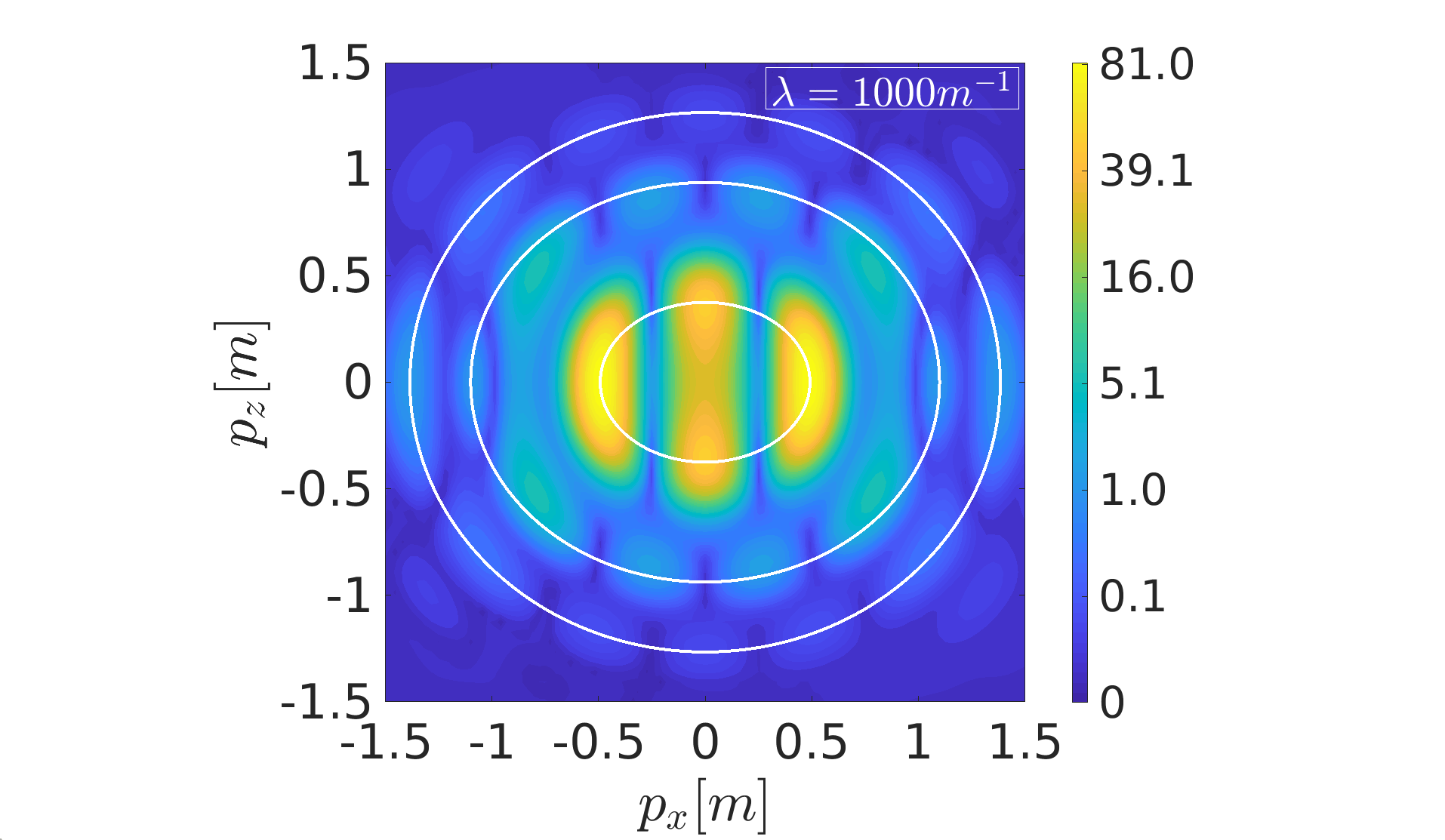} 
	\includegraphics[trim=3.5cm 0cm 3.5cm 0cm,clip,width=\figlenB]{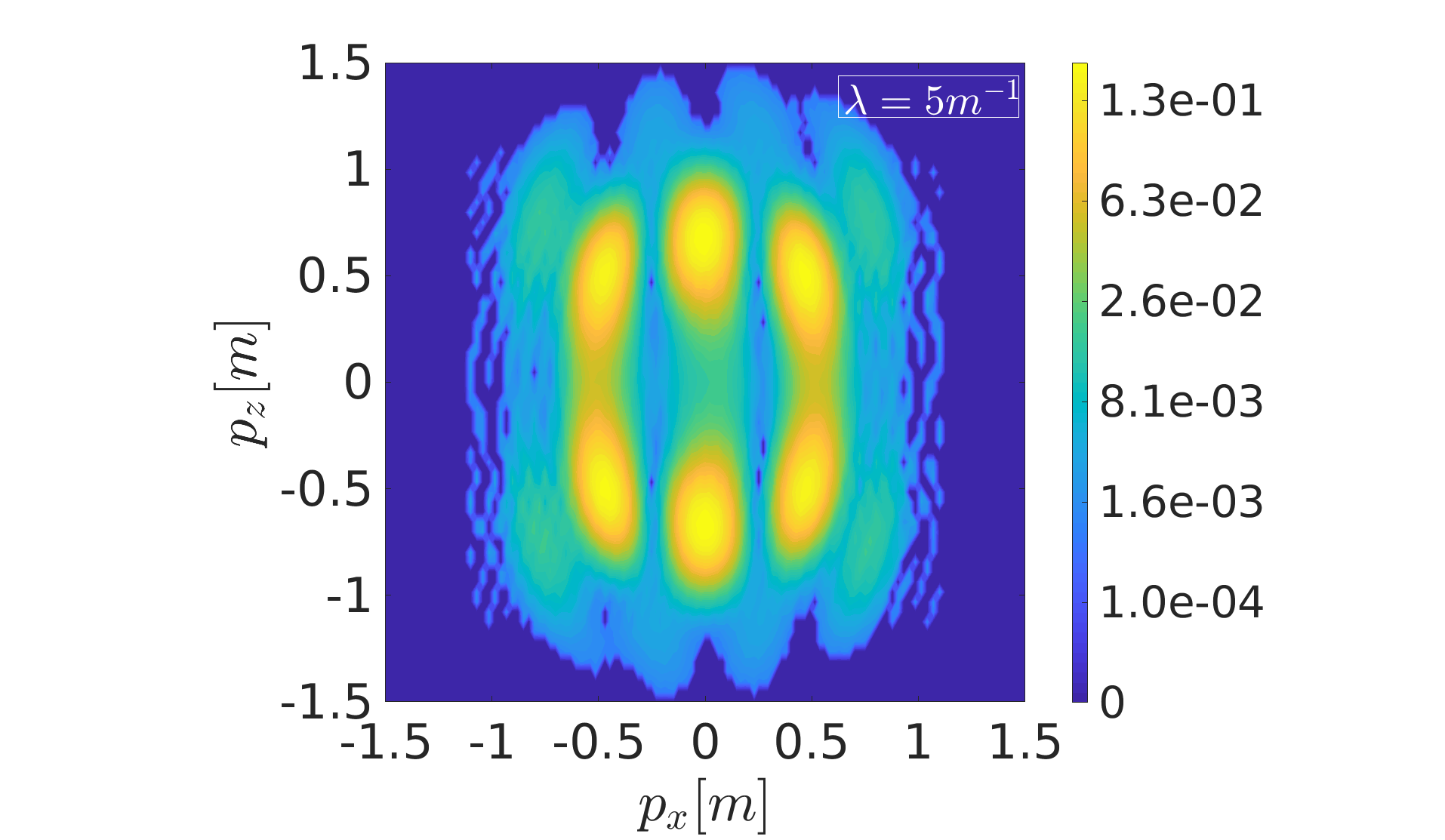} 
      \end{center}
      \caption{Non-linearly scaled particle momentum spectrum $f(p_x,p_z)$ for large (top) and small (bottom) spatial extent $\lambda$. Electric field strength $e\varepsilon = 0.5m^2$, 
        pulse length $\tau = 25m^{-1}$ and field frequency $\omega = 0.5m$ are fixed. Due to the presence of a strong field and the high photon energies 
        particles are predominantly created along ellipses. For a vanishing magnetic field (top) multiphoton peaks as well as a pronounced interference pattern
        are clearly visible. A strong magnetic field disturbs these patterns and additionally applies a strong force in perpendicular direction $\pm p_z$.
        } 
      \label{fig:fpxpz3}
      \end{figure}   

      \begin{figure}[t]
      \begin{center}
	\includegraphics[width=\figlenB]{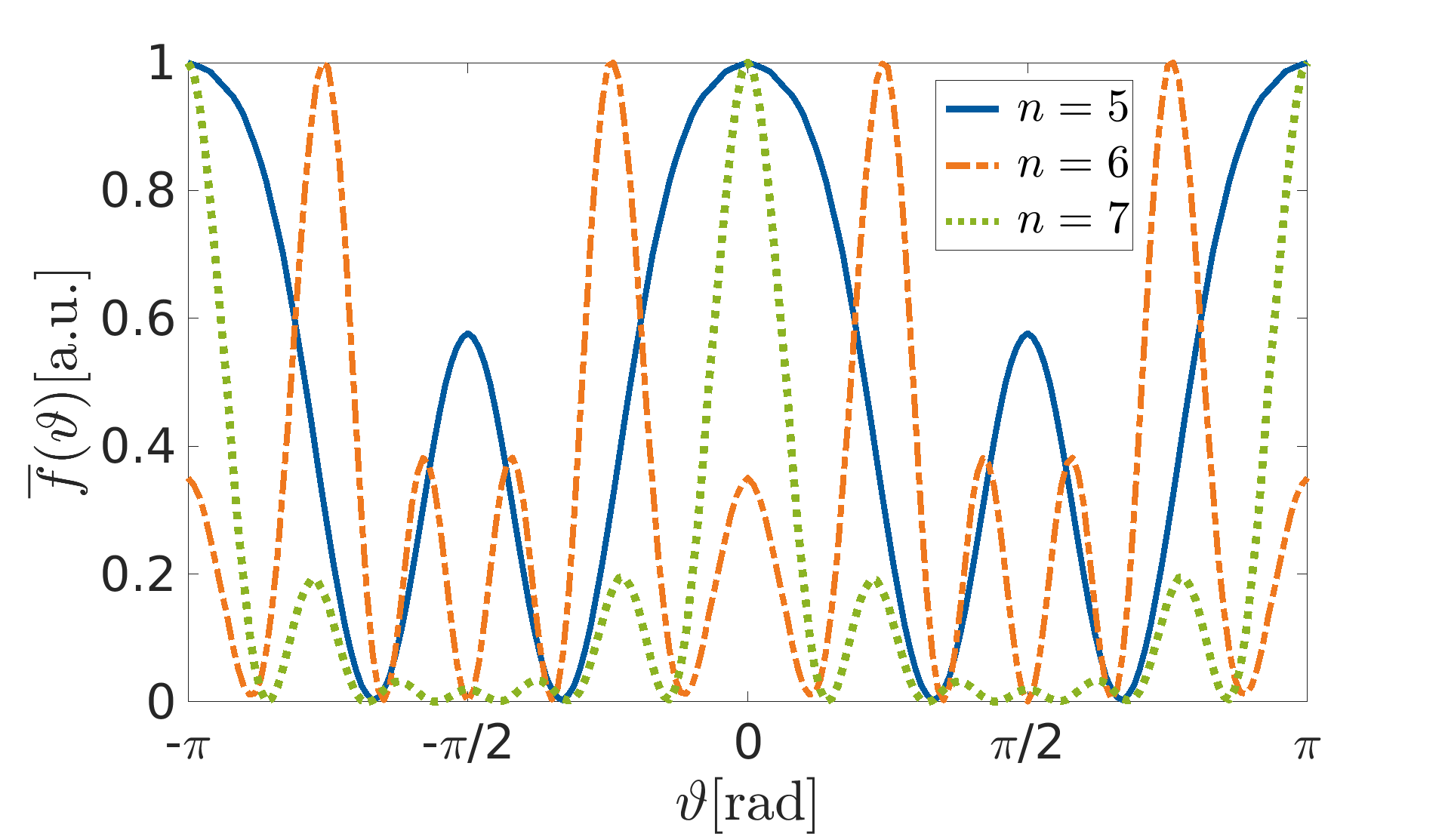} 
      \end{center}
      \caption{Normalized radial distribution function $\overline f(\vartheta)$ for various $n$-photon channels for a configuration with field frequency $\omega = 0.5m$,
        field strength $e\varepsilon = 0.5m^2$ and pulse length $\tau = 25m^{-1}$. The higher the photon number $n$ the more minima/maxima in the spectrum are visible. 
        In particular at $\vartheta = \pm \pi/2$ we have alternating extrema. 
        The variable $\vartheta$ rotates clockwise with starting point $(p_{x,n},0)$.} 
      \label{fig:fth2}
      \end{figure}        
            
Employing a background field with a field strength of $e\varepsilon=0.5m^2$ and a field frequency of $\omega=0.5m$ we find many characteristic traits
of photon absorption in the particle distribution function, see Fig. \ref{fig:fpxpz3}. These structures are thus best discussed in terms of energy conservation.
The only energy source is given by the absorption of photons $n \omega$. The total absorption energy must be equal
the total particle energy ${\cal E}_n$, which is given by the particles' rest energy plus their ponderomotive energy. The former can be stated easily as 
$E = \sqrt{ m^2 + p_x^2 + p_z^2}$, while the latter depends on the fields' strength, frequency and polarization direction, c.f. Ref. \cite{Otto:2014ssa} for a complete derivation.
Here, the oscillations in the electric field only lead to a modification of the particles' parallel momentum component
\begin{multline}
 {\cal E} \left(p_x, p_z \right) = \\
 \frac{\omega}{2 \pi} \int_{-\pi/\omega}^{\pi/\omega} {\rm d}t' \ \sqrt{ m^2 + \left( p_x + \frac{e\varepsilon}{\omega} \sin \left(\omega t' \right) \right)^2 + p_z^2}.
\end{multline}
As a result the various absorption channels form ellipses in the particle phase-space instead of circles as one would naively expect.

For the sake of a better understanding of the interference pattern we display the particle distribution along the ellipses in Fig.~\ref{fig:fth2}.
Note, that we have used linear interpolation techniques to illustrate the distribution function $f (p_x, p_z)$ as a function of the polar angle $\vartheta$,
where $\vartheta$ is defined as the ejection angle with respect to the fields' polarization direction. 
We immediately see, that lines with an odd number of photons are strongest at $\vartheta=0$ and $\vartheta=\pi$ ($p_z=0$) and show side maxima at 
$\vartheta=\pm \pi/2$ ($p_x=0$). In turn, an even photon number corresponds to minima at $\vartheta=-\pi/2$ and $\vartheta=+\pi/2$ (dotted orange line).

This behavior can be very well understood considering that the toy model given in Eq. \eqref{equ:A} still describes fields within the dipole approximation.
As parity as well as charge parity has to be conserved, the dipole approximation allows us to discuss the particles' angular distribution by 
simply counting the number of possible final quantum states. To be more specific, the intrinsic parity of an electron-positron pair is $(-1)$. The 
particles' orbital momentum contributes by an additional factor $(-1)^L$. Charge conjugation symmetry gives $(-1)^{L+S}$ depending on the particles' spin 
orientation ($S=0$ or $S=1$, cf. Ref. \cite{Kohlfurst}). For the incoming photons we have C-parity $(-1)^n$ and parity $(-1)$ due to the fact that only electric dipole transitions are possible.
Nevertheless, upon absorption every photon transports a unit of angular momentum to the pair $\Delta L = \pm 1$. The change in the magnetic quantum number is zero though, 
because Eq. \eqref{equ:A} only describes linearly polarized waves.

An $n$-photon process therefor requires a parity of $(+1)$ for an even number of photons and $(-1)$ for an odd number. Consequently, 
if $n$ is even $L$ has to be odd. Performing a partial wave analysis \cite{Kohlfurst} we find that the final particle state can 
be conveniently written in terms of Legendre polynomials
\begin{equation}
 \psi \left( \vartheta \right) = \sum_L^n ~ b_L ~ P_L \left( \cos \vartheta \right).
 \label{equ:PL}
\end{equation}
In this case, $L$ is odd thus the sum in Eq. \eqref{equ:PL} is over the Polynomials with $L$ odd only. At vanishing parallel momentum we have $\vartheta=\pm \pi/2$ 
for which all these remaining Legendre polynomials vanish. As a result, in a pure multiphoton absorption process with an even number of photons,
neither electrons nor positrons can be emitted in a $90$ degree angle.

Besides, these structures in the spectrum are very sensitive to an external magnetic field. With increasing magnetic field strength the particles are
accelerated in transversal direction similarly to the results displayed in section \ref{Sec:Schwinger}.
The most notable difference is the intact symmetry in $p_z$ even for extremely strong magnetic fields. The reason is that in this case the particle formation time is
much longer, thus it is impossible to attribute a peak in $E(t,z)$ with a peak in $f(p_x,p_z)$. Consequently, particle trajectories are not unevenly separated
and as a result the symmetry in $p_z$ is not broken.

\subsection{Multi-mechanism pair production}

In the following, we discuss field configurations that combine multi-cycle pulses with high field strengths to enhance Schwinger pair production via absorption effects and vice versa.
In order to do so we employ a super-Gaussian envelope function to ensure that the background operates close
to its maximum value for multiple field cycles. The consequences are twofold. At the one hand, chances for Schwinger pair production increase,
because also the ``side peaks'' can produce a sizable amount of particles. On the other hand, a higher amount of
significant field oscillations also increases the likelihood of $n$-photon absorption processes. 

\subsubsection{Quasi-homogeneous fields}

      \begin{figure}[t]
      \begin{center}
	\includegraphics[trim=3.5cm 0cm 3.5cm 0cm,clip,width=\figlenB]{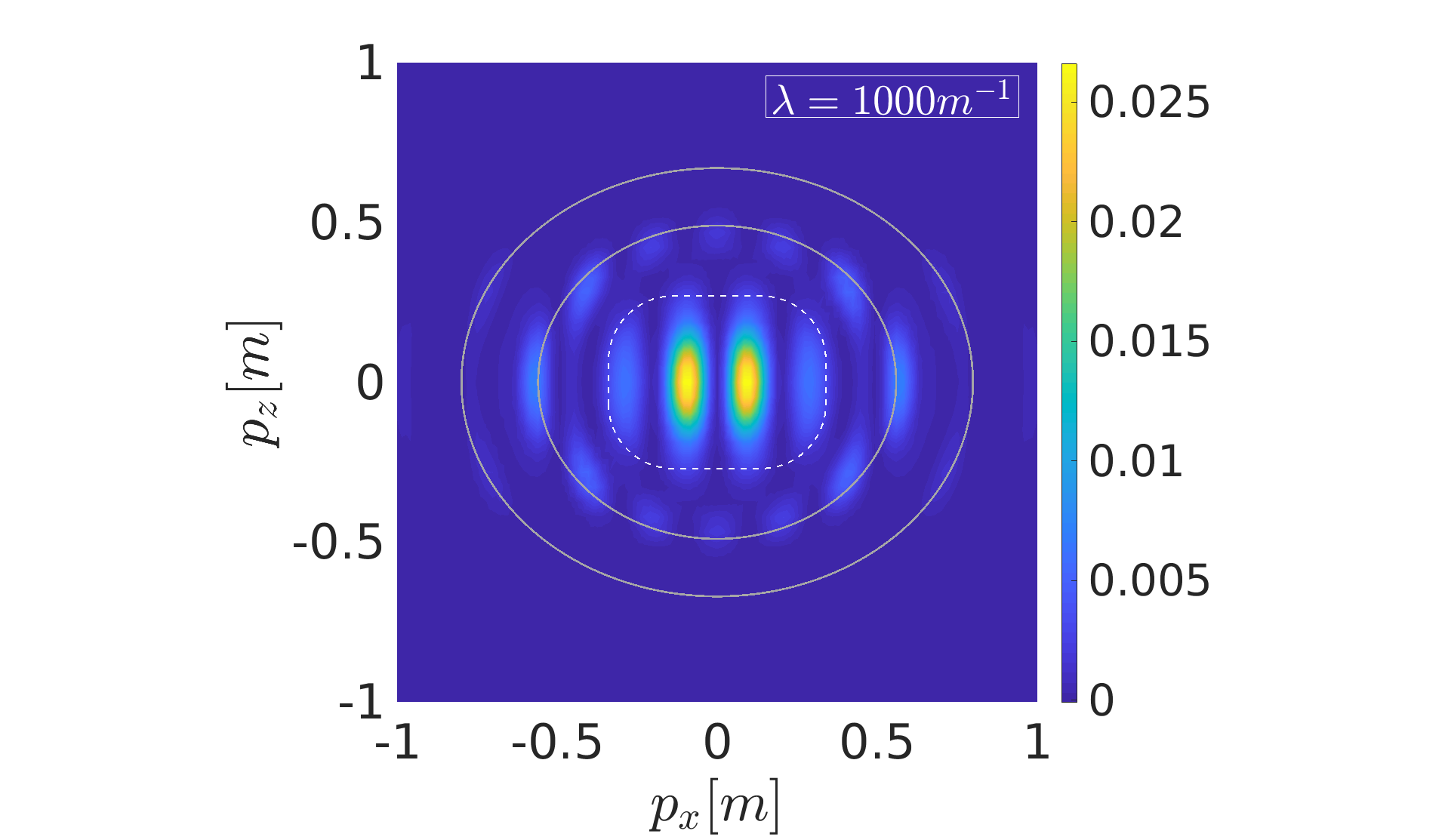} 
	\includegraphics[trim=3.5cm 0cm 3.5cm 0cm,clip,width=\figlenB]{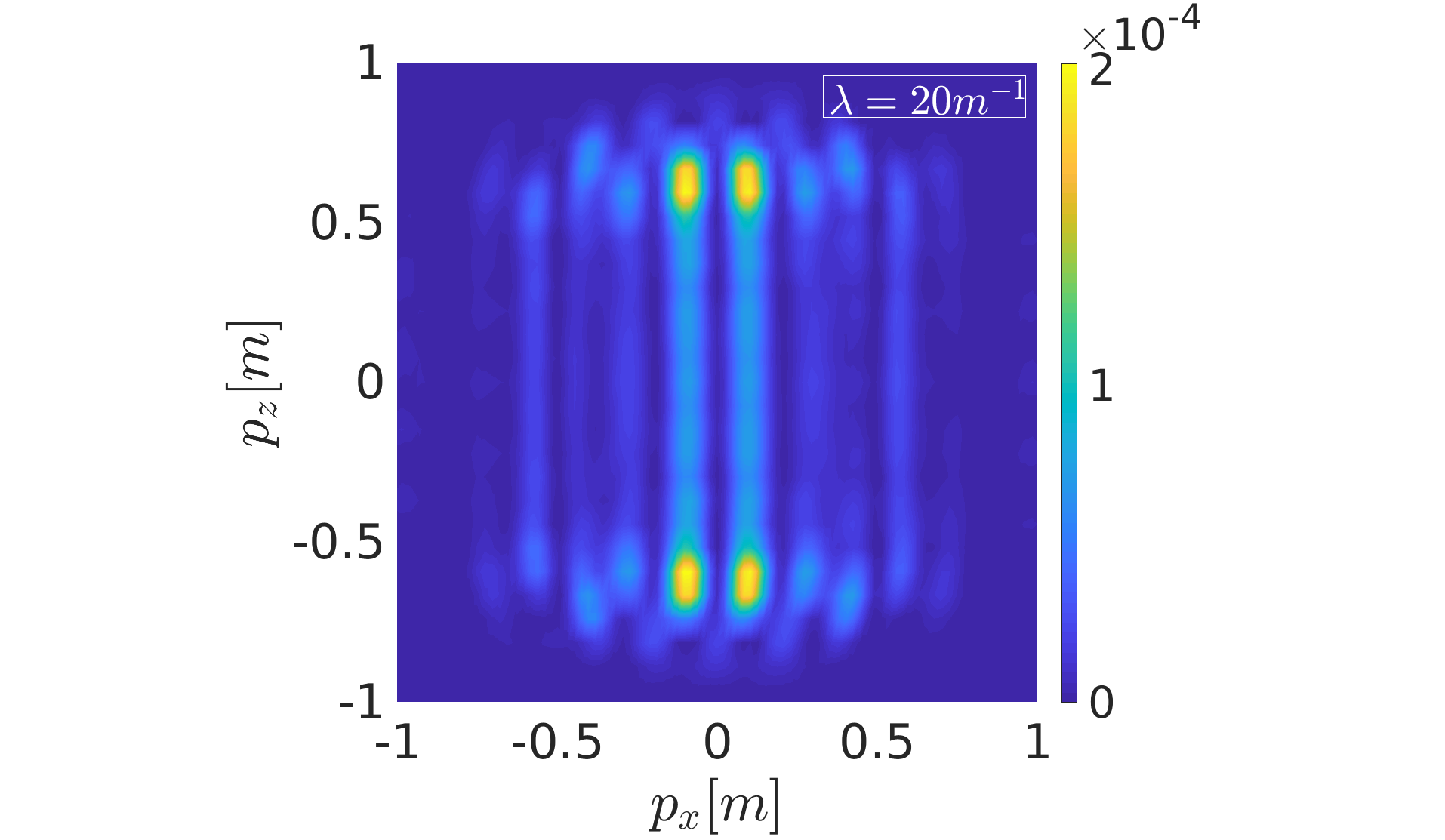} 
      \end{center}
      \caption{Particle distribution $f(p_x, p_z)$ for a background field featuring a field strength of $e\varepsilon = 0.2m^2$, 
        a super-Gaussian envelope with pulse length $\tau = 75m^{-1}$, a field frequency of $\omega = 0.2m$ and a spatial extent of 
        $\lambda = 1000m^{-1}$ (top) as well as $\lambda = 20m^{-1}$ (bottom). The spectrum shows a Schwinger-like distribution (within white dashed area)
        as well as multiphoton-like patterns (ellipses).} 
      \label{fig:FPxPz1}
      \end{figure}   

     \begin{figure}[t]
      \begin{center}
	\includegraphics[trim=3.5cm 0cm 3.5cm 0cm,clip,width=\figlenB]{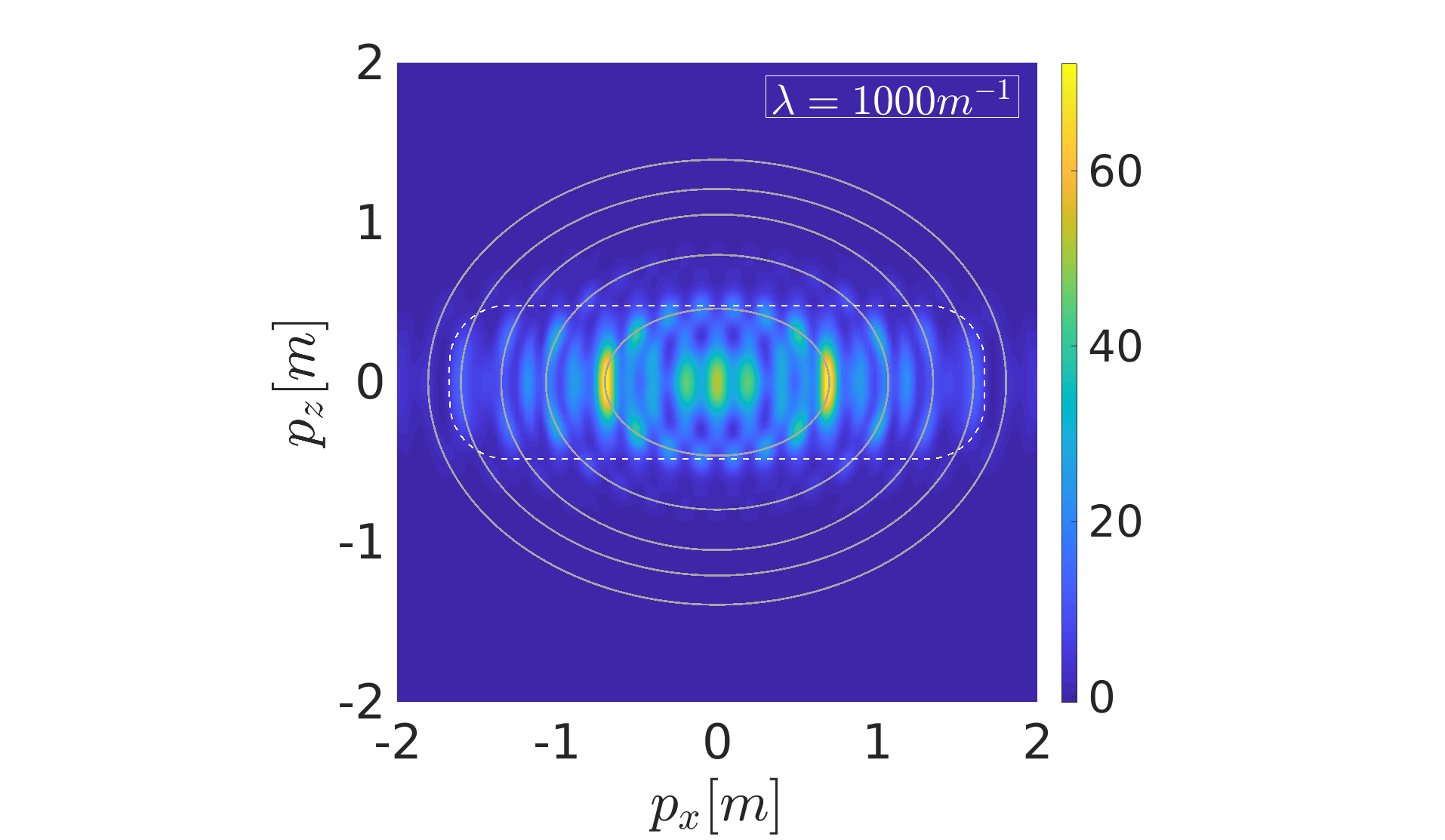} 
	\includegraphics[trim=3.5cm 0cm 3.5cm 0cm,clip,width=\figlenB]{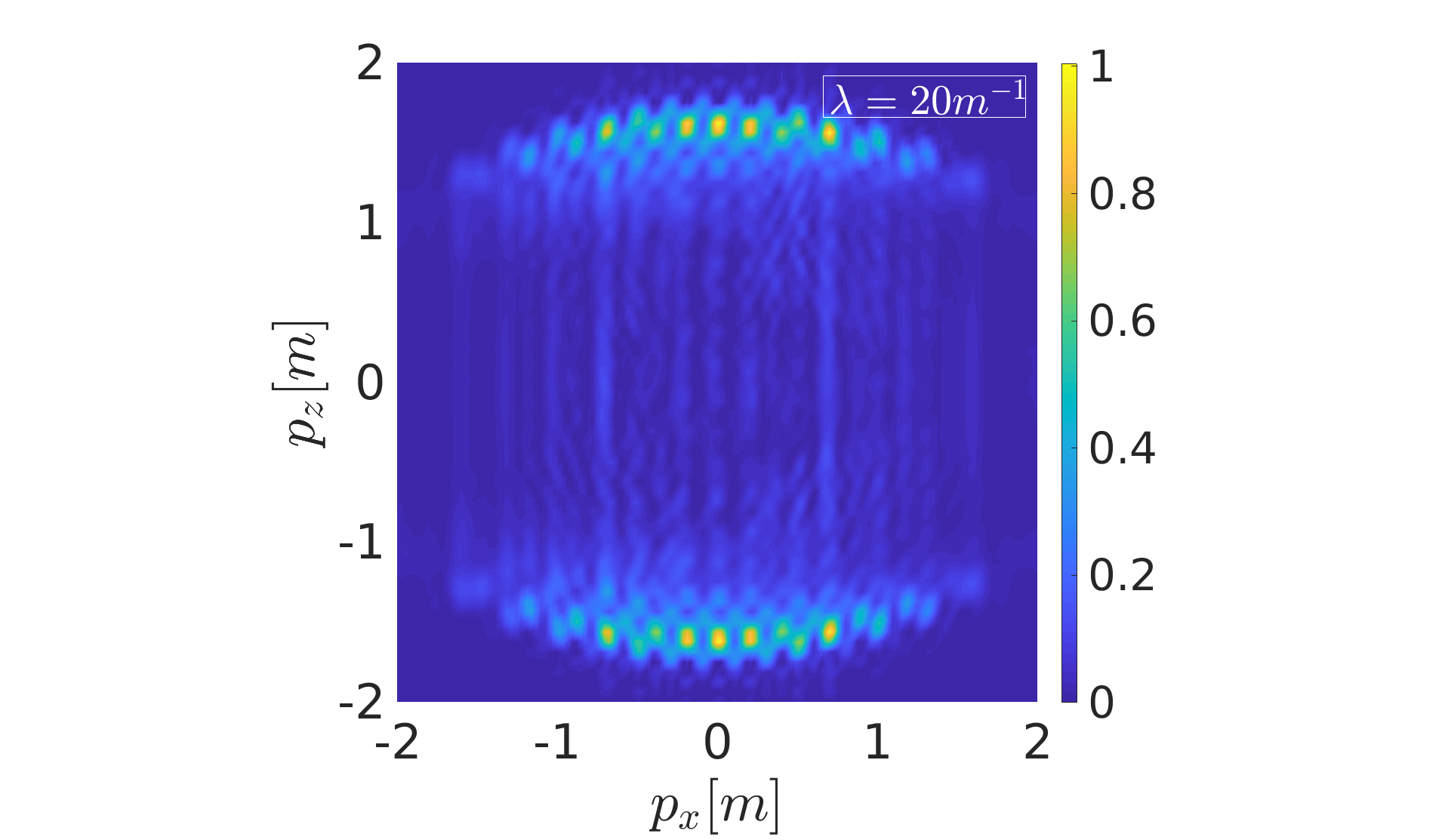} 
      \end{center}
      \caption{Particle distribution $f(p_x, p_z)$ for a background field featuring a field strength of $e\varepsilon = 0.5m^2$, 
        a super-Gaussian envelope with pulse length $\tau = 75m^{-1}$, a field frequency of $\omega = 0.2m$ and a spatial extent of 
        $\lambda = 1000m^{-1}$ (top) as well as $\lambda = 20m^{-1}$ (bottom). 
        The main structure (cigar shaped area) is superposed by multiple ellipses stemming from multiphoton pair production giving rise to additional quantum interferences.} 
      \label{fig:FPxPz2}
      \end{figure} 
      
      \begin{figure}[t]
      \begin{center}
	\includegraphics[width=\figlenB]{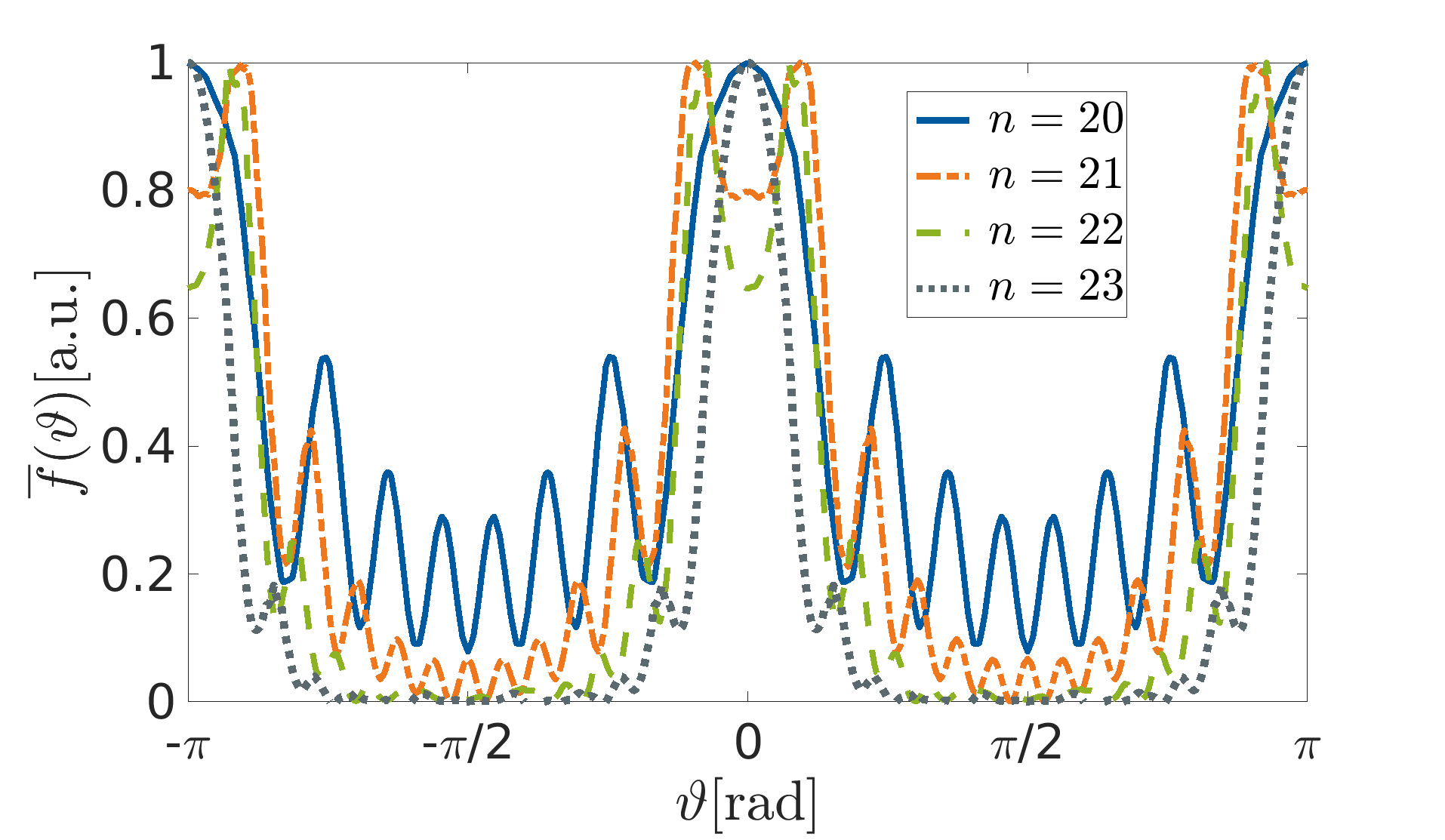} 
      \end{center}
      \caption{Normalized radial distribution function $\overline f(\vartheta)$ for various above-threshold signals in momentum space for
      a background field with peak field strength $e\varepsilon = 0.5m^2$, pulse length $\tau = 75m^{-1}$ (super-Gaussian envelope) and frequency $\omega = 0.2m$. 
      Due to the high intensities the threshold for 
      pair production is increased thus particles acquire less kinetic energy. Here, the photon numbers $n$ correspond to effective energies of ${\cal E}_{20}=0.65m$, ${\cal E}_{21}=1.06m$,
      ${\cal E}_{22}=1.36m$ and ${\cal E}_{23}=1.60m$.} 
      \label{fig:fth3}
      \end{figure}

\ctable[
caption = {Table showing the effective energy of the particles created through the absorption mechanism. In order to 
compare the outcome of the simulation with the effective energy model we
fit ellipses to the particle spectrum. This gives us the shape parameters $a$ and $b$. The parameters
$p_x^\ast$ and $p_z^\ast$ determine the ellipses of equal effective energy ${\cal E} \left(p_x^\ast, p_z^\ast \right)$.
Background field: $e\varepsilon = 0.2m^2$, $\tau = 75m^{-1}$ (super-Gaussian envelope), $\omega = 0.2m$.},
label   = Tab_Ell,
pos     = b,
width = \tablen,
]{lcccc}{
}{                                
\FL
$n$ \hspace{0.65cm} & \hspace{0.65cm} $a$ \hspace{0.65cm} & \hspace{0.65cm} $b$ \hspace{0.65cm} & \hspace{0.65cm} $p_x^\ast$ \hspace{0.65cm} & \hspace{0.65cm} $p_z^\ast$ \hspace{0.65cm}  \ML
13 & 0.56 & 0.49 & 0.52 & 0.45 \NN
14 & 0.80 & 0.67 & 0.78 & 0.68 \NN
15 & 0.98 & 0.83 & 0.97 & 0.87 \LL
}

\ctable[
caption = {Comparison of the effective energy model with the outcome of the simulation for particles created through the absorption mechanism.
We fit ellipses to the particle spectrum in order to obtain the shape parameters $a$ and $b$. The parameters
$p_x^\ast$ and $p_z^\ast$ determine the ellipses of equal effective energy ${\cal E} \left(p_x^\ast, p_z^\ast \right)$.
Background field: $e\varepsilon = 0.5m^2$, $\tau = 75m^{-1}$ (super-Gaussian envelope), $\omega = 0.2m$.},
label   = Tab_Ell2,
pos     = b,
width = \tablen,
]{lcccc}{
}{                                
\FL
$n$ \hspace{0.65cm} & \hspace{0.65cm} $a$ \hspace{0.65cm} & \hspace{0.65cm} $b$ \hspace{0.65cm} & \hspace{0.65cm} $p_x^\ast$ \hspace{0.65cm} & \hspace{0.65cm} $p_z^\ast$ \hspace{0.65cm}  \ML
20 & 0.70 & 0.46 & 0.64 & 0.44 \NN
21 & 1.07 & 0.80 & 1.06 & 0.75 \NN
22 & 1.35 & 1.05 & 1.36 & 0.98 \NN
23 & 1.60 & 1.21 & 1.60 & 1.18 \NN
24 & 1.80 & 1.39 & 1.80 & 1.36 \NN
25 & 1.99 & 1.45 & 1.98 & 1.52 \LL
}      

As one can see in Fig.~\ref{fig:FPxPz1} the calculated particle spectrum indeed displays a mixture of exponentially decaying as well as elliptical structures. 
At vanishing particle momentum we obtain an interference pattern typical for tunneling-enhanced pair production. To be more specific,
the field configuration under consideration (super-Gaussian envelope with $e\varepsilon=0.2m^2$ and $\omega = 0.2m$) is at the edge of seeing the $12$-particle channel directly.
Nevertheless, the absorption of $12$ photons creates a highly excited state. As this is the equivalent of lowering the threshold by the same amount 
subsequent tunneling is barely suppressed. In this way the onset of a new channel can be seen very well even below the threshold. On a side
note, the particle distribution falls off exponentially in transversal direction, which is a clear sign for tunneling.

We analyze this area using the simple man's model. As the model's output is a smooth particle density peaked at vanishing momenta
we can easily obtain the equipotential lines, where the distribution function holds $10\%$ of its maximal value. 
In case of $e\varepsilon=0.2m^2$ we obtain the shape parameters $a=0.375$ and $b=0.3$. 
This is extremely close to the values obtained from the DHW calculation: $p_x^{\ast}=0.34$ and $p_z^{\ast}=0.27$. 

Similar holds for the widespread particle distribution in $p_x$ for strong fields, see Fig. \ref{fig:FPxPz2}.
In terms of the simple man's model particles are produced around the main
peaks of the electric field $t_k = \dfrac{k \pi}{\omega}$ with $k=-2,-1,\ldots 2$. In a quasi-homogeneous setup, a particle created exactly at one of these peaks acquires nearly
no net momentum due to the almost flat envelope. However, the high field strength of $e\varepsilon = 0.5m^2$ allows for easy particle production at the slopes, too. For example,
at $\tilde t=2.25m^{-1}$ the electric field shows a local field strength of $E(\tilde t,0) = 0.4 E_{\rm cr}$. As the vector potential is highly nonzero at $\tilde t$, 
particles created at this instant in time are strongly accelerated up to a final momentum of $p_{x,f} \approx 1.09m$. As a result, a sizable amount of particles
can be found even at large parallel momenta, e.g. $f(p_x \approx 1.1m) \approx \frac{1}{4} f(0)$.

In addition, elliptical structures appear at higher particle momenta in Figs. \ref{fig:FPxPz1}-\ref{fig:FPxPz2}, 
which can be interpreted as $13+$ and $20+$ photon absorption processes, respectively. The corresponding analysis using fits to
determine the corresponding effective energy of the elliptic particle distributions in comparison with
the predictions from the effective energy model is given in Tabs. \ref{Tab_Ell}-\ref{Tab_Ell2}. The most surprising result is that 
simulation and model deviate slightly in assessing the particles final transversal momentum ($\Delta p_z ~ \sim 5\%$). Predictions
for the parallel momenta perfectly agree with the simulation for all above-threshold peaks, while for each of the lowest ellipses 
the model underestimates the final value by up to $\sim 9 \%$. It is very likely that for such low energies remnants of tunneling 
influence the final particle momentum.

      \begin{figure}[t]
      \begin{center}
	\includegraphics[width=\figlenB]{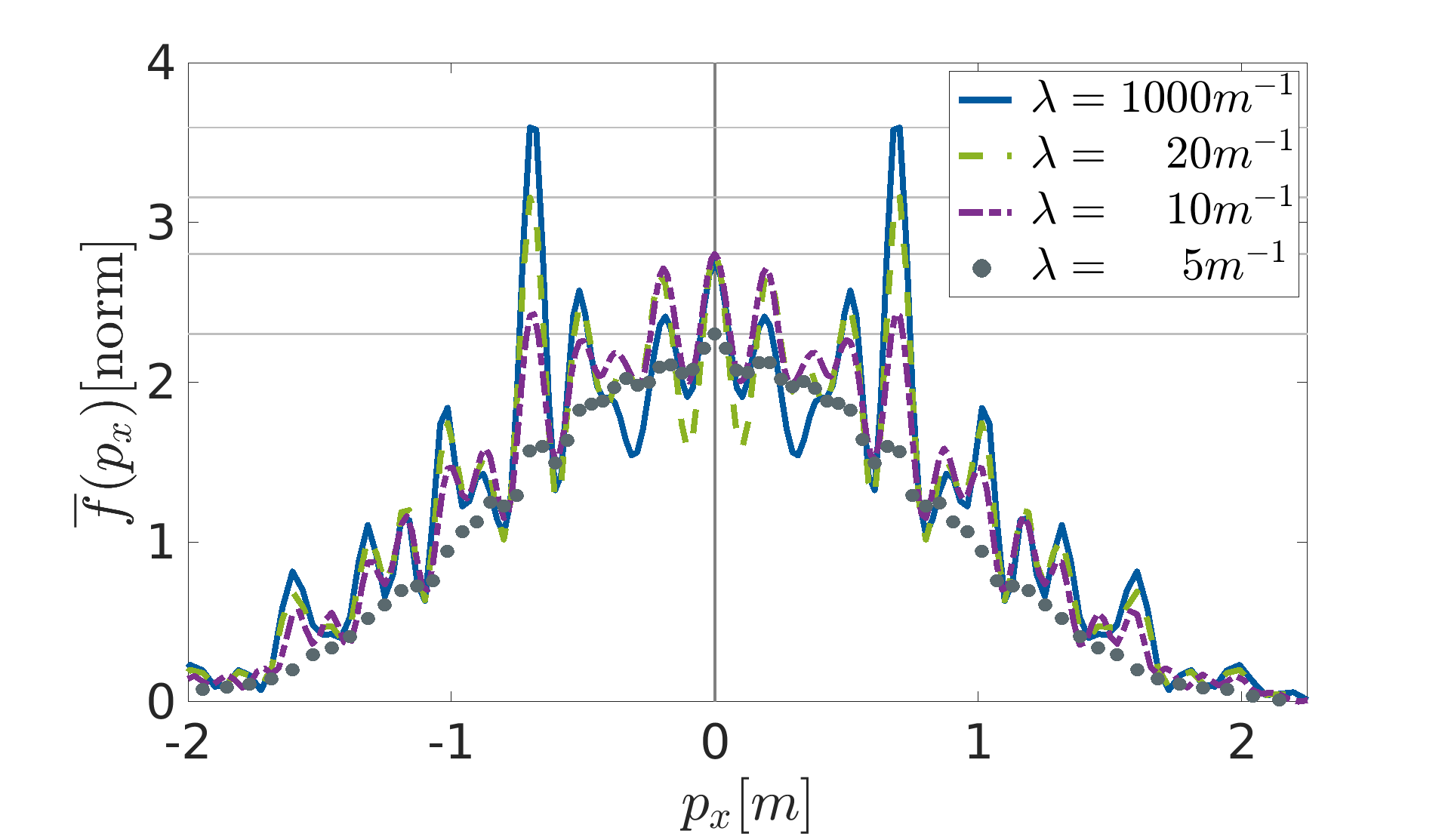} 
      \end{center}
      \caption{Normalized particle distribution as a function of parallel momentum $p_x$. The oscillations in $f(p_x)$ are composed of multiple frequencies,
        which vanish the smaller $\lambda$ gets and thus the stronger the applied magnetic field becomes. 
        Parameters: $e\varepsilon = 0.5m^2$, $\tau = 75m^{-1}$ (super-Gaussian envelope), $\omega = 0.2m$.} 
      \label{fig:fpxpz4a}
      \end{figure}  

Similarly to the previously presented case for highly oscillating fields, we display the multiphoton structures obtained for the strong field configuration 
in terms of radial distribution functions, see Fig. \ref{fig:fth3}. 
Due to the large photon count $n$ the maximal orbital angular momentum particles can acquire is much higher than in the previous case. As a result,
these functions show a large number of side peaks. Moreover, multiphoton and tunneling distributions interfere making an evaluation based on conservation laws difficult. 
The wild pattern around $\vartheta=\pi$ are a remnant of this superposition. 
Nevertheless, the considerations on parity and C-parity still hold approximately. Hence, 
the $n=20$ channel still exhibits local minima at $\vartheta = -\pi/2$ and $\vartheta = +\pi/2$.

\subsubsection{Strong magnetic fields} 

      \begin{figure}[t]
      \begin{center}
	\includegraphics[width=\figlenB]{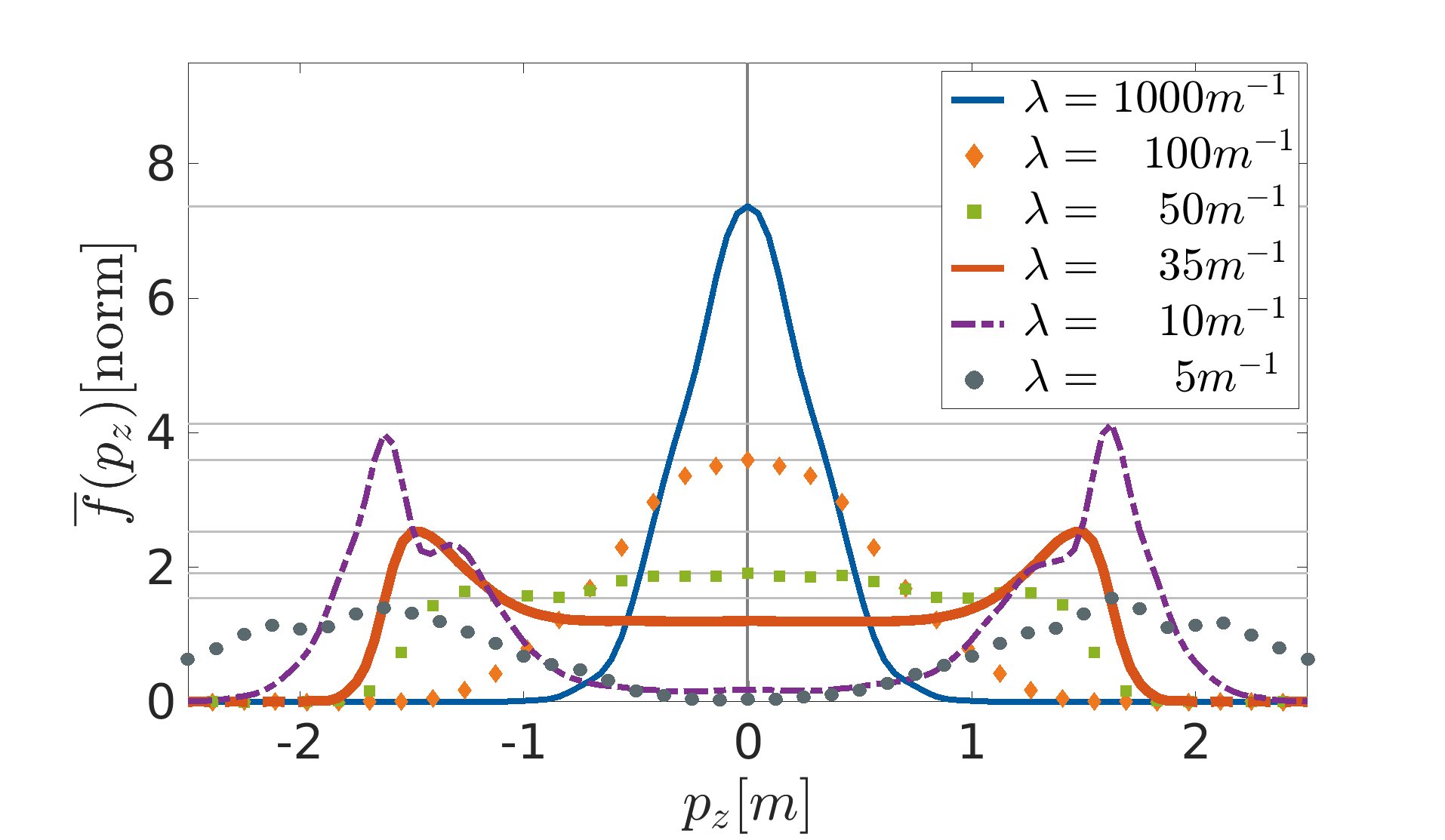} 
      \end{center}
      \caption{Normalized particle distribution function $\overline f(p_z)$ for multiple values of $\lambda$ (smaller $\lambda$ corresponds to higher magnetic field
        strength). The change in the distributions with increased magnetic field strength 
        can be understood as a two-step process. For weak magnetic fields particles that have already been accelerated by the electric field are deflected by
        the magnetic field in direction of $p_z$. The stronger the magnetic field becomes the stronger the effect. At a certain limit ($\lambda \sim 10m^{-1}$) the particle rate
        drops considerably and a strong magnetic field prevents a clear signal to form. Additionally, for high magnetic field strengths the symmetry in $p_z$ is broken. 
        Field parameters: $e\varepsilon = 0.5m^2$, $\tau = 75m^{-1}$ (super-Gaussian envelope), $\omega = 0.2m$.} 
      \label{fig:fpxpz4b}
      \end{figure}  
      
\ctable[
caption = {Table of total particle number $N$ as well as the normalized particle number $\overline N = N/N_{\rm cl}$ as a function of the spatial extent $\lambda$. 
All other field parameters have been kept
fixed at $e\varepsilon=0.5m^2$, $\tau=75m^{-1}$ (super-Gaussian envelope) and $\omega = 0.2m$. For comparison, for every value of $\lambda$ the deviation 
to an assumed linear dependency is shown $\Delta N_{\rm lin}$.},
label   = Tab_N,
pos     = h,
width = \tablen,
]{lcccc}{}{                                
\FL
& \hspace{0.5cm} $\lambda$ \hspace{0.5cm} & \hspace{0.5cm} $N$ \hspace{0.5cm} & \hspace{0.5cm} $N/N_{\rm cl}$ \hspace{0.5cm} & \hspace{0.5cm} $\Delta N_{\rm lin} [\%]$ \ML
& 1000 & 53.36 & 5.17 & 0 \NN
& 100 & 5.29 & 5.14 & 0.9 \NN
& 50 & 2.70 & 5.26 & 1.1 \NN
& 35 & 1.85 & 5.16 & 0.9 \NN
& 20 & 1.05 & 5.13 & 1.7 \NN
& 10 & 0.50 & 5.01 & 6.6 \NN
& 5 & 0.196 & 4.15 & 31.4 \LL
}              

In multi-cycle fields the particles' phase-space occupancy is much more involved, thus quantum
interferences form easily. Nevertheless, when exposed to strong forces due to the magnetic field, particle bunches are accelerated in direction of $p_z$. However,
they are boosted in such a way that their relative quantum phases hardly change. In fact, all the individual peaks in the particle distribution function for quasi-homogeneous
fields can still be linked to the peaks observable in the spectrum for strong magnetic fields, c.f. Fig. \ref{fig:FPxPz2}.
What changes are the positions of these peaks in the spectrum as well as their relative size.         

The distribution function for parallel momenta $f(p_x)$, see the blue solid line in Fig.~\ref{fig:fpxpz4a}, indicates that the Schwinger effect is the main source of particle production.
In section \ref{Sec:Schwinger} we have already established, that a smooth distribution function superposed by quantum interferences are typical signs of tunneling pair production.
The difference here is that we observe an irregular pattern on top of a broad spectrum. 
We interpret the data such that particles mostly tunnel through the Coulomb barrier, thus also the smooth decrease in $f(p_z)$ for small momenta $p_z$, c.f. Fig. \ref{fig:fpxpz4b}. The irregular oscillations that superpose the
smooth tunneling spectrum, c.f. Fig. \ref{fig:fpxpz4a}, are caused by multiphoton processes which includes assisted tunneling. These patterns are still visible if the spatial extent is chosen to
be small $\lambda=20m^{-1}$. Only at extreme values, $\lambda=5m^{-1}$, where pair production in general starts to break down, see Tab. \ref{Tab_N}, the interferences fade away.         
         
We complete this section by comparing the total particle yield $N$ for the configuration $e\varepsilon=0.5m^2$, $\tau=75m^{-1}$ (super-Gaussian envelope), $\omega = 0.2m$
for different values of $\lambda$. Similar investigations have already been performed for short, high-intensity fields \cite{Kohlfurst:2015niu}.
However, as the particle spectrum in this setup shows clear signatures of Schwinger as well as multiphoton pair production, c.f. Fig. \ref{fig:FPxPz2}, 
it is an ideal candidate for a study of the impact of magnetic fields on the general creation rate. Naively, one would expect a linear dependence on the 
spatial extent $\lambda$, given the particle yield scales linearly with the volume. This assumption is indeed true for wide fields $\lambda \ge 10m^{-1}$. 

For strongly focused fields, one might expect a faster-than-linear decrease due to the fact that the regions of significantly high effective field strength $E(t,z)^2-B(t,z)^2$
shrink non-linearly. However, studying the normalized yield $N/N_{\rm cl}$ in Tab. \ref{Tab_N} we find that starting with
$\lambda \approx 5m^{-1}$ the particle yield decreases much faster than expected. This might be a hint towards a critical point in a time-dependent, 
spatially inhomogeneous, high-intensity field \cite{Gies:2015hia}. Such an investigation, however, is beyond the scope of this article, and will be addressed elsewhere.

\section{Conclusion \& Outlook}
\label{Sec:Conclusion}

This study on pair production in inhomogeneous electromagnetic fields mark a significant step forward in demonstrating 
the capabilities of the Wigner formalism in general. Especially the possibility to perform calculations for almost arbitrarily focused 
backgrounds is a clear sign that quantum kinetic approaches are a valuable asset towards understanding quantum field theories in general.

To be more specific, we have adopted the Dirac-Heisenberg-Wigner formalism for large-scale computations by taking into account novel numerical methods. In this way, it was
possible to calculate the particle creation rates as well as their momentum spectra in inhomogeneous electromagnetic fields. In this regard,
we have significantly expanded the capabilities of quantum kinetic approaches such that we could finally study the impact of
a strongly inhomogeneous, time-dependent magnetic field on pair production processes even in long-pulsed fields. 

By thoroughly analyzing the so acquired data we were able to identify signatures of Schwinger as well as multiphoton effects
in an intermediate regime, where no creation mechanism is dominant. We further demonstrated how much impact the temporal envelope has
on the final particle distributions as we could easily enhance and suppress certain signatures by simply switching from a Gaussian to
a flat-top envelope. 

In the course of this study we also observed symmetry-breaking due to spin-field interactions in tunneling-dominated regions as well as vanishing above-threshold peaks
in absorption-dominated areas. For strong, multi-cycle field we were able to show that quantum interference patterns are 
preserved even in the presence of strong magnetic fields. Only if the spatial extent of the electric field is of the order of the Compton wavelength of the pair,
these interferences vanish as the pair production process breaks down independent of the regime.

\begin{acknowledgments}

We want to thank Holger Gies for many fruitful discussions.
Additionally, we want to thank Andr\'e Sternbeck for his support on
high-performance computing.
The work of CK is partially funded by the
Helmholtz Association through
the Helmholtz Postdoc Programme (PD-316) and by the BMBF under grant No. 05P15SJFAA (FAIR-APPA-SPARC).


\end{acknowledgments}

\onecolumngrid

\newpage

\section{Appendix}
\label{Sec:Appendix}

Collection of figures not suited for publication in the main body of the manuscript. Nevertheless, they carry interesting information,
thus we decided to present all data available. 

\vspace{1cm}

\noindent%
\begin{minipage}{\linewidth}
\begin{center}
	\includegraphics[trim=0cm 0cm 0cm 0cm,clip,width=\figlenB]{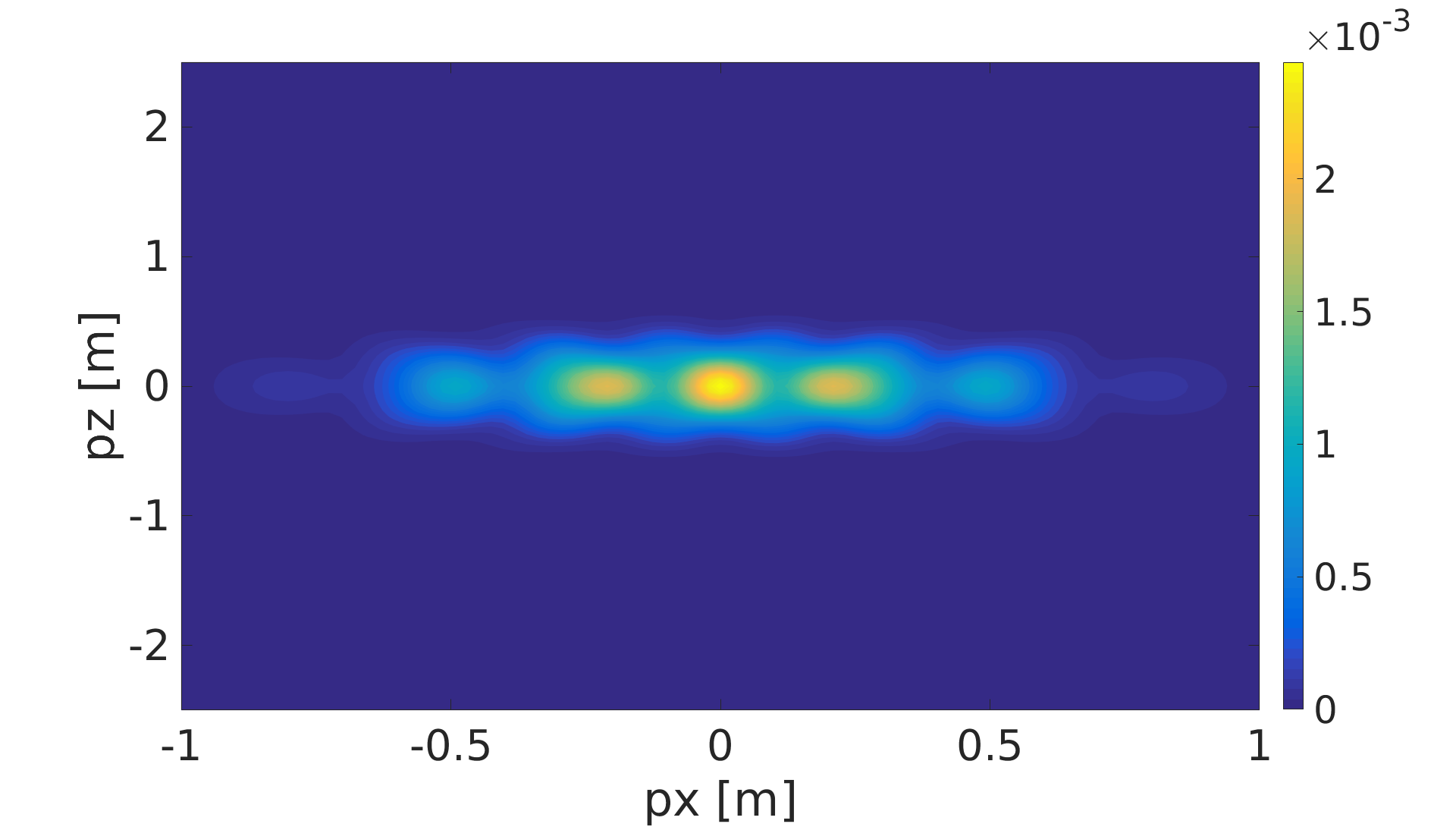} 
	\includegraphics[trim=0cm 0cm 0cm 0cm,clip,width=\figlenB]{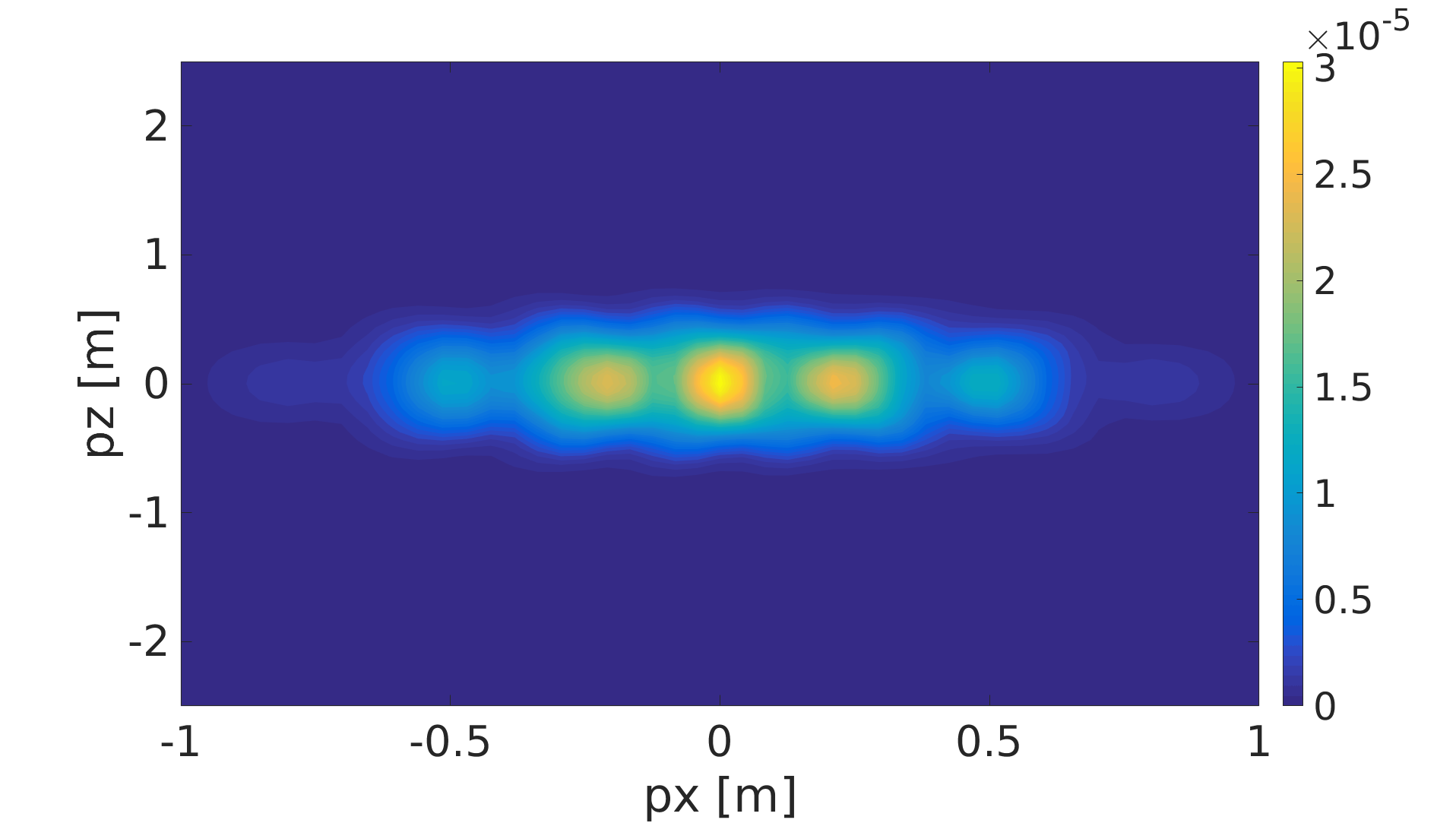} 	
	\includegraphics[trim=0cm 0cm 0cm 0cm,clip,width=\figlenB]{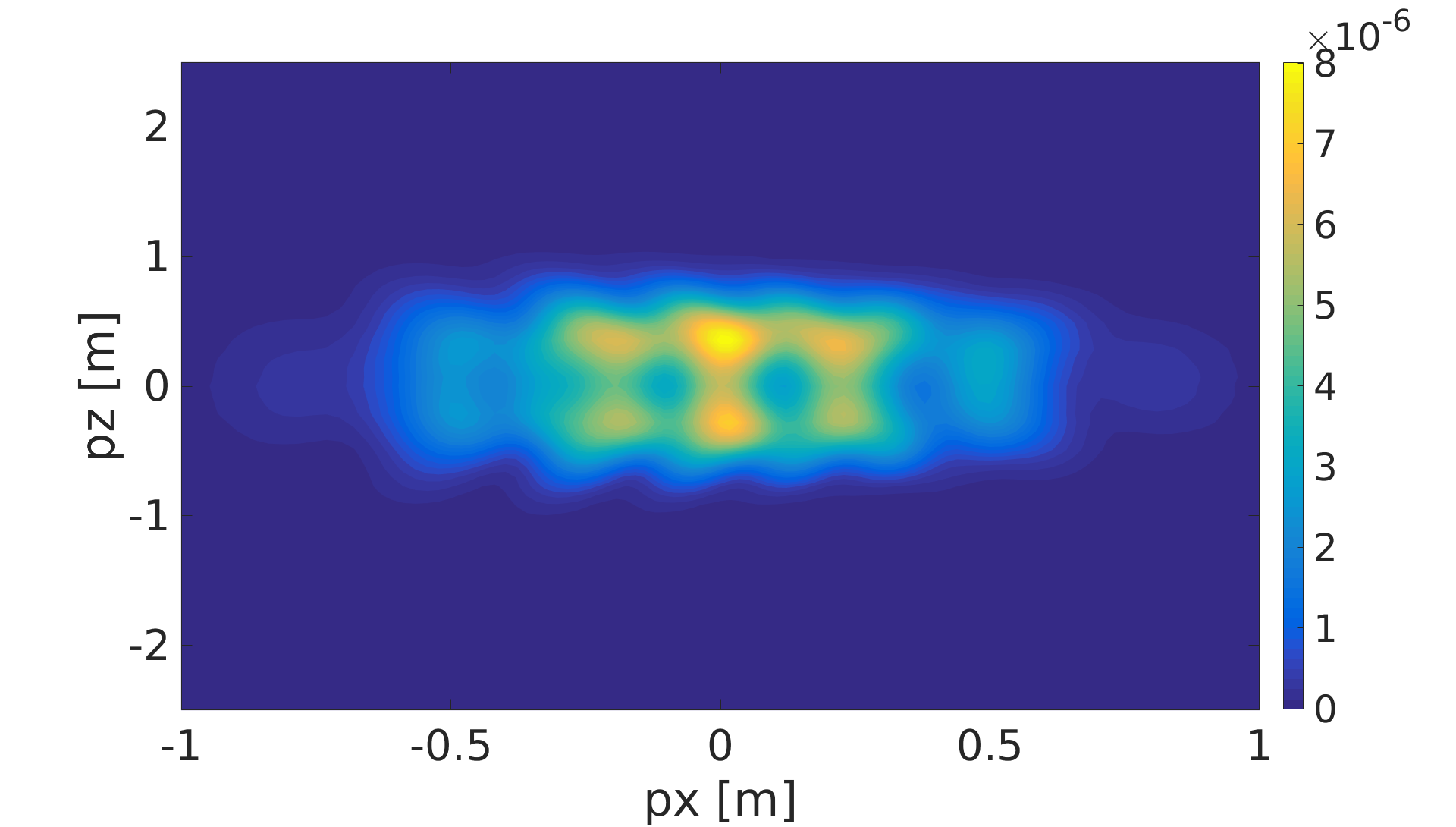} 
	\includegraphics[trim=0cm 0cm 0cm 0cm,clip,width=\figlenB]{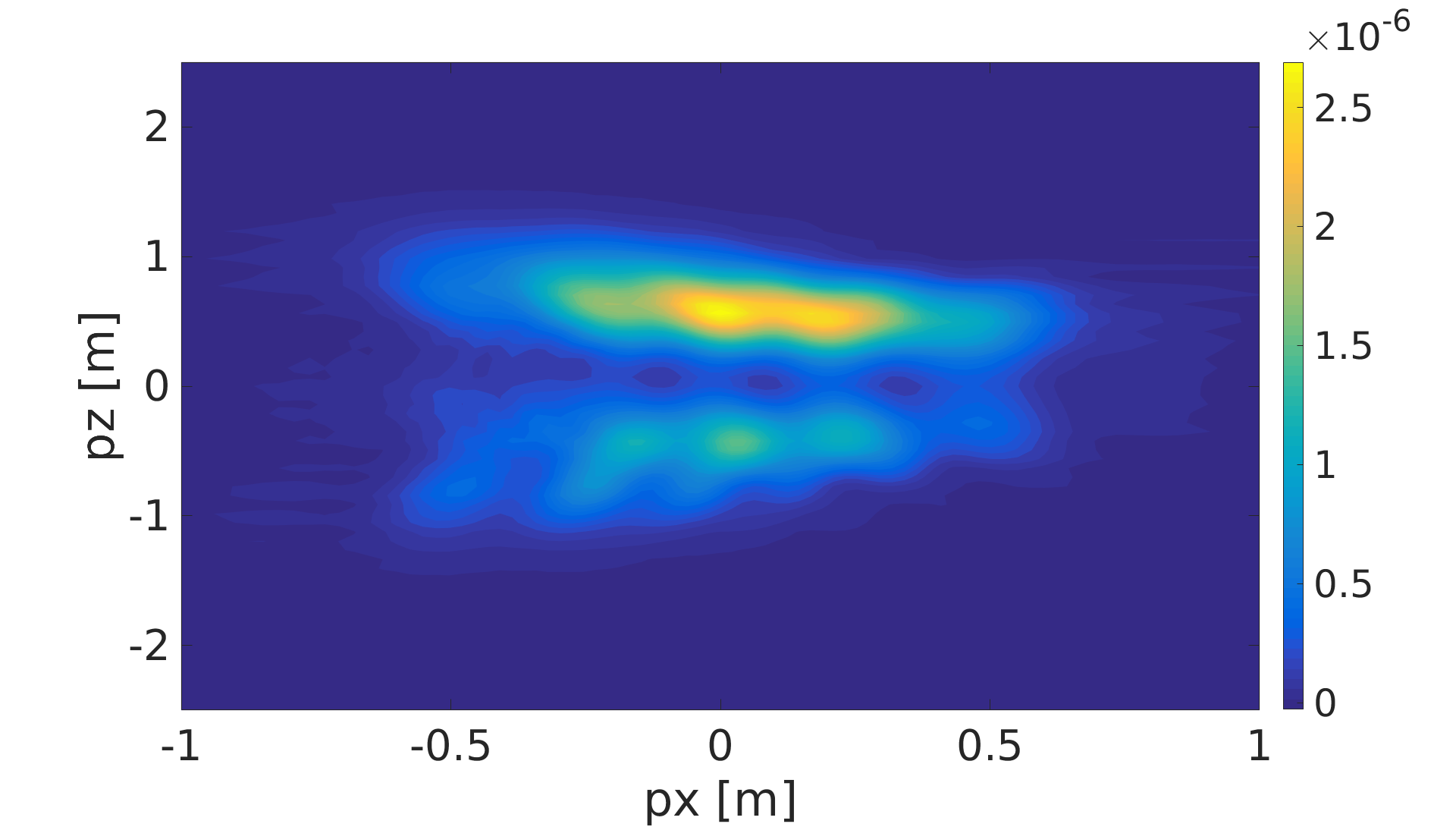} 
	\includegraphics[trim=0cm 0cm 0cm 0cm,clip,width=\figlenB]{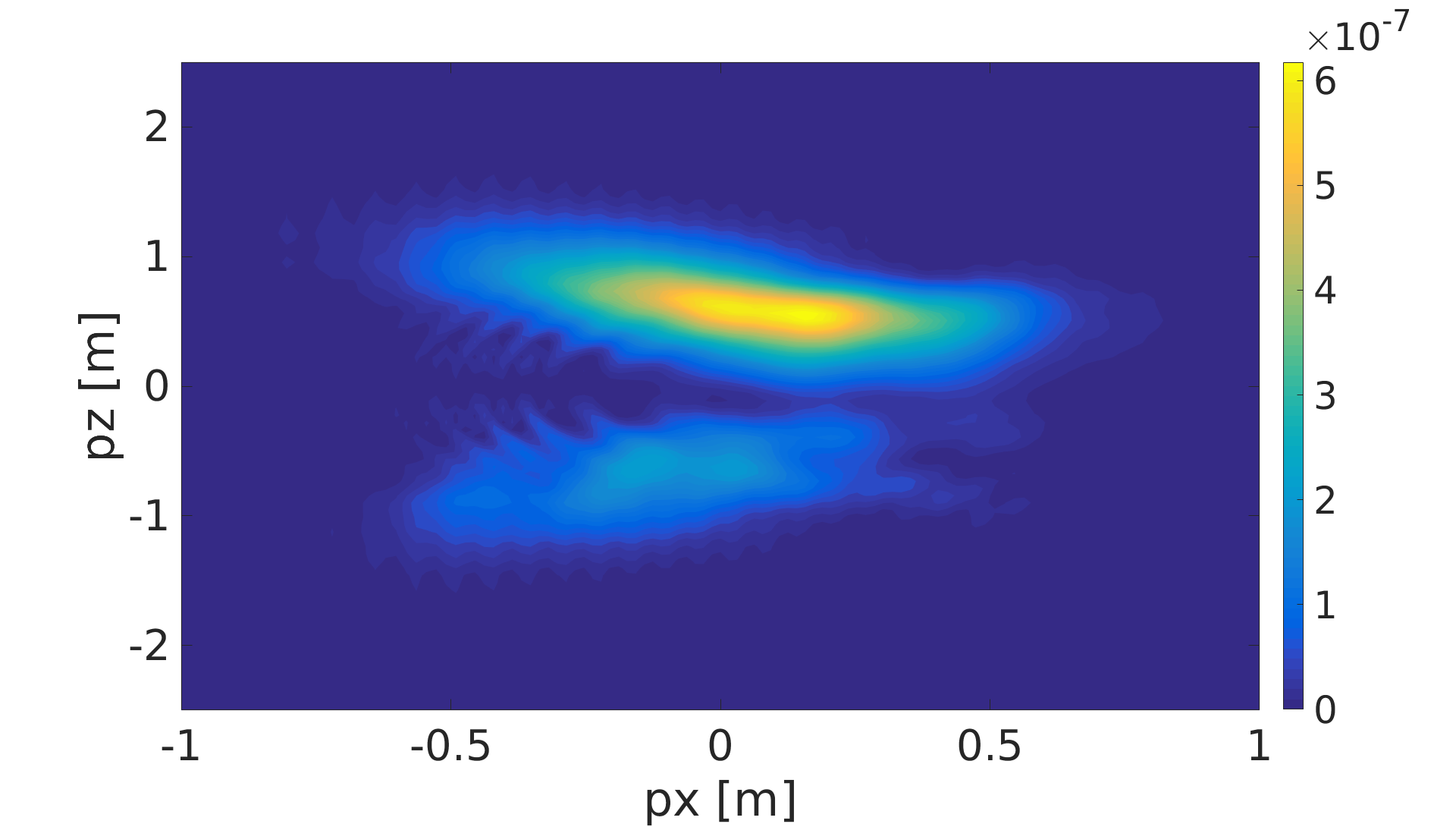}
\end{center}	
\begin{flushleft}
 \captionof{figure}{Particle momentum distribution as funtion of parallel ($p_x$) and perpendicular ($p_z$) momentum for various spatial extents $\lambda$. From top to bottom:
         $\lambda=1000m^{-1}$, $\lambda=20m^{-1}$, $\lambda=10m^{-1}$, $\lambda=5m^{-1}$ and $\lambda=3m^{-1}$. Particles are created due to the Schwinger effect
         and subsequently accelerated by electric and magnetic fields. The smaller $\lambda$ the stronger the magnetic field thus the more particles are pushed to
         non-vanishing transversal momenta.
 	Further parameters: $e\varepsilon = 0.2m^2$, $\tau = 25m^{-1}$ and $\omega = 0.2m$.}
\end{flushleft}
\end{minipage}   

\begin{figure}
	\includegraphics[trim=0cm 0cm 0cm 0cm,clip,width=\figlenB]{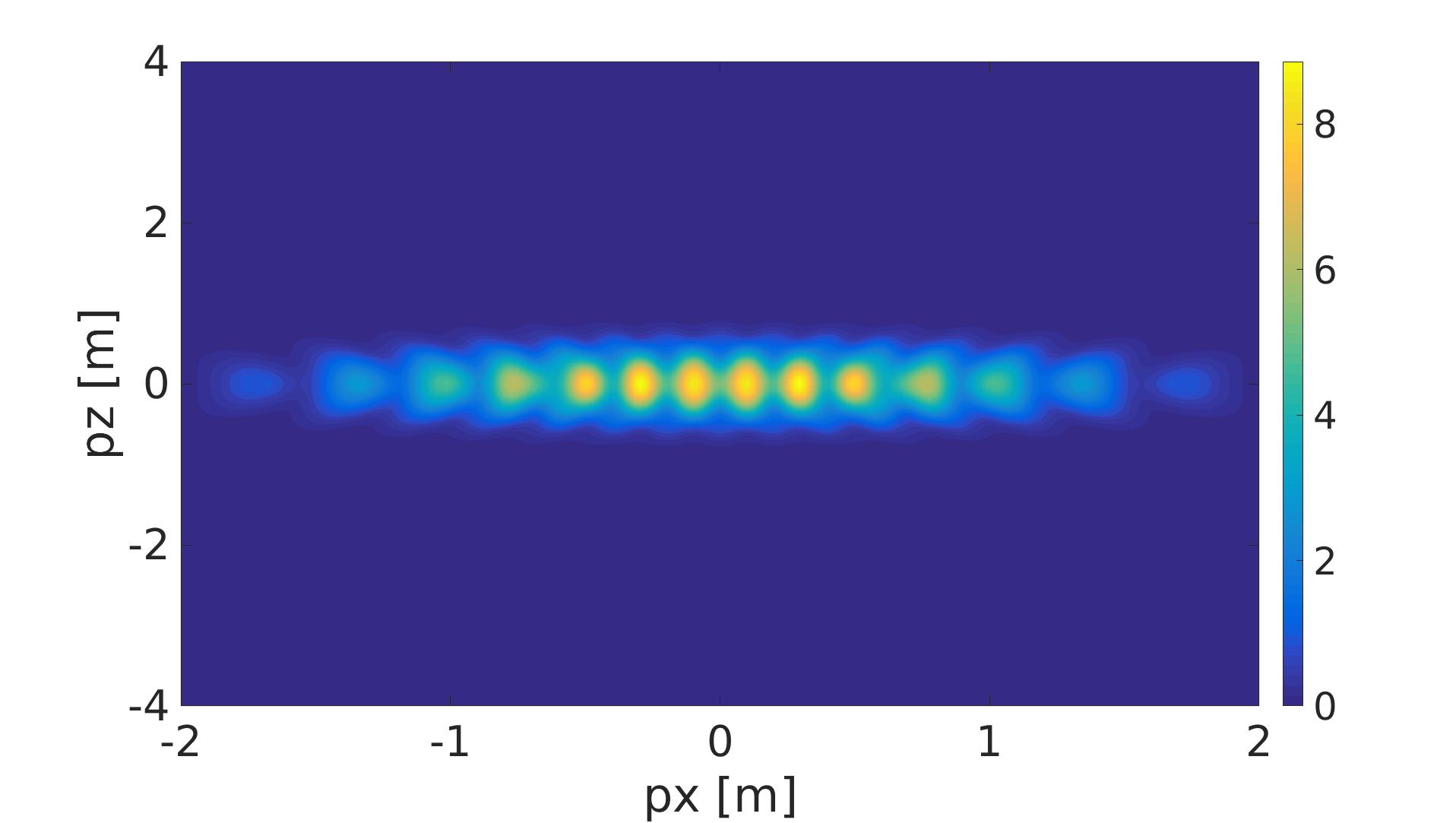} 
	\includegraphics[trim=0cm 0cm 0cm 0cm,clip,width=\figlenB]{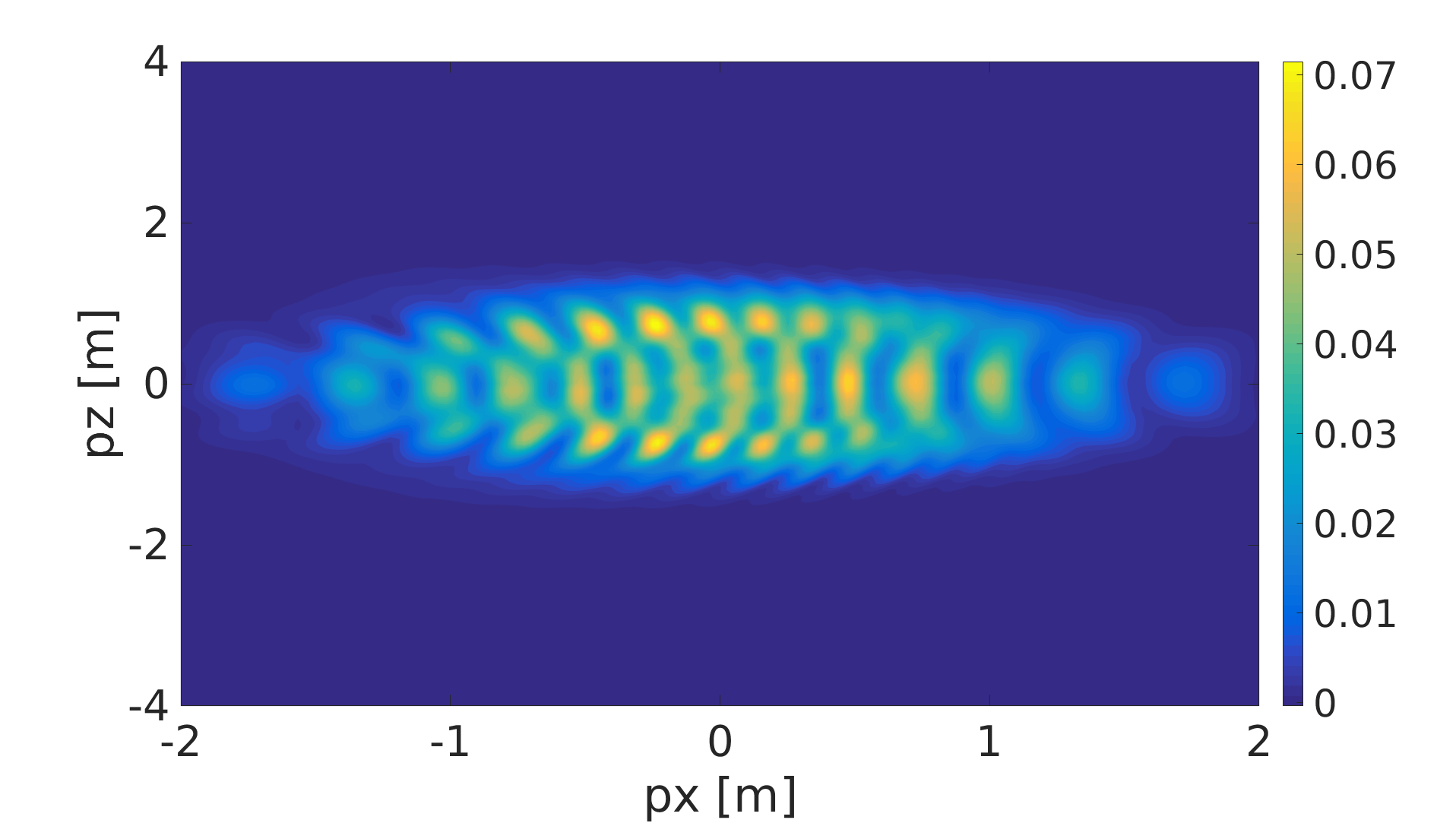} 
	\includegraphics[trim=0cm 0cm 0cm 0cm,clip,width=\figlenB]{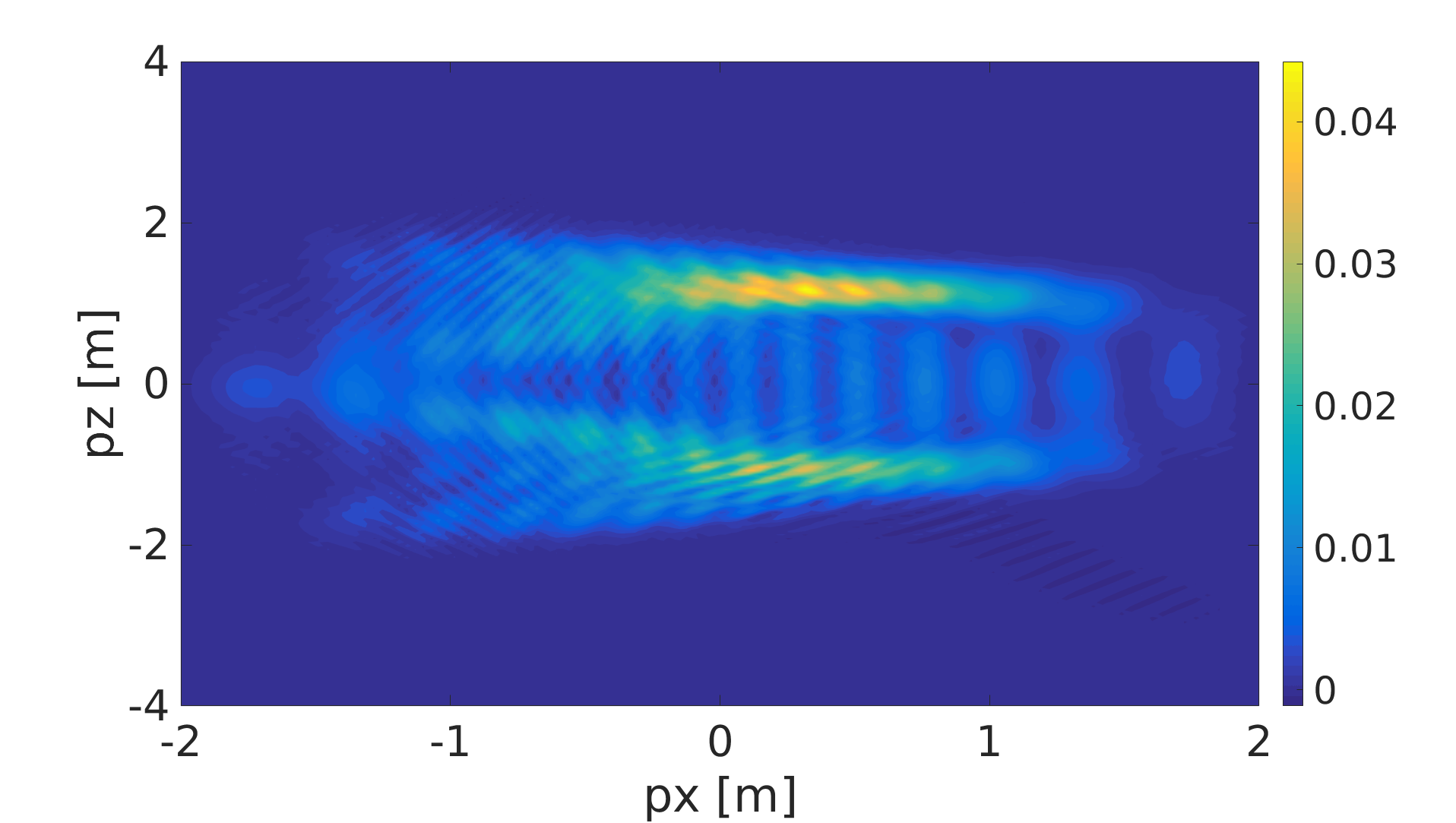} 	
	\includegraphics[trim=0cm 0cm 0cm 0cm,clip,width=\figlenB]{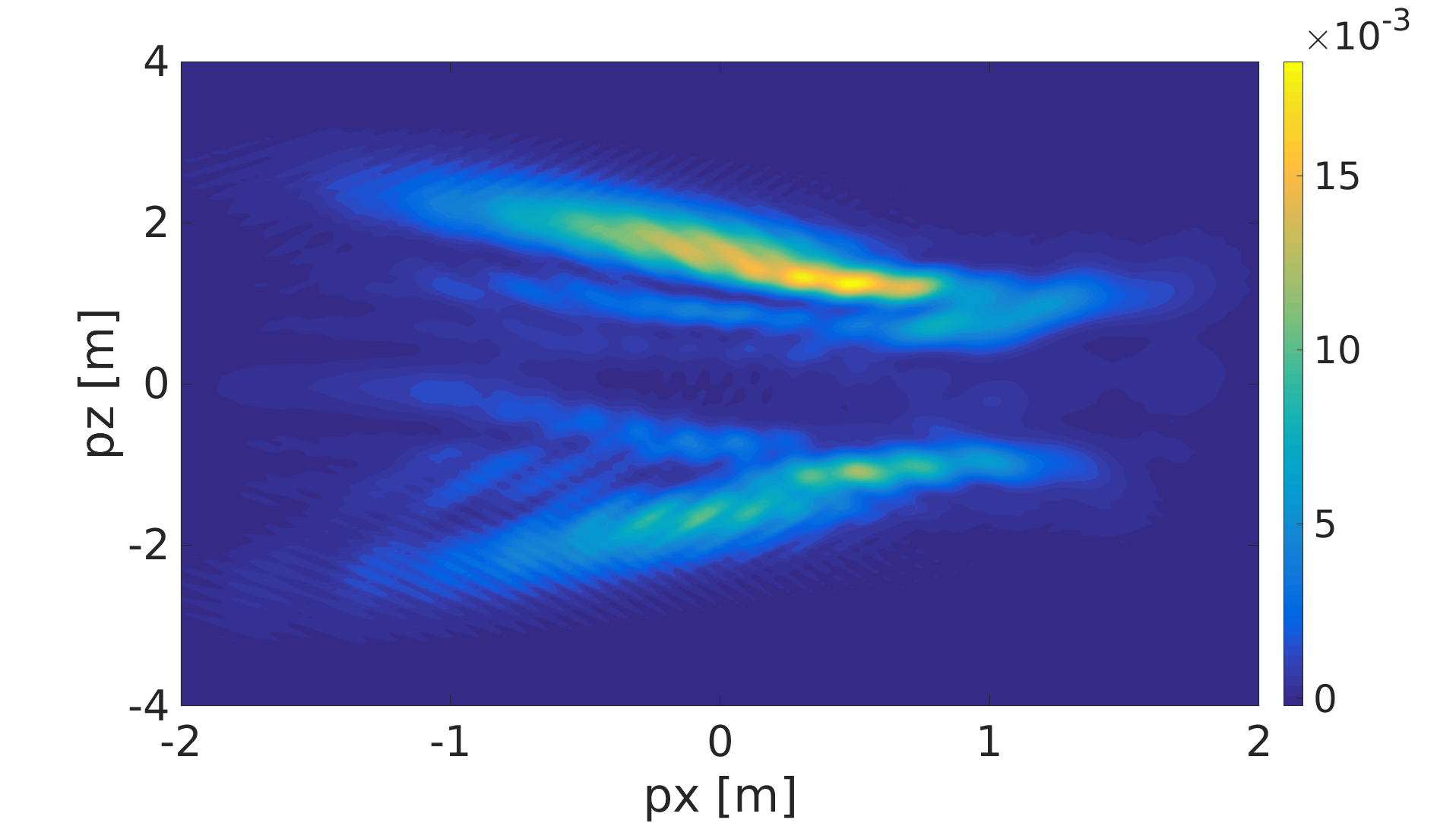} 
	\includegraphics[trim=0cm 0cm 0cm 0cm,clip,width=\figlenB]{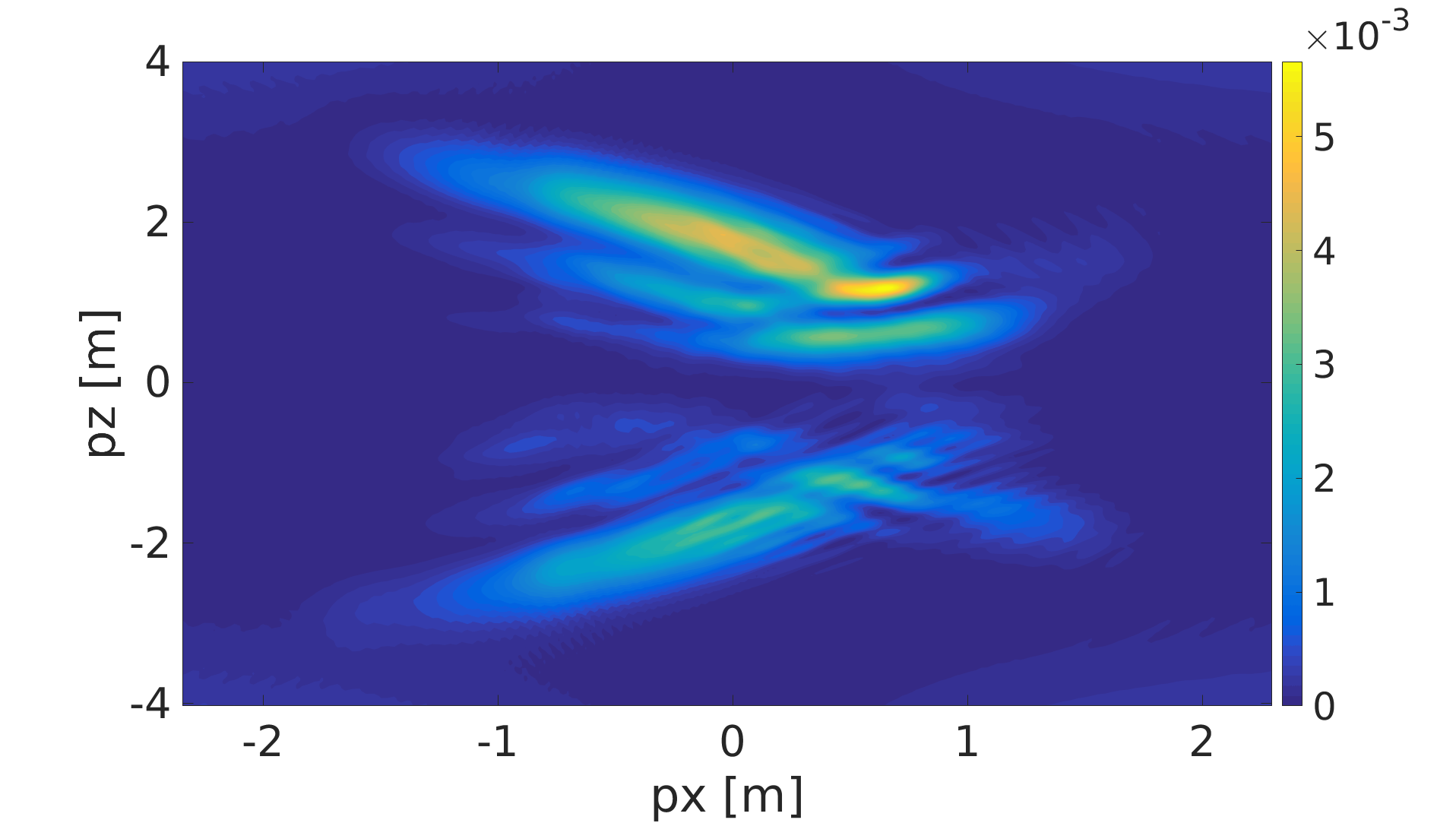} 
      \caption{Particle spectrum as a function of parallel ($p_x$) and perpendicular ($p_z$) momentum for various spatial extents $\lambda$. Particles are created mainly via the
        Schwinger effect. The smaller $\lambda$ the stronger the applied magnetic fields thus the stronger the distortion. Additionally, quantum interference effects vanish.
	Parameters: $e\varepsilon = 0.5m^2$, $\tau = 25m^{-1}$, $\omega = 0.2m$ and, in terms of appearance, $\lambda = 1000m^{-1}$ (top), $20m^{-1}$, $10m^{-1}$, $5m^{-1}$
	and $3m^{-1}$ (bottom). To improve readability we only show absolute values in the last plot.}
\end{figure}    

\begin{figure}
	\includegraphics[width=\figlenB]{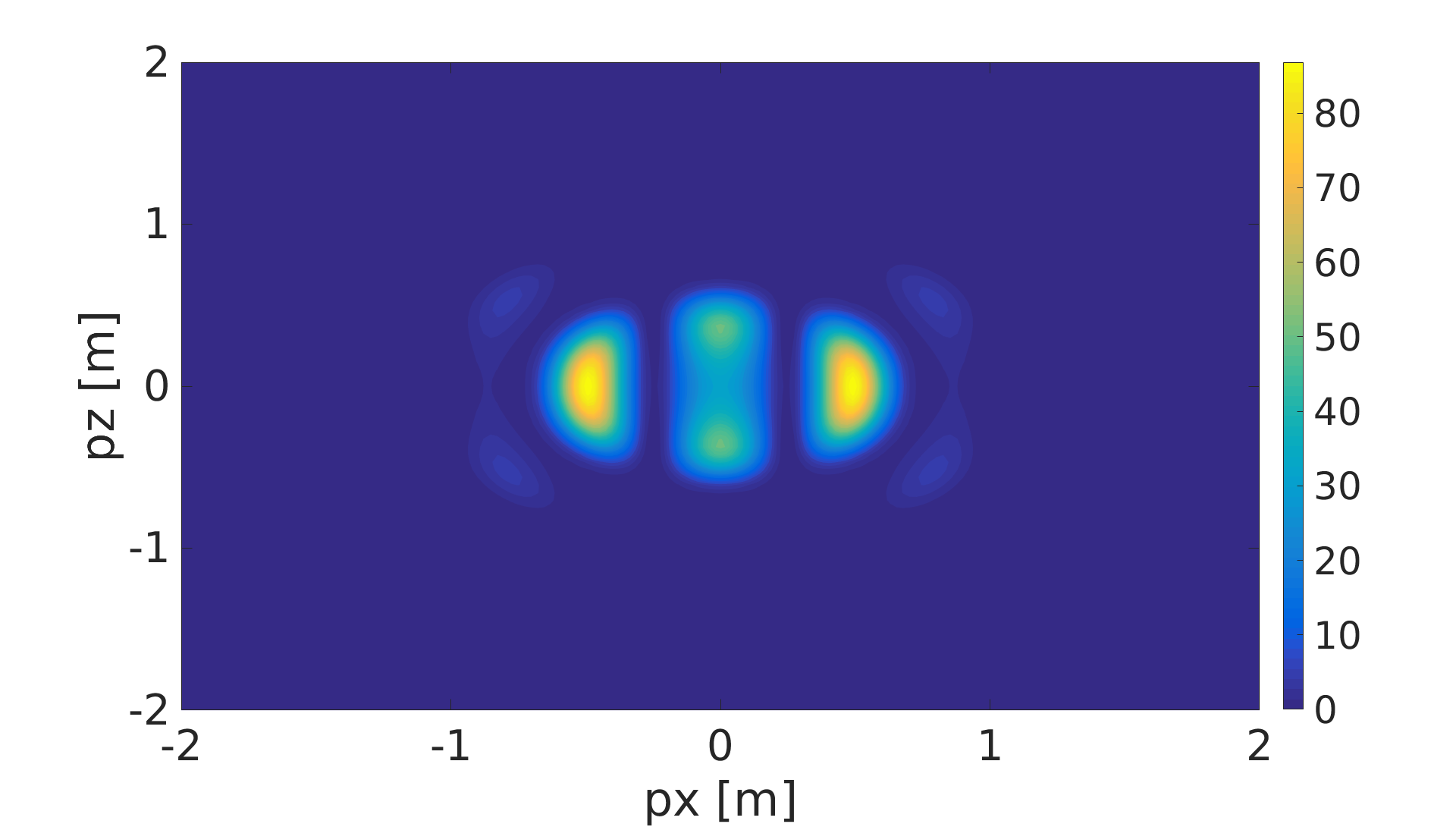} 
	\includegraphics[width=\figlenB]{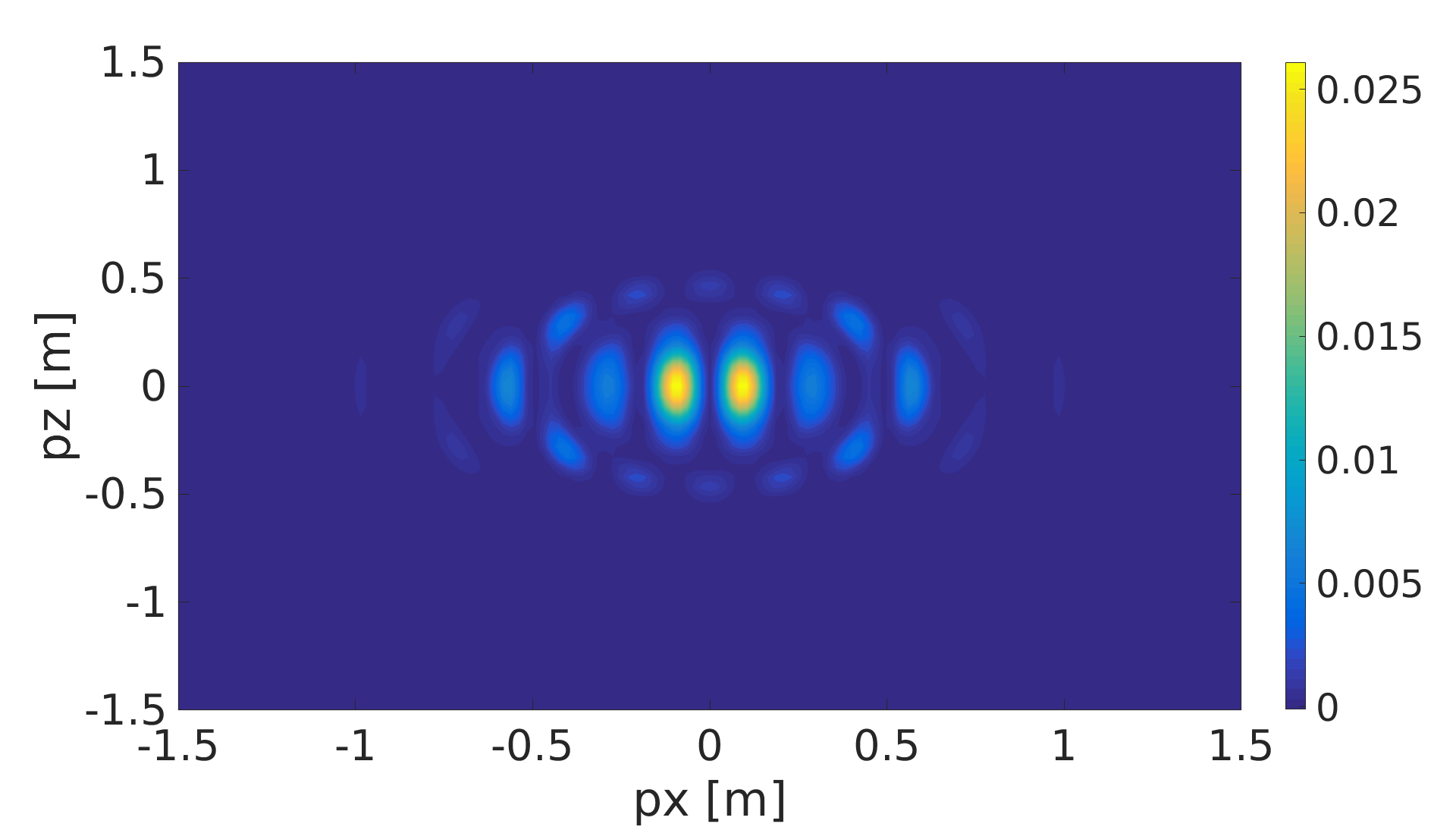} 
	\includegraphics[width=\figlenB]{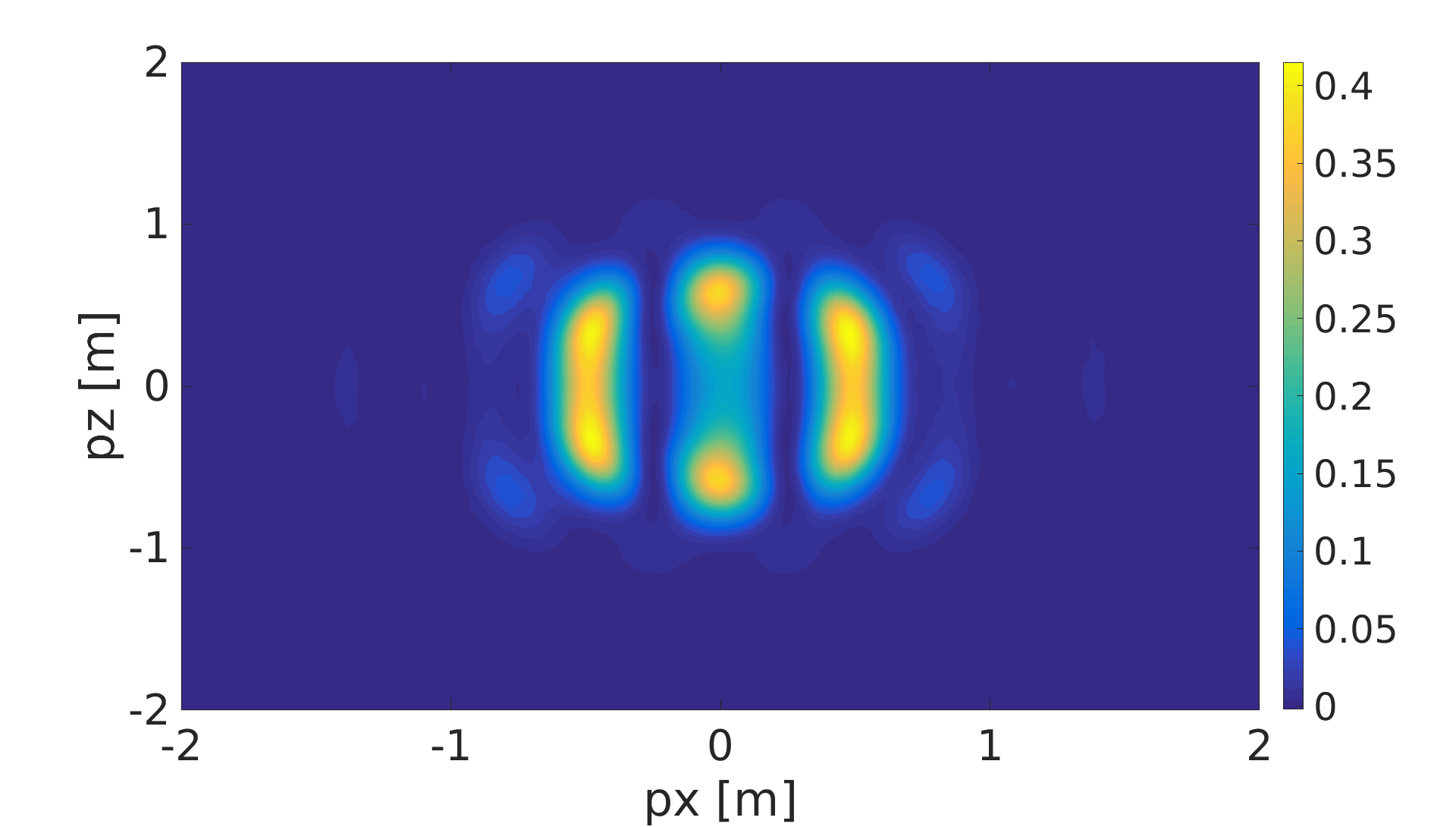}
	\includegraphics[width=\figlenB]{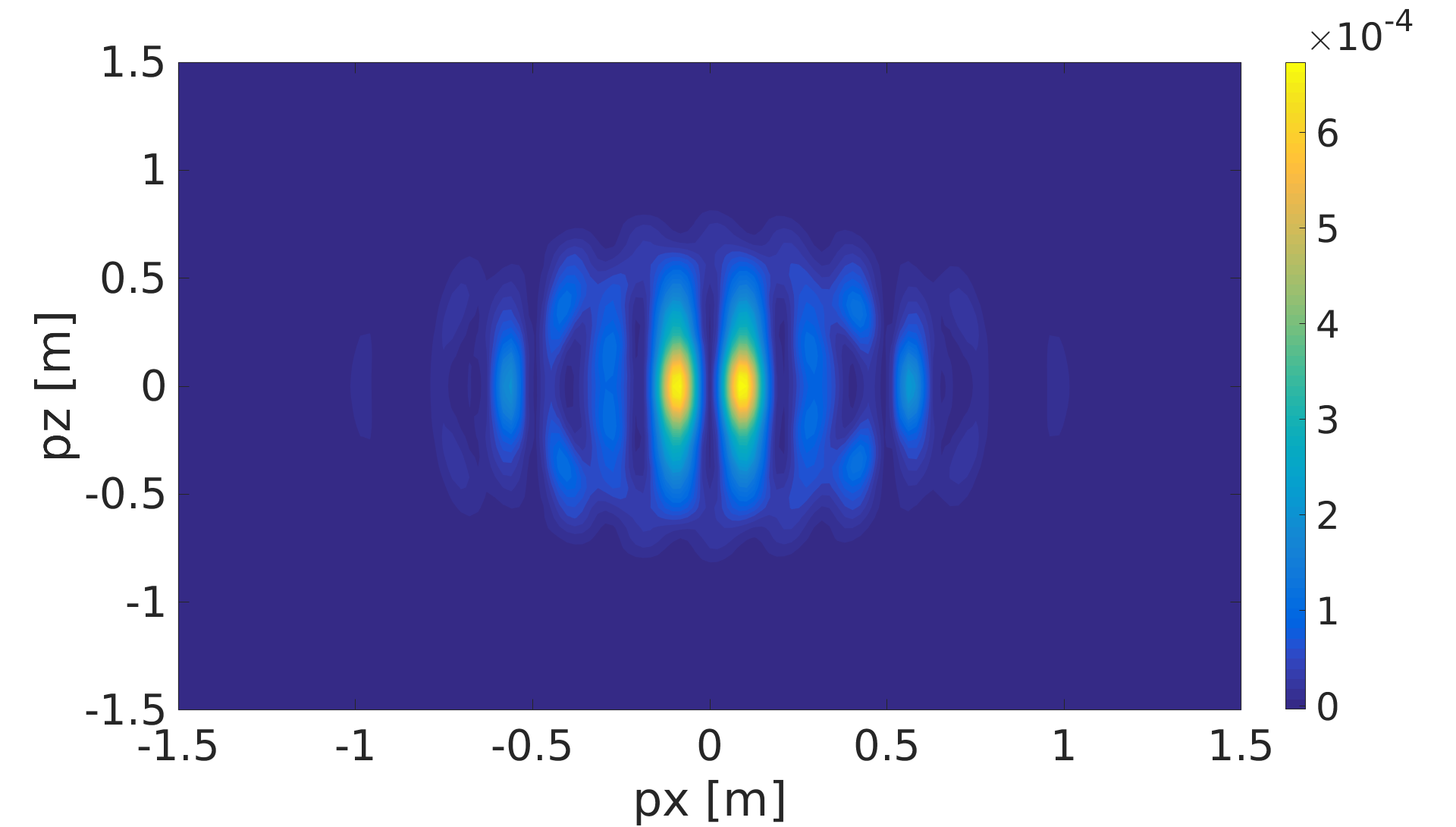} 	
	\includegraphics[width=\figlenB]{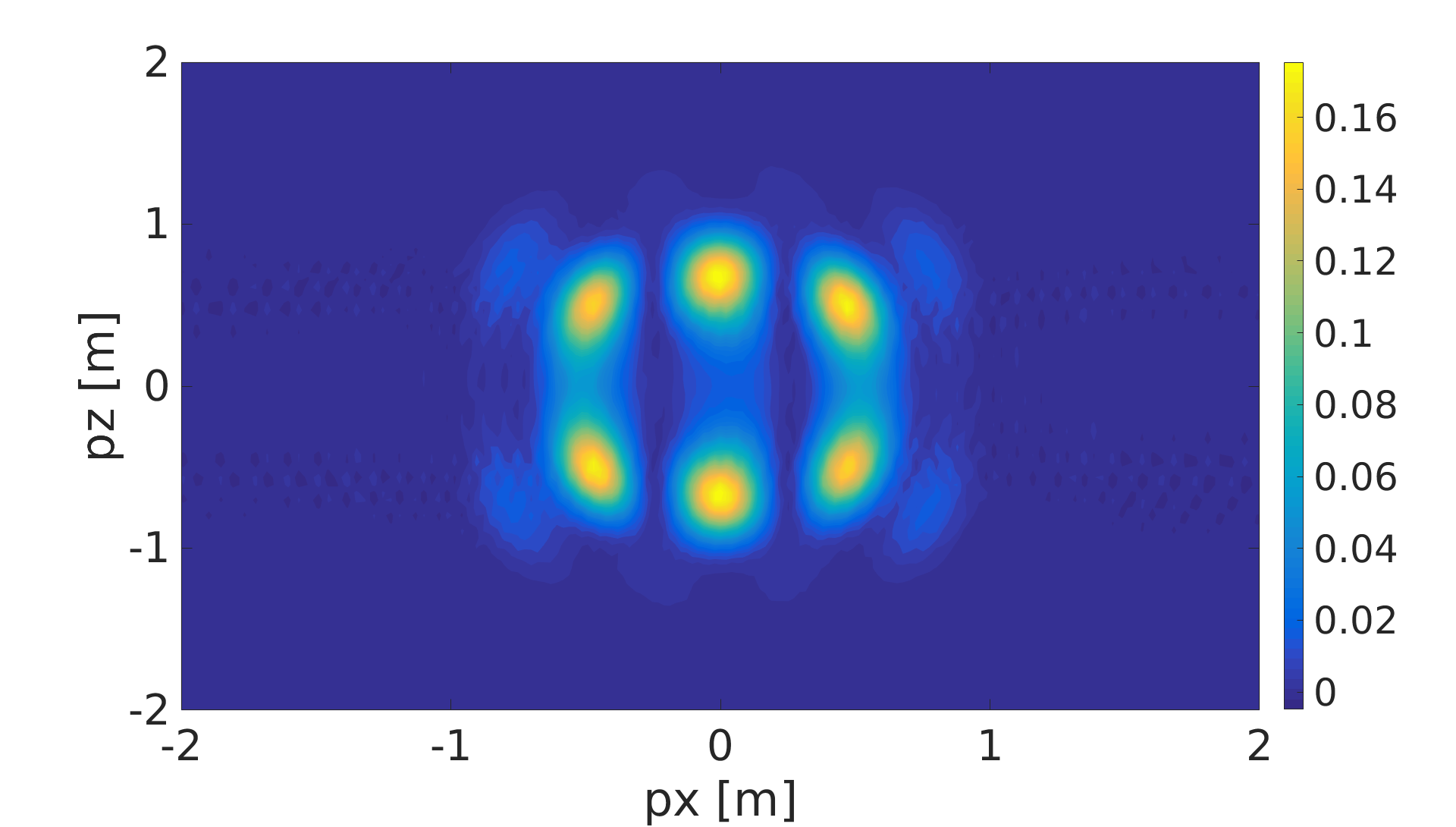}
	\includegraphics[width=\figlenB]{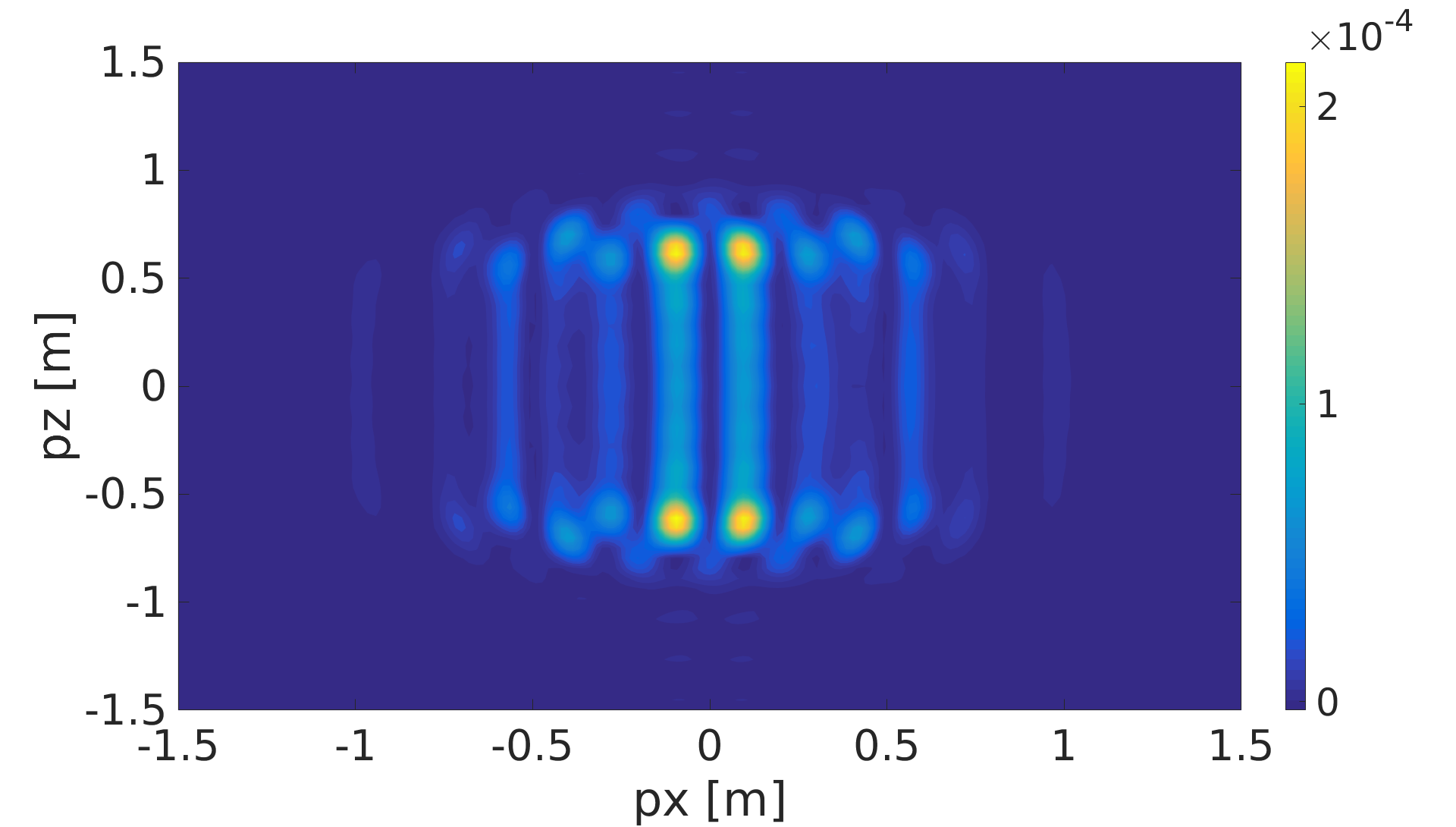} 
	\includegraphics[width=\figlenB]{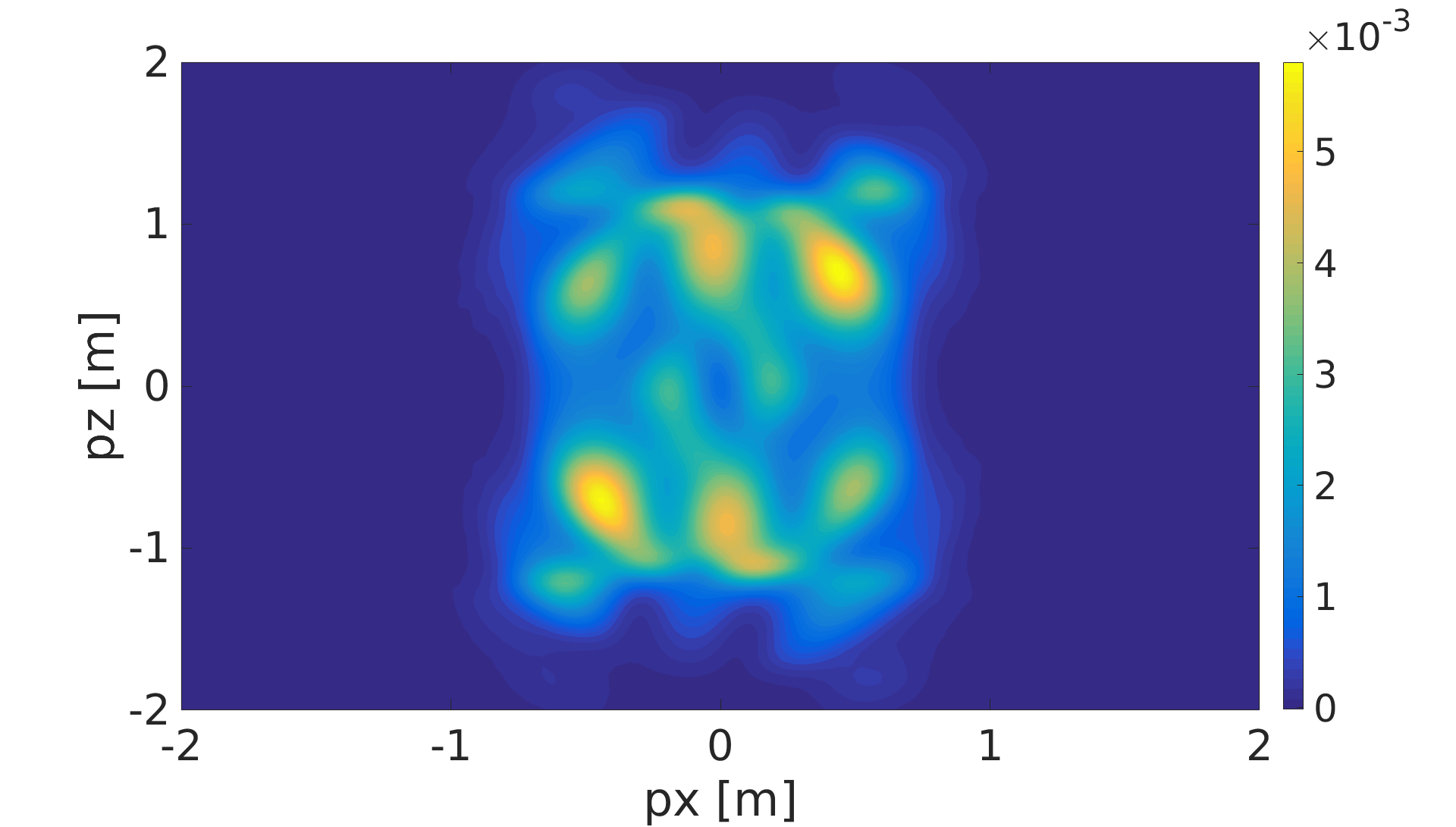}
	\includegraphics[width=\figlenB]{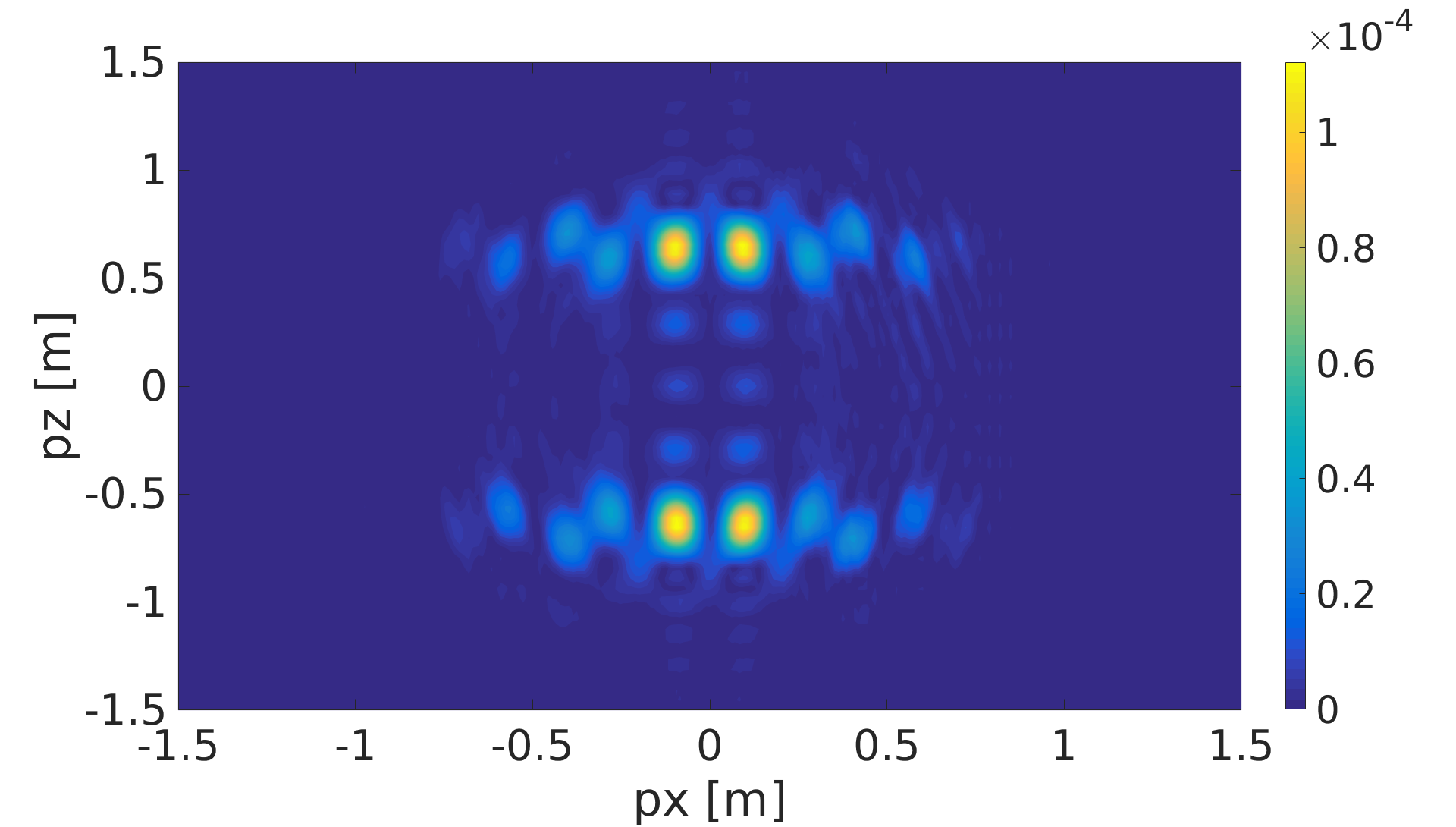} 
      \caption{Left: Particle distribution $f(p_x, p_z)$ for various values of the spatial extent $\lambda$ (and therefore the magnetic peak field strength). We have chosen a peak
        field strength of $e\varepsilon = 0.5m^2$, a pulse length of $\tau = 25m^{-1}$, a field frequency of $\omega = 0.5m$ and a spatial envelope parameter
        of $\lambda = (1000, 10, 5, 1.5)m^{-1}$ (top to bottom). Schwinger as well as
        multiphoton pair production are visible (strong peaks, elliptical shape of the distribution function). A strong magnetic field can deform the particle structure and
        break $p_z$ symmetry. \\
        Right: Particle distribution $f(p_x, p_z)$ for various values of the spatial extent $\lambda = (1000, 50, 20, 10)m^{-1}$ (top to bottom) for 
        field strength $e\varepsilon = 0.2m^2$, pulse length $\tau = 75m^{-1}$ (super-Gaussian envelope) and field frequency $\omega = 0.2m$. Due to the appearance of multiple strong subcycles 
        in the electric field signatures of multiphoton pair production are clearly visible (ring superposed by quantum interferences). A strong magnetic field can destroy the rings,
	but the multipeak structure is still retained. 
	To improve readability we only show absolute values in the last plot.}
\end{figure}
      
      \begin{figure*}[t]
      \begin{center}
	\includegraphics[width=\figlenB]{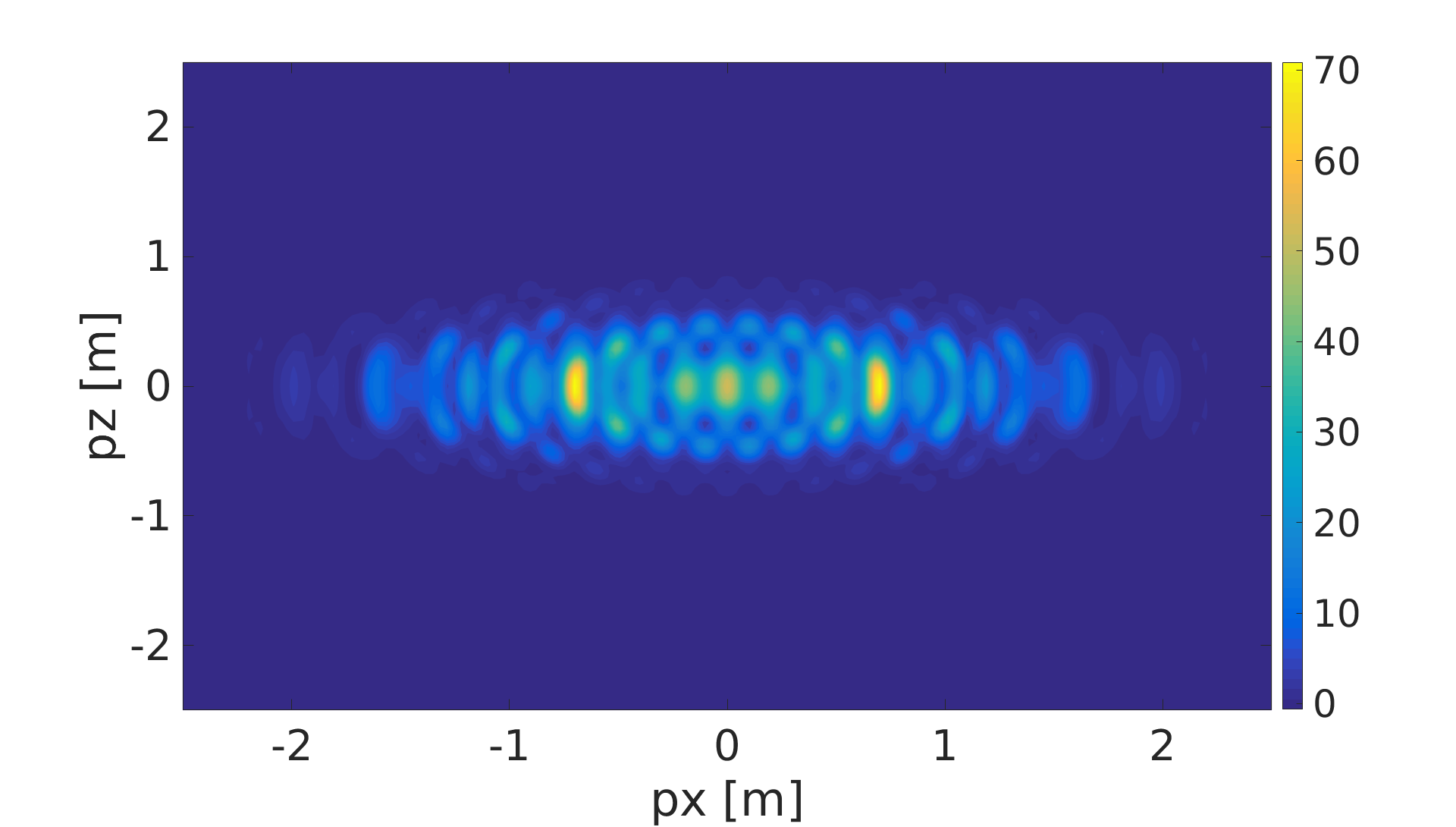} 
	\includegraphics[width=\figlenB]{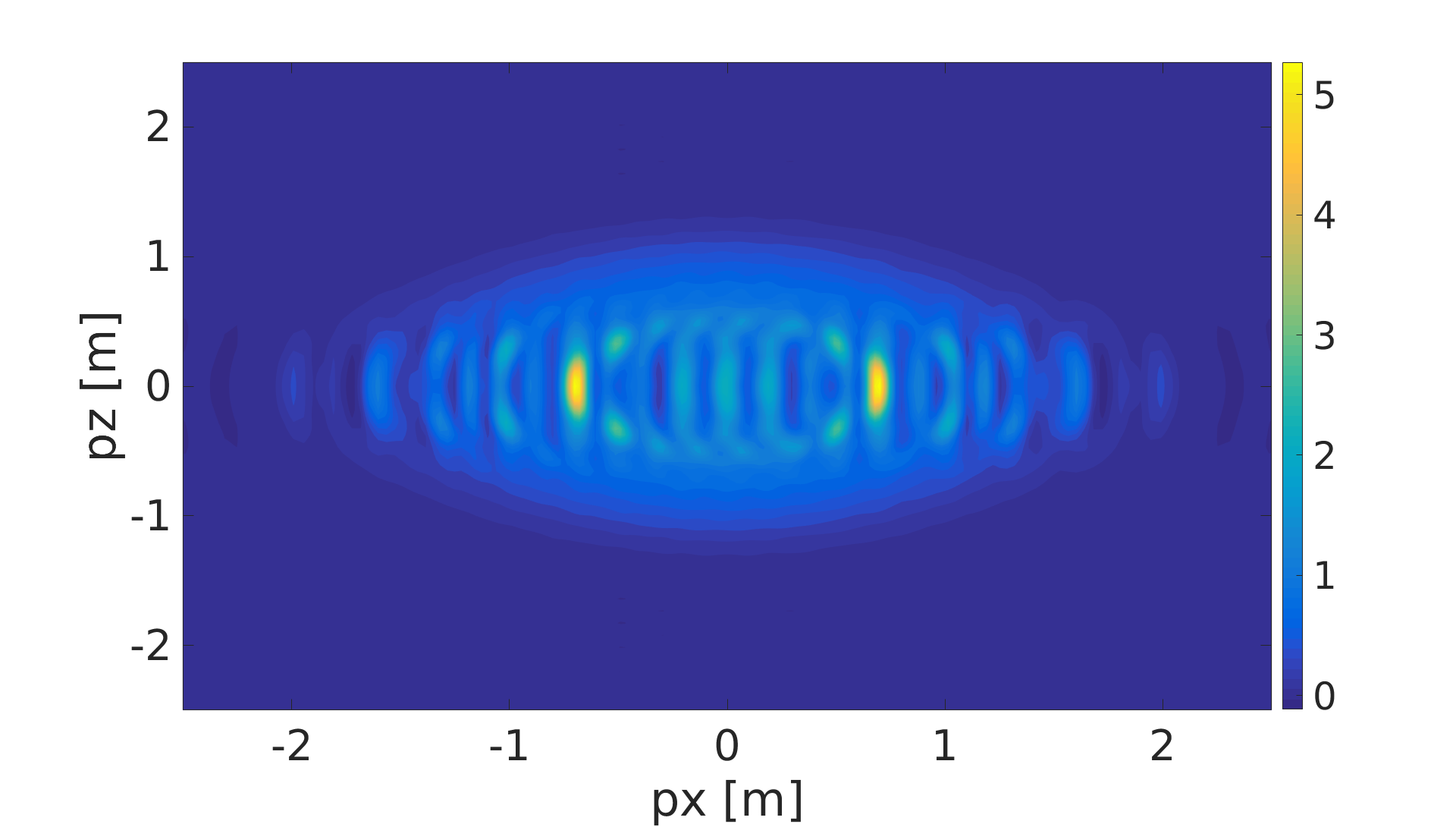} 
	\includegraphics[width=\figlenB]{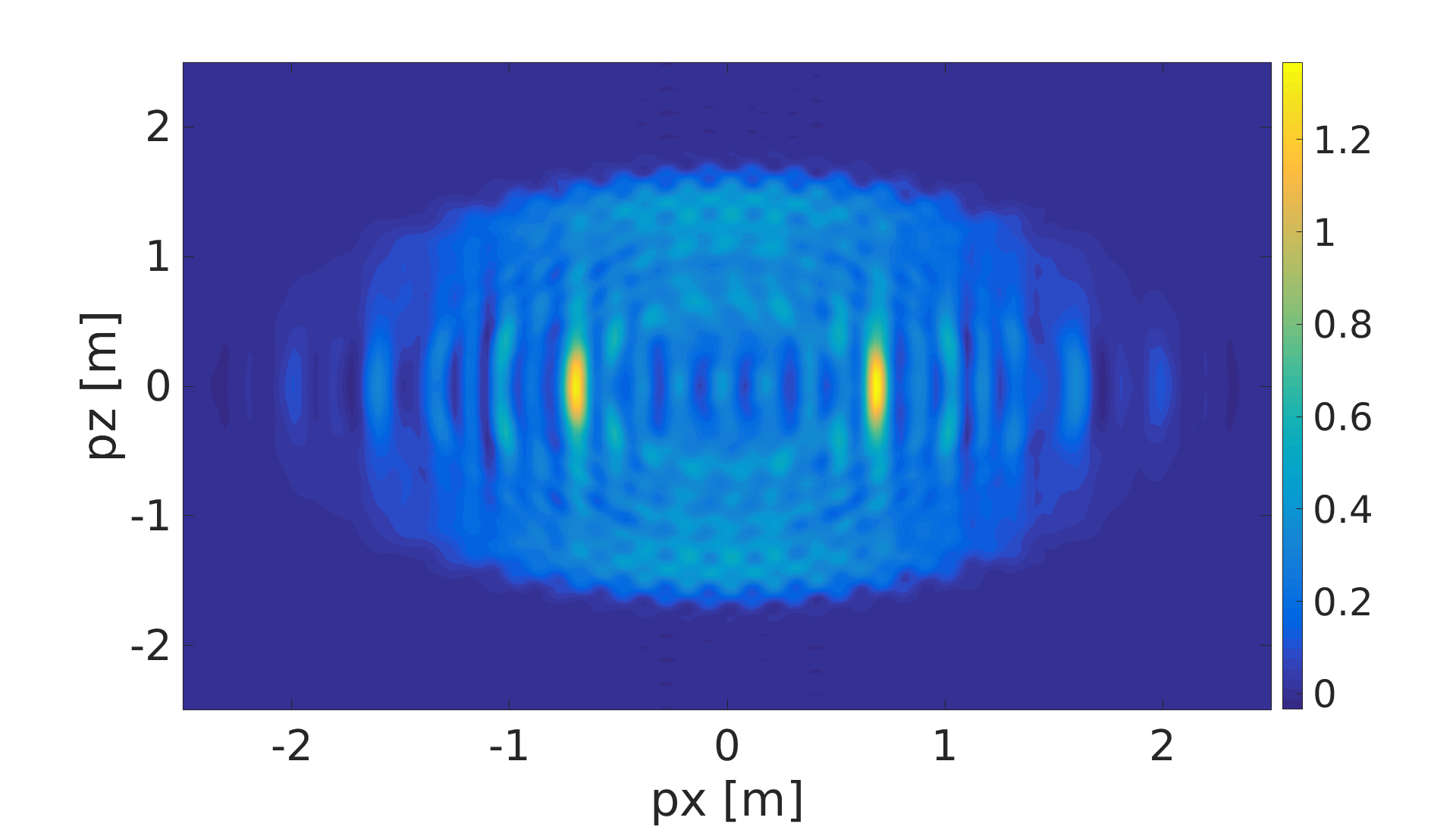} 	
	\includegraphics[width=\figlenB]{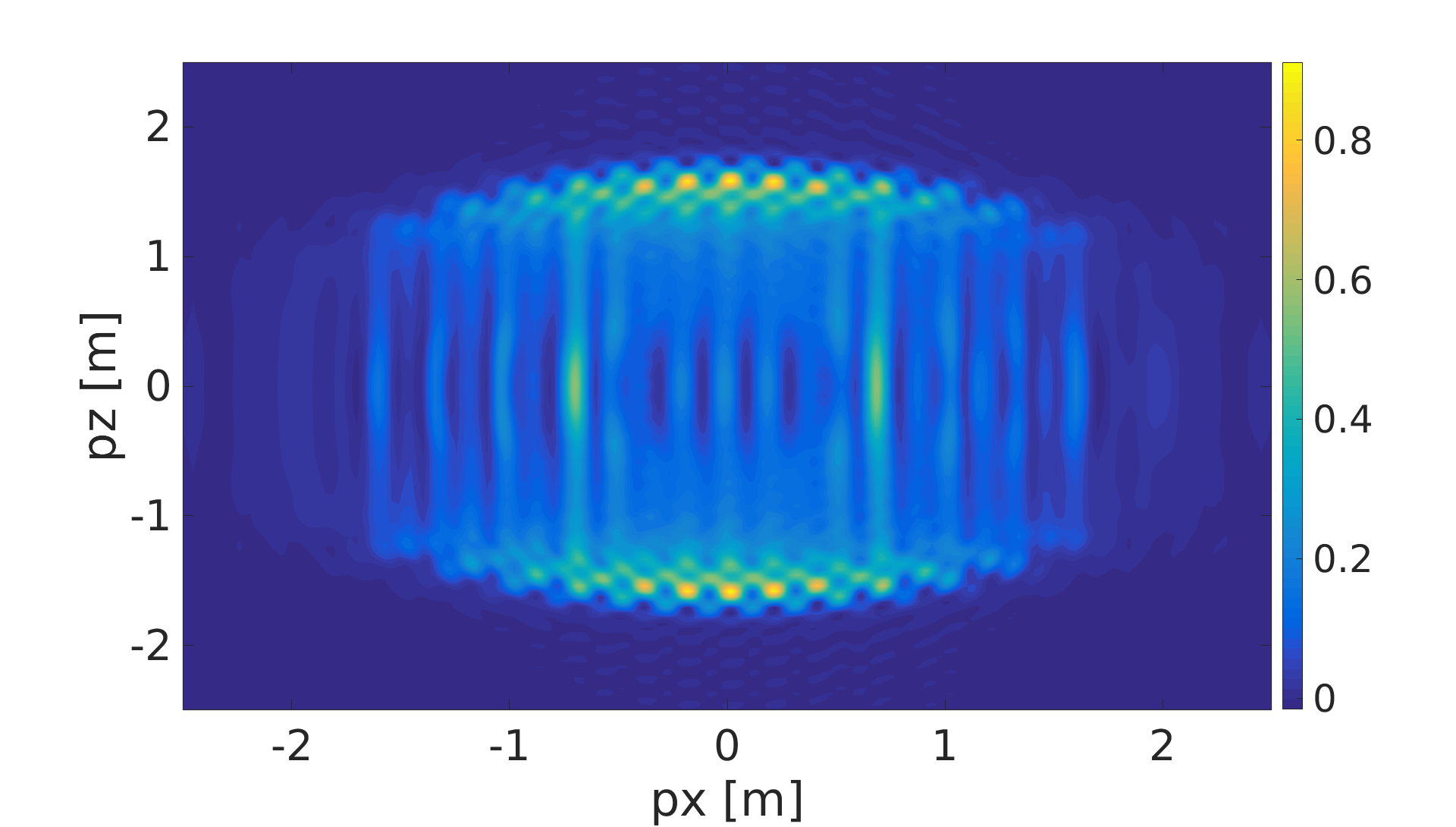} 
	\includegraphics[width=\figlenB]{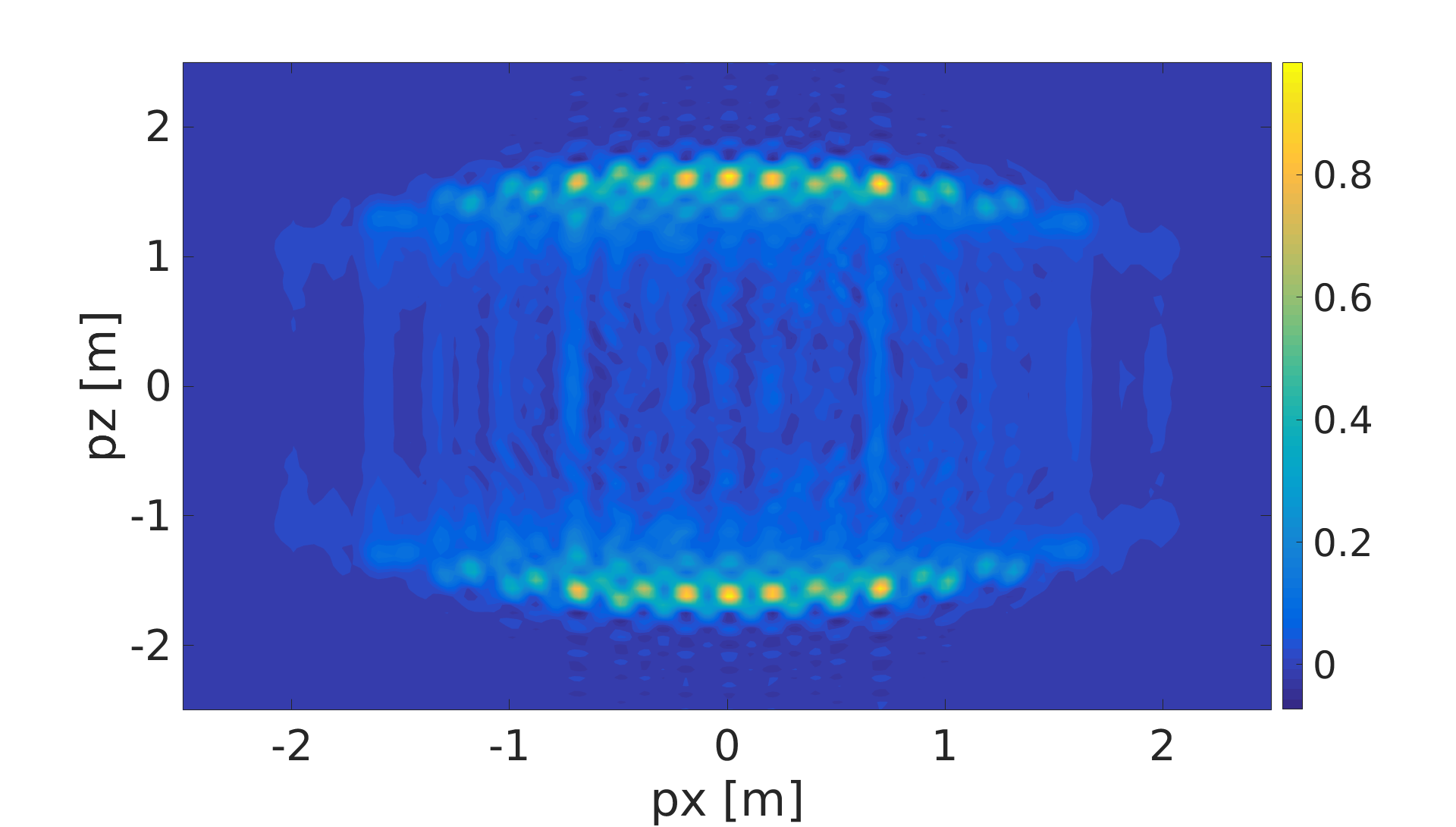} 	
	\includegraphics[width=\figlenB]{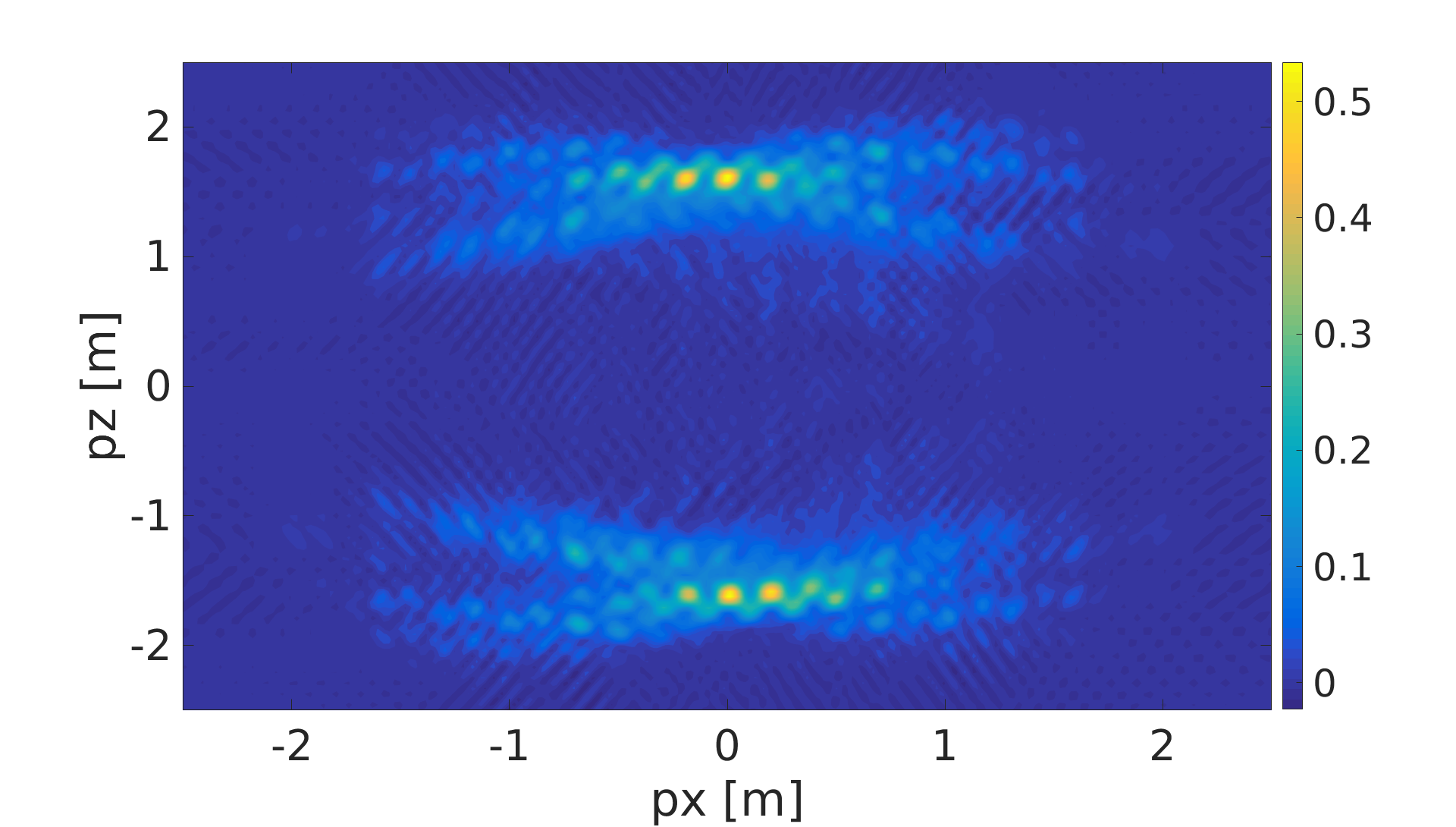} 
	\includegraphics[width=\figlenB]{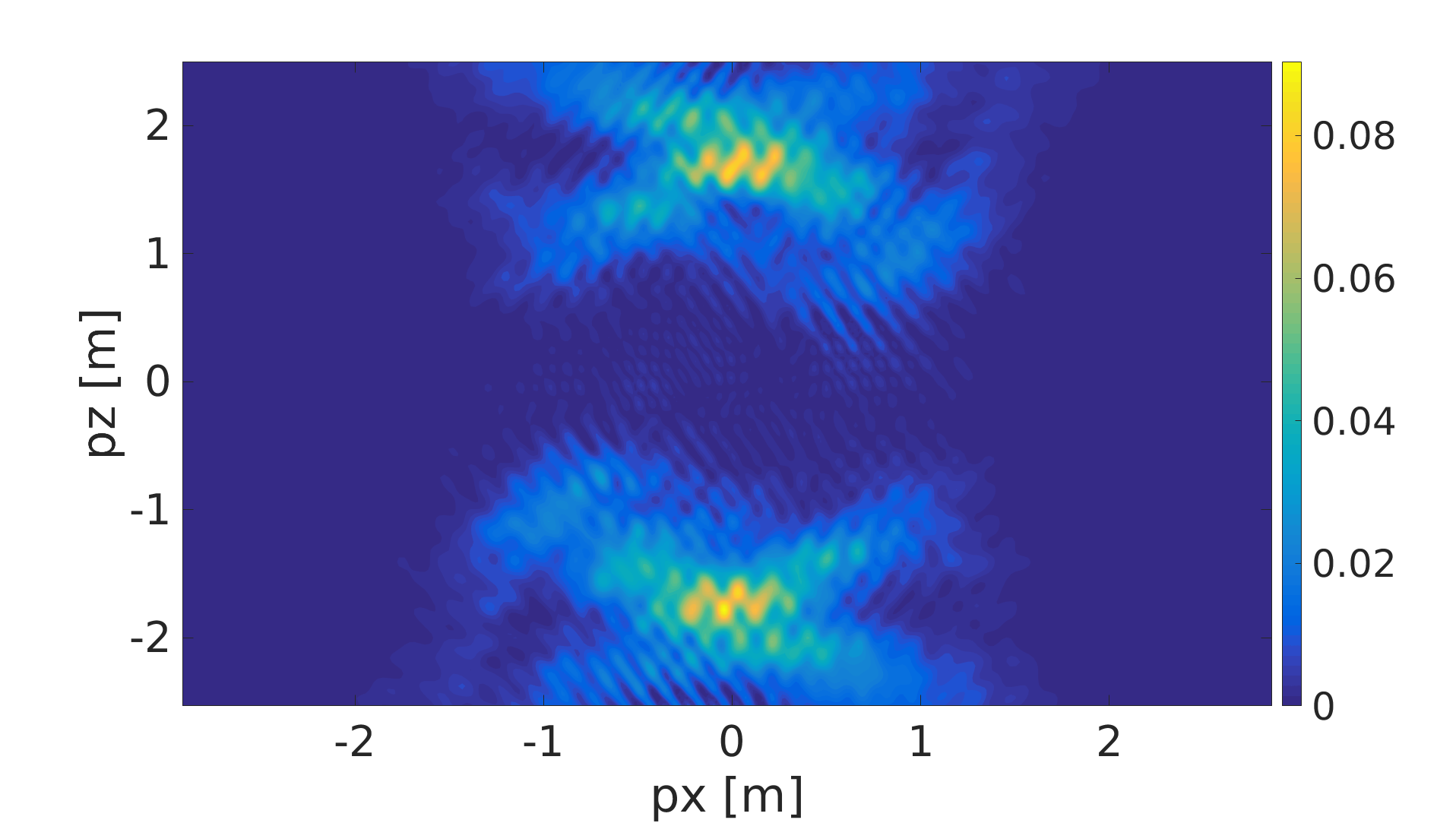} 
      \end{center}
      \caption{Particle distribution $f(p_x, p_z)$ in the intermediate regime for various values of the spatial extent $\lambda = (1000, 100, 50, 35, 20, 10, 5)m^{-1}$ (left to right, top to bottom) for 
        field strength $e\varepsilon = 0.5m^2$, pulse length $\tau = 75m^{-1}$ (super-Gaussian envelope) and field frequency $\omega = 0.2m$. 
        The stronger the magnetic field the more the particles are accelerated in transversal direction $p_z$. 
	To improve readability we only show absolute values in the last plot.} 
      \label{fig:MomPxPz4}
      \end{figure*}              

\end{document}